\documentclass{aa}
\usepackage{graphicx}
\usepackage{natbib}
\bibpunct{(}{)}{;}{a}{}{,}
\usepackage{txfonts}
\usepackage{xcolor}
\newcommand{\kpc}{\,{\rm kpc}}
\newcommand{\ha}{H$\alpha$}
\newcommand{\hb}{H$\beta$}

\newcommand{\kms}{\,km\,s$^{-1}$}  
\newcommand{\myr}{\,$M_{\sun}\,{\rm yr}^{-1}$}
\newcommand{\ro}{\,$R_{\sun}$}
\newcommand{\mo}{\,$M_{\sun}$}
\newcommand{\lo}{\,$L_{\sun}$}
\newcommand{\cmt}{\,cm$^{-3}$}

\newcommand{\cmd}{\,cm$^{-2}$}

\newcommand{\es}{$\rm\,erg\,s^{-1}$}
\newcommand{\ecs}{$\rm\,erg\,s^{-1}\,cm^{-2}$}
\newcommand{\ecsa}{$\rm\,erg\,s^{-1}\,cm^{-2}\,\AA^{-1}$}
\begin{document}

\titlerunning{On the path to Z~And-type outbursts: The case of V426~Sge}


\title{The path to Z~And-type outbursts: 
       The case of V426~Sagittae (HBHA~1704-05)}

\author{A.~Skopal               \inst{\ref{asusav}}
     \and S.~Yu.~Shugarov       \inst{\ref{asusav},\ref{sai}}
     \and U.~Munari             \inst{\ref{asiago}}
\and N.~Masetti    \inst{\ref{mas1},\ref{mas2}}
\and E.~Marchesini \inst{\ref{mar1},\ref{mar2},\ref{mar3},\ref{mar4},\ref{mas1}}
     \and R.~M.~Kom\v{z}\'{\i}k \inst{\ref{asusav}}
     \and E.~Kundra             \inst{\ref{asusav}}
     \and N.~Shagatova          \inst{\ref{asusav}}
     \and T.~N.~Tarasova        \inst{\ref{crao}}
     \and C.~Buil               \inst{\ref{toloson}}
     \and C.~Boussin            \inst{\ref{ocr_fr}}
     \and V.~I.~Shenavrin       \inst{\ref{sai}}
     \and F.-J.~Hambsch         \inst{\ref{anscoll}}
     \and S.~Dallaporta         \inst{\ref{anscoll}}
     \and A.~Frigo              \inst{\ref{anscoll}}
     \and O.~Garde              \inst{\ref{garde}}
     \and A.~Zubareva           \inst{\ref{iaras},\ref{sai}}
     \and P.~A.~Dubovsk\'y      \inst{\ref{kolonica}}
     \and P.~Kroll              \inst{\ref{kroll}}
}
\institute{Astronomical Institute, Slovak Academy of Sciences, 
           059\,60 Tatransk\'{a} Lomnica, Slovakia\label{asusav}
\and
Sternberg Astronomical Institute, Moscow State University, 
Universitetskij pr., 13, Moscow, 119991, Russia\label{sai}
\and
INAF - Osservatorio Astronomico di Padova, 36012 Asiago (VI), 
Italy\label{asiago}
\and
INAF - Osservatorio di Astrofisica e Scienza dello Spazio, 
via Gobetti 93/3, I-40129, Bologna, Italy\label{mas1}
\and
Departamento de Ciencias F\'isicas, Universidad Andr\'es Bello, 
Fern\'andez Concha 700, Las Condes, Santiago, Chile\label{mas2}
\and
Dipartimento di Fisica, Universit\`a degli Studi di Torino, 
via Pietro Giuria 1, I-10125 Torino, Italy\label{mar1}
\and
INFN - Istituto Nazionale di Fisica Nucleare, Sezione di Torino, 
via Pietro Giuria 1, I-10125 Turin, Italy\label{mar2}
\and
Facultad de Ciencias Astron\'omicas y Geof\'isicas, Universidad 
Nacional de La Plata, Paseo del Bosque, B1900FWA, La Plata, 
Argentina\label{mar3}
\and
Instituto de Astrof\'isica de La Plata, CONICET-UNLP, CCT La Plata, 
Paseo del Bosque, B1900FWA, La Plata, Argentina\label{mar4}
\and
Scientific Research Institute, Crimean Astrophysical Observatory, 
298409 Nauchny, Crimea\label{crao}
\and
Castanet Tolosan Observatory, 6 place Clemence Isaure,
31320 Castanet Tolosan, France\label{toloson}
\and
Observatoire de l'Eridan et de la Chevelure de B\'er\'enice, 
02400 Epaux-B\'ezu, France\label{ocr_fr}
\and
ANS Collaboration, c/o Astronomical Observatory, 36012 Asiago (VI), 
Italy\label{anscoll}
\and 
Observatoire de la Tourbi\`ere, 38690 Chabons, France\label{garde}
\and
Institute of Astronomy, Russian Academy of Sciences, Russia\label{iaras}
\and
Vihorlat Astronomical Observatory, Mierov\'a 4, 066\,01 Humenn\'e, 
Slovakia\label{kolonica}
\and
Sonneberg Observatory, Sternwartestr. 32, 96515 Sonneberg, 
Germany\label{kroll}
}
\date{Received / Accepted}

\abstract
 {
The star V426~Sge (HBHA~1704-05), originally classified as an 
emission-line object and a semi-regular variable, brightened 
at the beginning of August 2018, showing signatures of 
a symbiotic star outburst. 
 }
 {
We aim to confirm the nature of V426~Sge as a classical symbiotic 
star, determine the photometric ephemeris of the light minima, 
and suggest the path from its 1968 symbiotic nova outburst 
to the following 2018 Z~And-type outburst. 
 }
 {
We re-constructed an historical light curve (LC) of V426~Sge 
from approximately the year 1900, and used original 
low- (R$\sim$500--1\,500; 330--880\,nm) and high-resolution 
(R$\sim$11\,000--34\,000; 360--760\,nm) spectroscopy 
complemented with {\em Swift}-XRT and UVOT, optical 
$UBVR_{\rm C}I_{\rm C}$ and near-infrared $JHKL$ photometry 
obtained during the 2018 outburst and the following quiescence. 
 }
 {
The historical LC reveals no symbiotic-like activity from 
$\sim$1900 to 1967. In 1968, V426~Sge experienced a symbiotic 
nova outburst that ceased around 1990. From approximately 1972, 
a wave-like orbitally related variation with a period of 
$493.4\pm 0.7$ days developed in the LC. This was interrupted 
by a Z~And-type outburst from the beginning of August 2018 
to the middle of February 2019. At the maximum of the 2018 
outburst, the burning white dwarf (WD) increased its 
temperature to $\ga 2\times 10^5$\,K, generated a luminosity 
of $\sim7\times 10^{37}\,(d/3.3\,{\rm kpc})^2$\es\, and blew 
a wind at the rate of $\sim3\times 10^{-6}$\myr. 
Our spectral energy distribution models from the current quiescent 
phase reveal that the donor is a normal M4-5\,{\small III} giant 
characterised with 
$T_{\rm eff}\sim 3\,400$\,K, 
$R_{\rm G}\sim 106\,(d/3.3\,{\rm kpc})$\ro\ and 
$L_{\rm G}\sim 1350\,(d/3.3\,{\rm kpc})^2$\lo\ 
and the accretor is a low-mass $\sim$0.5\mo\ WD. 
 }
 {
During the transition from the symbiotic nova outburst to 
the quiescent phase, a pronounced sinusoidal variation along 
the orbit develops in the LC of most symbiotic novae. 
The following eventual outburst is of Z~And-type, when 
the accretion by the WD temporarily exceeds the upper limit 
of the stable burning. 
At this point the system becomes a classical symbiotic star. 
}
\maketitle
%
%
\section{Introduction}
\label{s:intro}
Symbiotic stars are the widest interacting binaries. Their
orbital periods run from hundreds of days to a few tens or 
even hundreds of years, or are unknown 
\citep[e.g.][]{belczynski+00,parimucha+02,schmid+schild02,matt+karov06}. 
This type of binary consists of a cool giant as the donor star 
and a white dwarf (WD), accreting from the giant's wind 
\citep[e.g.][]{boy67,kenyon86,ms99}. 
The accretion process heats up the WD to $\ga 10^5$\,K and increases 
its luminosity to $\sim 10^1-10^4$\lo\ \citep[][]{mu+91,sk05a}, which 
ionises a fraction of the wind from the giant giving rise to 
the nebular emission \citep[e.g.][]{boy+66,stb,nussvog87}. 
This configuration represents the so-called {\it quiescent phase}, 
during which symbiotic systems releases their energy at 
an approximately constant rate and spectral energy distribution (SED). 
The presence of an extended and partially optically thick nebula 
in the rotating binary gives rise to the wave-like variation 
of the light as a function of the orbital phase, which represents 
the most distinctive feature in the light curves (LCs) of symbiotic 
stars during their quiescent phases 
\citep[see][ and references therein]{sk01,sekeras+19}. 
In most cases, high luminosities of a few times $10^3$\lo\ 
are generated by stable hydrogen burning on the WD surface at 
accretion rates of a few times ($10^{-8} - 10^{-7})$\myr, 
depending on the WD mass \citep[e.g.][]{paczyt78,shen+bild07}. 
In some cases we measure low luminosities of $\sim 10^1 - 10^2$\lo, 
which are generated by the accretion process onto the WD, 
when its gravitational potential energy is converted into radiation 
by the disc \citep[e.g.][]{pringle81}. 

%
Sometimes, symbiotic systems undergo outbursts indicated by 
brightening of a few magnitudes in the optical. Outbursts, 
resulting from a prolonged accretion by the WD until ignition 
of a thermonuclear event, can be grouped into two classes: (i) 
The {\it symbiotic nova} outbursts occur when the nuclear 
burning at the WD surface turns on in non-degenerate conditions, 
is non-explosive, and proceeds in thermal equilibrium 
\citep[see][]{fujimoto82a,fujimoto82b}. 
The LC is characterised by a moderate amplitude (a few magnitudes) 
and slow (a few years) rise to maximum and long duration 
(from decades to centuries or more). More details can be found in 
\cite{allen80}, \cite{munari97}, \cite{mn94}, and \cite{munari19}. 
(ii) The other class occur when the material is accreting onto 
the WD under degenerate conditions, and the thermonuclear 
outburst is explosive. The outburst reaches peak luminosity within 
a few minutes, ejects a large amount of mass at high velocities, 
and lasts for weeks or months. If the accretion runs at a high 
rate onto a massive WD, a thermonuclear outburst can be recorded 
more than once during the history of observations. These events 
are called {\it recurrent (symbiotic) novae} 
\citep[e.g.][ for a review]{bode08,starrfield+16}. 
Finally, there are `Z-And type' outbursts that can result 
from an increase in the accretion rate above that sustaining 
the stable burning, which leads to expansion of the burning 
envelope simulating an A--F type pseudophotosphere 
\citep[e.g.][]{pacrud80} and/or blowing optically thick wind 
from the WD \citep[][]{hachisu+96}. The latter was recently 
demonstrated for the 2015 outburst of AG~Peg \citep[][]{skopal+17}. 
This type of outburst shows 1--3\,mag brightening in the optical 
evolving on a timescale of weeks to years, often prolonged 
with multiple re-brightenings 
\citep[see e.g. Fig.~1 of][]{skopal+18}, and signatures of 
enhanced mass outflow \citep[e.g.][]{f-c+95}. This stage is 
called the {\it active phase} of the symbiotic 
binary. 

Based on disentangling the composite continuum of symbiotic stars, 
\cite{sk05a} classified Z~And-type outbursts into two classes
depending on the orbital inclination: 
Systems with a high orbital inclination show 
the 1st-type outburst; their spectra consist of a low-temperature 
($1-2\times 10^4$\,K) warm pseudophotosphere and a strong nebular 
continuum. Systems with a low orbital inclination exhibit 
the 2nd-type outburst; their spectra consist of a high-temperature 
($\approx 165\,000$\,K) stellar component and a strong nebular 
continuum. 
This classification is based on a disc-like structure around 
the burning WD created during Z~And-type outbursts at the orbital 
plane \citep[][]{sk05a,cs12}. The flared outer rim of the disc 
(which is the warm pseudophotosphere) then occults the central 
ionising source on the line of sight for systems with a high 
orbital inclination, which gives rise to the spectrum observed 
during the 1st-type outbursts. On the other hand, for low-inclination 
systems we can directly see the hot central source, and thus 
measure spectrum with characteristics of the 2nd-type 
outbursts. The nebula above and below the disc is observable 
in both cases. 

The star V426~Sge \citep[see][ for the given name]{kazarovets+19}
was originally classified as an emission-line star 
HBHA~1704-05 by \cite{1999A&AS..134..255K}. Based on measurements 
obtained within The All Sky Automated Survey for SuperNovae 
\citep[ASASSN, see][]{2014ApJ...788...48S,2018MNRAS.477.3145J}, 
the object was catalogued in the Variable 
Star Index as a semi-regular variable with a periodicity of around 
418 days\footnote{\tiny{
https://www.aavso.org/vsx/index.php?view=detail.top\&oid=572102}}.
According to "Transient Object Followup Reports" of CBAT
\footnote{\tiny{
http://www.cbat.eps.harvard.edu/unconf/followups/J19544251+1722281.html}}
the star TCP J19544251+1722281 brightened from $V = 12.0$ on 
July 31.945, 2018, to $V = 10.7$ on August 8.938, 2018. 
The brightening was confirmed by the ASASSN measurements for 
the SR variable ASASSN-V J195442.95+172212.6 as a possible 
symbiotic star in outburst. 
On August 11, 2018, \cite{2018ATel11937....1M} reported that both 
these stars are coincident with the position of 
HBHA~1704-05, a symbiotic star undergoing a `hot-type' 
outburst\footnote{\tiny{
i.e., consistent with the 2nd-type outbursts as described above}}.

Here, we explore the historical LC of V426~Sge, determine 
characteristics of its 2018 outburst and, with the aid 
of similar behaviour of other symbiotic novae, we suggest 
the path from the symbiotic nova outburst to the first Z~And-type 
outburst. For this purpose, we re-construct the historical LC 
from approximately the year 1900 to the present and use our 
high-cadence optical spectroscopy complemented with {\em Swift}-XRT 
and UVOT, optical $UBVR_{\rm C}I_{\rm C}$ and near-infrared (NIR) 
$JHKL$ photometry. 
We introduce our observations in Sect.~\ref{s:obs}. We describe 
our analysis and present our results in Sect.~\ref{s:analysis}. 
A discussion and summary are found in 
Sects.~\ref{s:dis} and \ref{s:sum}, respectively. 
%
%
\begin{table}
\caption{Log of {\em Swift}-XRT observations of V426~Sge during 
its 2018 outburst and following quiescent phase. Count rates and 
fluxes are in 10$^{-3}$cts\,s$^{-1}$ and 10$^{-13}$\ecs, respectively 
(see Sect.~\ref{sss:xphot}). 
}
\label{tab:resx}
\begin{center}
\begin{tabular}{ccccc}
\hline
\hline
\noalign{\smallskip}
Date & $T_{\rm start}$ & $t_{\rm exp}$ & Count-rate & Flux \\
     & (UT)            & (ks)          &           &      \\
\noalign{\smallskip}
\hline
\noalign{\smallskip}
Aug. 28, 2018 & 17:03 & 1.0 & 15.6$\pm$4.9 & 3.7$\pm$1.2 \\
Sep. 11, 2018 & 10:47 & 0.8 &  9.7$\pm$4.2 & 2.3$\pm$1.0 \\
Sep. 18, 2018 & 02:23 & 1.1 & 12.0$\pm$3.8 & 2.9$\pm$0.9 \\
Oct. 06, 2018 & 10:25 & 2.0 &  6.7$\pm$2.2 & 1.6$\pm$0.5 \\
Oct. 11, 2018 & 01:46 & 2.3 &  9.0$\pm$2.3 & 2.2$\pm$0.6 \\
Mar. 03, 2019 & 04:18 & 2.8 &       $<$2.2 &      $<$0.6 \\
\noalign{\smallskip}
\hline
\end{tabular}
\end{center}
\end{table}
%
%
\begin{table}
\caption{As in Table~\ref{tab:resx}, but for {\it Swift}-UVOT 
  observations and fluxes in 10$^{-14}$\ecsa. The reported 
  magnitudes are in the Vega system. }
\label{tab:resuv}
\begin{center}
\begin{tabular}{@{~~}c@{~~~}c@{~~~}c@{~~~}r@{~~~}c@{~~~}c@{~~}}
\hline
\hline
\noalign{\smallskip}
Filter & Date & $T_{\rm start}$ & $t_{\rm exp}$ & Mag. & Flux \\
       &      & (UT)            & (s)           &      &      \\
\noalign{\smallskip}
\hline
\noalign{\smallskip}
UVW2 & Aug. 28, 2018 & 17:19 &   81 & 11.49$\pm$0.03 & 13.6$\pm$0.4  \\
UVW2 & Sep. 11, 2018 & 10:50 &  800 & 11.77$\pm$0.02 & 10.5$\pm$0.2  \\
UVW2 & Sep. 18, 2018 & 02:25 & 1032 & 11.85$\pm$0.02 &  9.7$\pm$0.2   \\
UVW2 & Oct. 06, 2018 & 10:28 &  620 & 12.06$\pm$0.02 & 8.04$\pm$0.16 \\
UVW2 & Oct. 11, 2018 & 02:03 &  173 & 12.10$\pm$0.03 & 7.72$\pm$0.18 \\
UVW2 & Mar. 03, 2019 & 04:18 &  550 & 13.67$\pm$0.04 & 1.82$\pm$0.06 \\
\noalign{\smallskip}
\hline
\noalign{\smallskip}
UVM2 & Aug. 28, 2018 & 16:58 &  128 & 11.58$\pm$0.03 & 10.8$\pm$0.3  \\
UVM2 & Oct. 06, 2018 & 10:34 &  620 & 12.16$\pm$0.02 & 6.37$\pm$0.14 \\
UVM2 & Oct. 11, 2018 & 01:41 &  312 & 12.15$\pm$0.03 & 6.42$\pm$0.15 \\
\noalign{\smallskip}
\hline
\noalign{\smallskip}
UVW1 & Oct. 06, 2018 & 10:39 &  669 & 11.40$\pm$0.02 & 10.5$\pm$0.2 \\
\noalign{\smallskip}
\hline
\end{tabular}
\end{center}
\end{table}
\section{Observations and data reduction}
\label{s:obs}
\subsection{Photometry}
\label{ss:phot}
\subsubsection{X-ray and ultraviolet observations with {\em Swift}}
\label{sss:xphot}
The {\em Neil Gehrels Swift Observatory} 
\citep[][ hereafter {\em Swift}]{gehrels+04} observed the source 
V426~Sge in six instances between August 28, 2018, and March 3, 2019, 
with the XRay Telescope \citep[][ hereafter XRT]{burrows+05} 
and the UltraViolet-Optical Telescope 
\citep[][ hereafter UVOT]{roming+05}. The XRT allows coverage of 
the 0.3-10\,keV band, whereas UVOT data were collected using 
the ultraviolet filters UVW2 ($\lambda$ = 1928\,\AA), UVM2 
($\lambda$ = 2246\,\AA), and UVW1 ($\lambda$ = 2600\,\AA) 
\citep[see][ in detail]{poole+08}. 
In a few cases, multiple UVOT data in different filters 
were collected during the same XRT pointing. 

All observations were reduced within the {\sc ftools} environment 
\citep[][]{1995ASPC...77..367B}. The XRT data reduction was performed 
using the {\sc xrtdas} standard data pipeline package 
({\sc xrtpipeline} v. 0.13.4) in order to produce screened event 
files. All data were extracted only in the photon counting (PC) 
mode \citep[][]{hill+04} adopting the standard grade filtering 
(0-12 for PC) according to the XRT nomenclature, and using an 
extraction radius of 47$''$ (20 pixels) centred at the optical 
coordinates of the source. 

The XRT count rates were then 
measured using the routines {\sc xselect} and {\sc image} assuming 
a threshold S/N=3 for the detections. The rate 
correction factors were determined using the {\sc xrtlccorr} task. 
Although, the X-ray detections were at low significance 
(around 3-4$\sigma$), it was possible to recognise that the bulk 
of photons ($>$92\%) falls below 2\,keV. 

X-ray fluxes were determined using the {\sc webpimms} online 
tool.\footnote{{\tt https://heasarc.gsfc.nasa.gov/cgi-bin/Tools/
w3pimms/w3pimms.pl}}
A crude spectral modelling for a bremsstrahlung emission with 
characteristic temperature $kT_{\rm br} < 1$\,keV or, alternatively, 
a blackbody spectrum with temperature $kT_{\rm bb} <0.35$\,keV, 
imply a count rate-to-flux conversion factor of 
$\sim$2.4$\times$10$^{-11}$\,erg\,cm$^{-2}$\,s$^{-1}$\,cts$^{-1}$. 
In modelling, we used a hydrogen column density corresponding 
to $E_{\rm B-V} = 0.2$\,mag (see the end of Sect.~\ref{ss:spec}) 
according to the formula of \cite{predehl+95}. 

Count rates on the UVOT images at the position of V426~Sge 
were measured through aperture photometry using 5$''$ apertures, 
whereas the corresponding background was evaluated for each image 
using a combination of several circular regions in source-free 
nearby areas. The data were calibrated using the UVOT 
photometric system described by \cite{poole+08}. Table~\ref{tab:resx} lists timing and results of the XRT 
observations, while Table~\ref{tab:resuv} introduces the 
same, but for the UVOT pointings. 
\subsubsection{Ground-based multicolour photometry}
\label{sss:multi}
Table~\ref{tab:phot} summarises basic information about 
instrumentation used to obtain our optical $UBVR_{\rm C}I_{\rm C}$ 
and NIR $JHKL$ photometry during the 2018 outburst and 
the following quiescence, as well as the historical photographic 
LC re-constructed from approximately 1900 to 1995. 
Data are available in Tables~\ref{tab:ubvri}, \ref{tab:asn}, 
\ref{tab:jhkl}, and \ref{tab:mpg}. 
Photometric evolution of V426~Sge is shown in Fig.~\ref{fig:hlc} 
and analysed in Sect.~\ref{ss:phot_evol}. 

Multicolour photometry was also used to convert the relative 
and/or arbitrary fluxes of our spectra to absolute fluxes. 
To obtain photometric flux-points of the true continuum, we 
determined corrections for emission lines from our low-resolution 
spectra using the procedure of \cite{sk07}. 
$UBV$ and $R_{\rm C}I_{\rm C}$ magnitudes were converted to fluxes 
according to the calibration of \cite{hk82} and \cite{bessel79}, 
respectively. 
%
%
\begin{table}
\caption[]{NIR $JHKL$ photometry of V426~Sge.}
\begin{center}
\begin{tabular}{ccccc}
\hline
\hline
\noalign{\smallskip}
    HJD       &  $J$  &  $H$   &  $K$   & $L$   \\
-2\,458\,000  &       &        &        &       \\
\noalign{\smallskip}
\hline
\noalign{\smallskip}
655.5& 7.95$\pm$0.01 &7.01$\pm$0.01 &6.69$\pm$0.00 &6.37$\pm$0.03 \\
656.4& 7.94$\pm$0.01 &7.00$\pm$0.01 &6.69$\pm$0.00 &6.35$\pm$0.03 \\
678.5& 7.89$\pm$0.01 &7.01$\pm$0.01 &6.64$\pm$0.01 &6.36$\pm$0.03 \\
707.4& 7.84$\pm$0.01 &6.91$\pm$0.01 &6.60$\pm$0.01 &6.32$\pm$0.04 \\
720.4& 7.84$\pm$0.01 &6.91$\pm$0.01 &6.59$\pm$0.01 &6.29$\pm$0.02 \\
\noalign{\smallskip}
\hline
\end{tabular}
\end{center}
\label{tab:jhkl}
\end{table}
%
%
\begin{table*}
\caption[]{
Log of observatories and instrumentations used to obtain our 
ground-based photometry. 
}
\begin{center}
\begin{tabular}{cccccc}
\hline
\hline
\noalign{\smallskip}
Observatory               &
Telescope$^{\dagger}$     & 
Detector                  &
Filters                   &
Reference$^{\star}$       &
Table                    \\
%
\noalign{\smallskip}
\hline
\noalign{\smallskip}
%
SLO,G2$^{a}$&  60 &CCD:~FLI~ML3041&$U,B,V,R_{\rm C},I_{\rm C}$& 1 & \ref{tab:ubvri} \\
SLO,G1$^{b}$&  18 &CCD:~SBIG~ST-10~MXE&$B,V,R_{\rm C},I_{\rm C}$ & 1 & \ref{tab:ubvri} \\
ANS Collab.$^{c}$& 30, 40 &CCD: SBIG ST-8, MG4-9000 &$B,V,R_{\rm C},I_{\rm C}$ &2,3&\ref{tab:asn}\\
AOKS$^{d}$&  35.6 &CCD:~G2-1600   &$B,V,R_{\rm C},I_{\rm C}$ & 4 & \ref{tab:ubvri} \\
SAI$^{e}$& 125 & InSb photometer & $J,H,K,L$ & 5 & \ref{tab:jhkl} \\
SAI      &40,16,50& photographic plate& $m_{\rm pg}$ & 6,7,8 & \ref{tab:mpg} \\ 
SO$^{f}$& 30,40 & photographic plate & $m_{\rm pg}$, $m_{\rm pv}$ & 6,7,8 & \ref{tab:mpg} \\
\noalign{\smallskip}
\hline
\end{tabular}
\end{center}
{\bf Notes.} 
$^{(a)}$\,Star\'a Lesn\'a Observatory -- pavilion G2, 
$^{(b)}$\,SLO -- pavilion G1, 
$^{(c)}$\,Asiago Novae and Symbiotic stars Collaboration, 
$^{(d)}$\,Astronomical Observatory at the Kolonica Saddle, 
$^{(e)}$\,Sternberg Astronomical Institute of the Moscow University, 
$^{(f)}$\,Sonneberg Observatory archive: 
http://www.4pisysteme.de/observatory/observatory$\_$4$\_$2$\_$en.html.
$^{\star}$\,1 -- \cite{sekeras+19},
2 -- \cite{2012BaltA..21...13M},
3 -- \cite{2012BaltA..21...22M},
4 -- \cite{kudzej+19},
5 -- \cite{shenavrin+11},
6 -- \cite{bacher+05},
7 -- \cite{sokolovsky+16},
8 -- http://scan.sai.msu.ru/vast/.
$^{\dagger}$\,diameter of the primary mirror in cm. 
\label{tab:phot}
\end{table*}
%
%
\begin{table*}
\caption[]{
Log of observatories and instrumentations used to obtain our 
spectroscopic observations. 
}
\begin{center}
\begin{tabular}{ccccrccc}
\hline
\hline
\noalign{\smallskip}
Observatory               &
Telescope$^{\dagger}$     & 
Spectrograph              & 
Camera                    &
Res.$^{\ddagger}$         &
Observer                  &
Reference$^{\star}$       &
Label$^{\ast}$           \\
\noalign{\smallskip}
\hline
\noalign{\smallskip}
%
CrAO$^{a}$  & 260 & slit SPEM        & SPEC-10          &  1000 & Tarasova
            &  1  & (iv)  \\
Asiago$^{b}$& 182 & echelle          & EEV~CCD47-10     & 21000 & Munari
            &  2  & (iii) \\
SPO$^{c}$   & 130 & echelle          & Andor~iKon-L~936 & 34000 & Kom\v{z}\'{\i}k, Seker\'a\v{s}
            &3,4,5& (i)   \\
Asiago      & 122 & Boller\&Chivens  & Andor~iDus~DU440A&  1000 & Munari
            & 6,7 & (iii) \\
SLO,G1      &  60 & echelle          & ATIK\,460EX      & 11000 & Shugarov, Kundra
            &  4  & (ii)  \\
Durtal      &  51 & echelle          & KAF-3200ME       & 11000 & Charbonnel
            & 8,9 & (x) \\
la Tourbi\`ere& 40& echelle          & ATIK\,460EX      & 11000 & Garde
            & 1,9 & (viii) \\
DCO$^{d}$   & 35 & LISA              & ATIK\,460EX      &  1000 & Sims 
            & 1,9 & (xiii) \\
Mill Ridge  & 31 & echelle           & ASI 1600mm CMOS  & 12000 & Lester
            & 1,9 & (ix) \\
AOKS        & 28 & LISA              & ATIK\,460EX      &   680 & Dubovsk\'y
            & 10 &  (v) \\
CTO$^{e}$   & 25 & UVEX              & SWO~CMOS         &  1200 & Buil 
            & 9,11&  (vi) \\
OCT-FR$^{f}$& 20 & slit Alpy~600     & ATIK~314L+       &   500 & Boussin 
            &  9 &  (vii) \\
L'Aquila$^{g}$& 20 & slit Alpy~600   & SBIG~ST-8300M    &  1600 & Sollecchia 
            &  9 &  (xi) \\
Balzaretto$^{h}$& 20 & slit Alpy~600 & SXVF-M7          &  530  & Franco
            &  9 &  (xii) \\
\noalign{\smallskip}
\hline
\end{tabular}
\end{center}
{\bf Notes.} 
$^{(a)}$\,Crimean Astrophysical Observatory, 
$^{(b)}$\,Asiago Astrophysical Observatory, 
$^{(c)}$\,Skalnat\'e Pleso Observatory, 
$^{(d)}$\,private Desert Celestial Observatory in Gilbert Arizona, 
$^{(e)}$\,Castanet Tolosan Observatory, 
$^{(f)}$\,Observatory de l'Eridan, 
$^{(g)}$\,private station in L'Aquila, 
$^{(h)}$\,A81 Balzaretto Observatory. 
$^{\star}$\,1 -- \cite{skopal+17}, 
2 -- \cite{1992PASP..104..121M}, 
3 -- \cite{baud+92}, 
4 -- \cite{pribulla+15}, 
5 -- \cite{dohring+19}, 
6 -- \cite{ciori+14}, 
7 -- \cite{siviero14}, 
8 -- \cite{skopal+14}, 
9 -- \cite{teyssier19}, 
10 -- \cite{kudub14}, 
11 -- http://www.astrosurf.com/buil/UVEX$\_$project$\_$us/.
$^{\dagger}$\,diameter of the primary mirror in cm, 
$^{\ddagger}$\,average resolution, 
$^{\ast}$\,label of the observatory in Tables~\ref{tab:low} and 
           \ref{tab:med}. 
\label{tab:spec}
\end{table*}
\subsection{Spectroscopy}
\label{ss:spec}
Spectroscopic observations were secured at 13 observatories 
and/or private stations. Basic information about instrumentation 
for the spectra acquisition is introduced in Table~\ref{tab:spec}. 

Low-resolution spectra (Table~\ref{tab:low}) were used 
to model the SED. Their flux calibration was verified with 
the aid of the near-simultaneous emission-line-free 
$UBVR_{\rm C}I_{\rm C}$ photometry. 
High- and medium-resolution spectra (Table~\ref{tab:med}) 
served to analyse variations in the line profiles and fluxes. 

The distance of $3.3^{+0.7}_{-0.6}$\,kpc was determined by 
\cite{bailer+18} using the parallax of V426~Sge published in 
the second {\em Gaia} data release \citep[][]{gaia+18}. 
According to a 3D map 
of interstellar dust reddening published by \cite{green+18}, 
we estimated the colour excess $E_{\rm B-V} \sim 0.2$\,mag 
in direction to V426~Sge for 3.3\,kpc. 
Modelling the {\em Swift-}UVOT and optical continuum fluxes 
allowed us to specify the colour excess to 
$E_{\rm B-V} = 0.2\pm0.02$\,mag (see Sect.~\ref{sss:sed}), 
which, together with the extinction curve of \cite{c+89}, 
we used to deredden our observations. 
%
%
%
\begin{table}
\caption[]{Log of low-resolution spectroscopic observations}
\begin{center}
\begin{tabular}{cccrc}
\hline
\hline
\noalign{\smallskip}
Date$^{a}$ & JD~2\,4...  & Range &$T_{\rm exp}$& Obs.$^{b}$\\
yyyy/mm/dd.ddd &         &  [nm] &   [s]       &           \\
\noalign{\smallskip}
\hline
\noalign{\smallskip}
 2018/08/10.851 & 58341.351 & 330 - 788 &  120 &  (iii)  \\
 2018/08/10.878 & 58341.378 & 375 - 737 & 3649 &  (xi)   \\
 2018/08/10.932 & 58341.432 & 350 - 507 & 6167 &  (vi)   \\
 2018/08/11.915 & 58342.415 & 370 - 757 & 3926 &  (vii)  \\
 2018/08/11.972 & 58342.472 & 339 - 457 & 5495 &  (vi)   \\
 2018/08/15.888 & 58346.388 & 370 - 757 & 4532 &  (vii)  \\
 2018/08/18.896 & 58349.396 & 370 - 757 & 4833 &  (vii)  \\
 2018/08/19.895 & 58350.395 & 327 - 796 &  120 &  (iii)  \\
 2018/08/21.899 & 58352.399 & 330 - 757 & 1200 &  (iv)   \\
 2018/08/28.860 & 58359.360 & 383 - 723 & 3600 &  (xii)  \\
 2018/09/01.828 & 58363.328 & 330 - 757 & 1200 &  (iv)   \\
 2018/09/07.886 & 58369.386 & 370 - 757 & 4525 &  (vii)  \\
 2018/09/11.903 & 58373.403 & 370 - 757 & 3621 &  (vii)  \\
 2018/09/15.911 & 58377.411 & 370 - 757 & 3621 &  (vii)  \\
 2018/09/19.827 & 58381.327 & 336 - 580 & 9970 &  (vi)   \\
 2018/09/20.888 & 58382.388 & 330 - 788 &  120 &  (iii)  \\
 2018/09/22.808 & 58384.308 & 580 - 879 & 4266 &  (vi)   \\
 2018/09/28.911 & 58390.411 & 370 - 757 & 2109 &  (vii)  \\
 2018/10/07.712 & 58399.212 & 330 - 757 &  900 &  (iv)  \\
 2018/10/17.732 & 58409.232 & 330 - 757 & 1200 &  (iv)  \\ 
 2018/11/18.708 & 58441.208 & 333 - 803 &  120 &  (iii) \\
 2018/12/08.681 & 58461.181 & 330 - 794 &  120 &  (iii) \\
 2018/12/17.704 & 58470.204 & 370 - 791 &  120 &  (iii) \\
\noalign{\smallskip}
\hline
\noalign{\smallskip}
\multicolumn{5}{c}{Quiescent phase} \\
\noalign{\smallskip}
\hline
\noalign{\smallskip}
 2019/06/23.872 & 58658.372 & 384 - 758 & 6000 &  (v)   \\
 2019/06/25.931 & 58660.431 & 384 - 758 & 6000 &  (v)   \\
 2019/08/19.832 & 58715.332 & 330 - 781 &  240 &  (iii) \\
 2019/09/07.191 & 58733.691 & 371 - 730 & 3323 &  (xiii)\\
 2019/10/05.112 & 58761.612 & 371 - 730 & 3994 &  (xiii)\\
\noalign{\smallskip}
\hline
\end{tabular}
\end{center}
{\bf Notes.}
  $^{(a)}$ Start of the observation in UT,
  $^{(b)}$ label of the observatory from Table~\ref{tab:spec}. 
\label{tab:low}
\end{table}
%
%
%
\begin{table}
\caption[]{Log of high-resolution spectroscopic
           observations}
\begin{center}
\begin{tabular}{cccrc}
\hline
\hline
\noalign{\smallskip}
Date$^{a}$     & JD~2\,4... & Range &$T_{\rm exp}$& Obs.$^{b}$ \\
yyyy/mm/dd.ddd &            & [nm]  &   [s]       &            \\
\noalign{\smallskip}
\hline
\noalign{\smallskip}
2018/08/10.845 & 58341.345 & 420-759 & 10800 & (viii) \\
2018/08/11.826 & 58342.326 & 420-759 & 12000 & (viii) \\
2018/08/12.872 & 58343.372 & 580-720 &  2700 & (i)    \\
2018/08/13.814 & 58344.314 & 425-720 &  2700 & (i)    \\
2018/08/13.827 & 58344.327 & 392-710 &  3600 & (ii)   \\
2018/08/15.853 & 58346.353 & 420-759 & 12000 & (viii) \\
2018/08/17.070 & 58347.570 & 403-795 &  9600 & (ix)   \\
2018/08/17.896 & 58348.396 & 392-710 &  3600 & (ii)   \\ 
2018/08/18.823 & 58349.323 & 392-710 &  3600 & (ii)   \\
2018/08/19.842 & 58350.342 & 392-710 &  3600 & (ii)   \\
2018/08/20.060 & 58350.560 & 403-795 & 10800 & (ix)   \\
2018/08/20.868 & 58351.369 & 425-720 &  3600 & (i)    \\
2018/08/22.833 & 58353.333 & 360-712 &   600 & (iii)  \\
2018/08/22.840 & 58353.340 & 425-720 &  4500 & (i)    \\
2018/08/27.796 & 58358.296 & 425-720 &  2700 & (i)    \\
2018/09/07.867 & 58369.367 & 425-720 &  3600 & (i)    \\
2018/09/19.821 & 58381.321 & 425-720 &  4500 & (i)    \\
2018/09/20.837 & 58382.337 & 425-720 &  4500 & (i)    \\
2018/09/20.882 & 58382.382 & 360-712 &   900 & (iii)  \\
2018/10/04.840 & 58396.340 & 392-710 &  3600 & (ii)   \\ 
2018/10/05.724 & 58397.224 & 392-710 &  5100 & (ii)   \\
2018/10/10.788 & 58402.288 & 392-710 &  4800 & (ii)   \\
2018/10/11.835 & 58403.335 & 392-710 &  6000 & (ii)   \\ 
2018/10/16.713 & 58408.213 & 425-720 &  5400 & (i)    \\
2018/10/18.797 & 58410.297 & 404-759 &  6159 & (x)    \\
2018/10/20.757 & 58412.257 & 360-712 &   600 & (iii)  \\
2018/11/08.741 & 58431.241 & 392-710 &  3600 & (ii)   \\
2018/11/10.773 & 58433.273 & 392-710 &  2700 & (ii)   \\ 
2018/11/12.803 & 58435.303 & 392-710 &  2400 & (ii)   \\
2018/11/23.729 & 58446.229 & 425-720 &  3600 & (i)    \\
2018/12/17.701 & 58470.201 & 360-711 &   900 & (iii)  \\
2019/11/13.760 & 58801.260 & 360-711 &   900 & (iii)  \\ 
\noalign{\smallskip}
\hline
\end{tabular}
\end{center}
{\bf Notes.}
$^{(a)}$\,Start of the observation in UT, 
$^{(b)}$\,label of the observatory from Table~\ref{tab:spec}. 
\label{tab:med}
\end{table}

\section{Analysis and results}
\label{s:analysis}
\subsection{Photometric evolution}
\label{ss:phot_evol}
Figure~\ref{fig:hlc} shows the historical LC of V426~Sge 
from 1900 to the present that we measured on photographic 
plates (Table~\ref{tab:mpg}), supplemented with ASASSN $V$ 
magnitudes and our $UBVR_{\rm C}I_{\rm C}$ photometry 
(Sect.~\ref{sss:multi}). 
The LC shows three particularly distinctive features: 
(i) a symbiotic nova outburst in 1968, 
(ii) development of wave-like orbitally related variations, and 
(iii) a Z~And-type outburst in 2018. 
We describe these events as follows. 

\subsubsection{Symbiotic nova outburst in 1968}
\label{sss:nova}
\begin{figure}
\begin{center}
\resizebox{\hsize}{!}{\includegraphics[angle=-90]{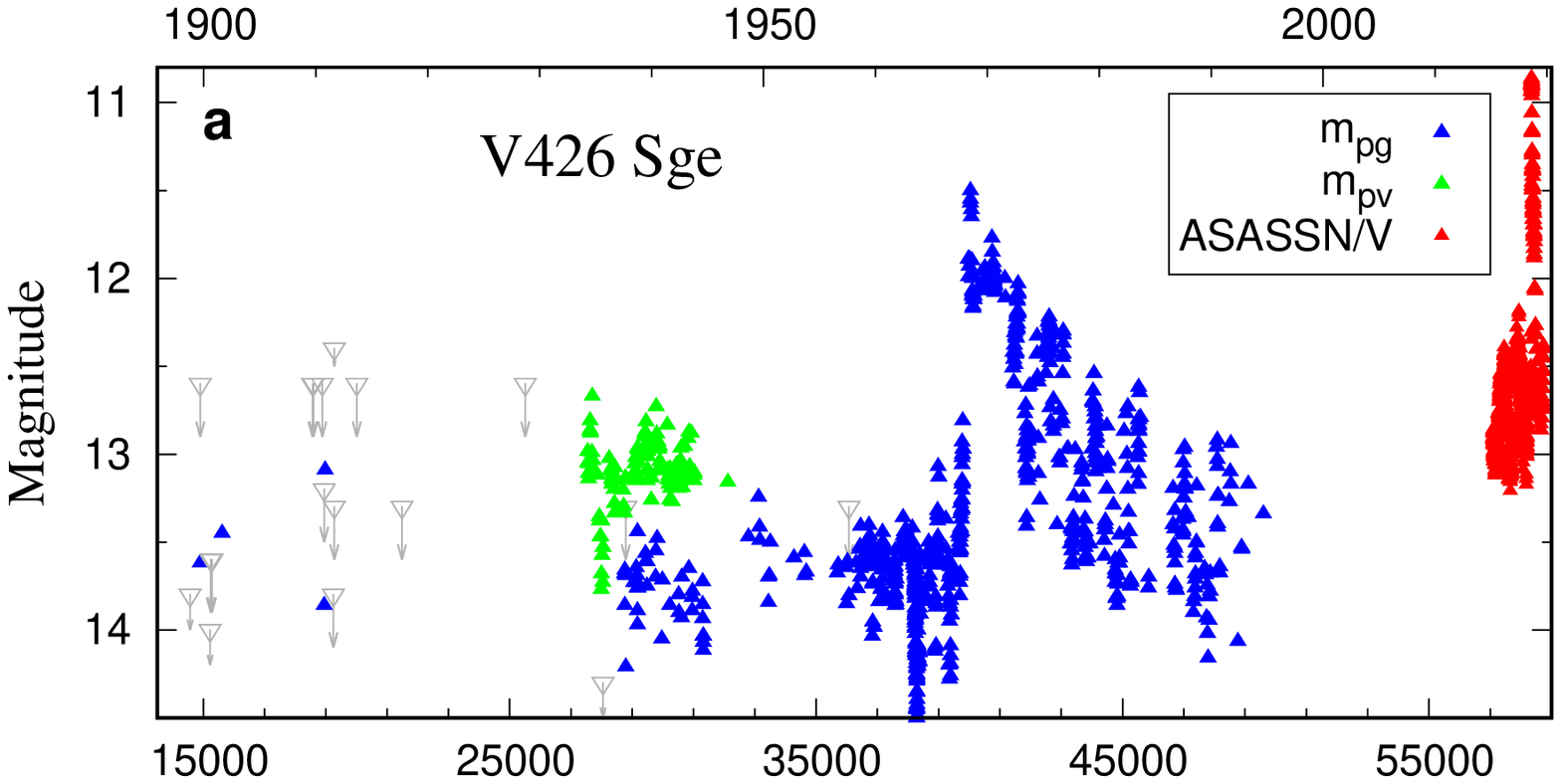}}
\resizebox{\hsize}{!}{\includegraphics[angle=-90]{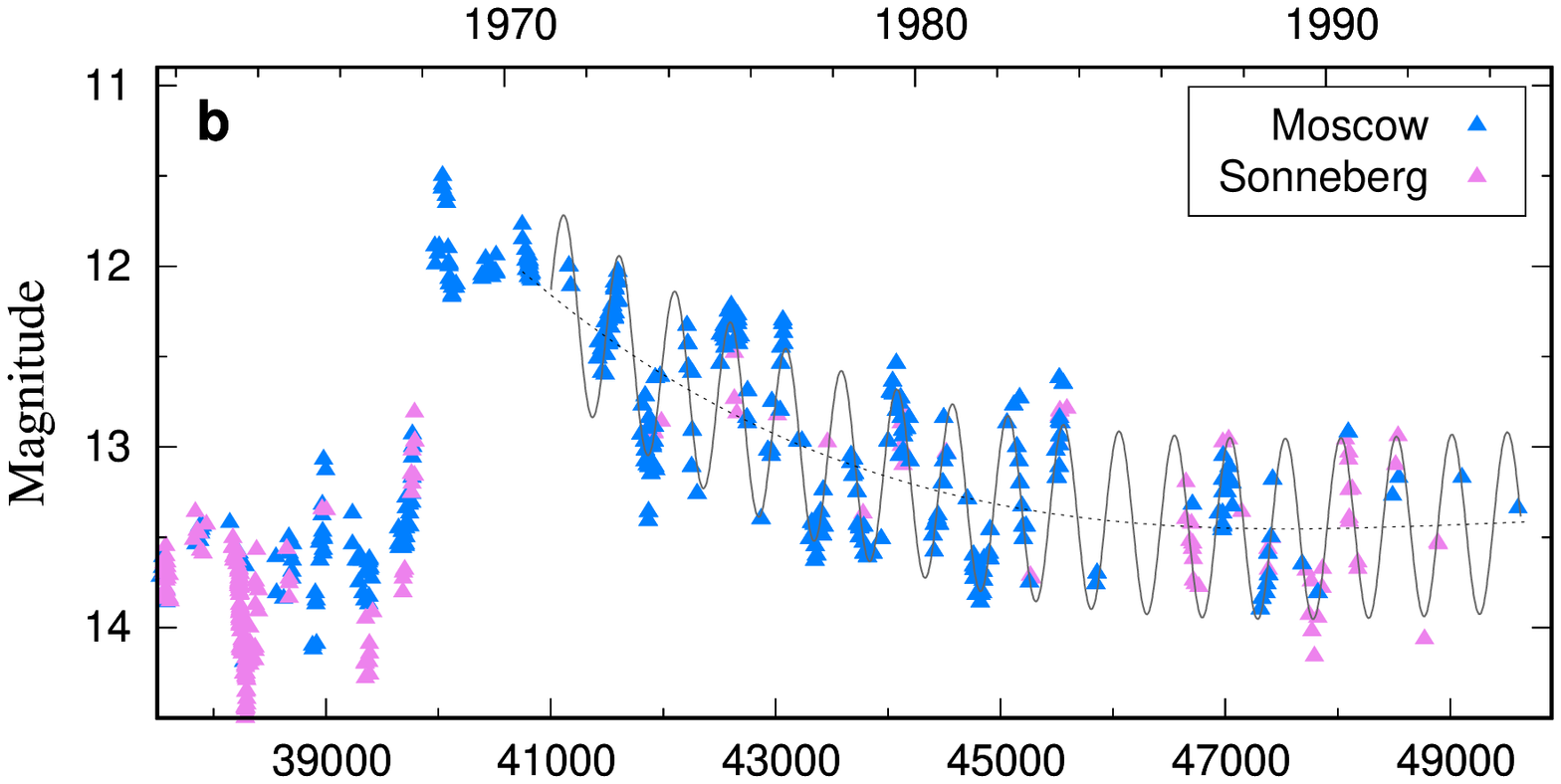}}
\resizebox{\hsize}{!}{\includegraphics[angle=-90]{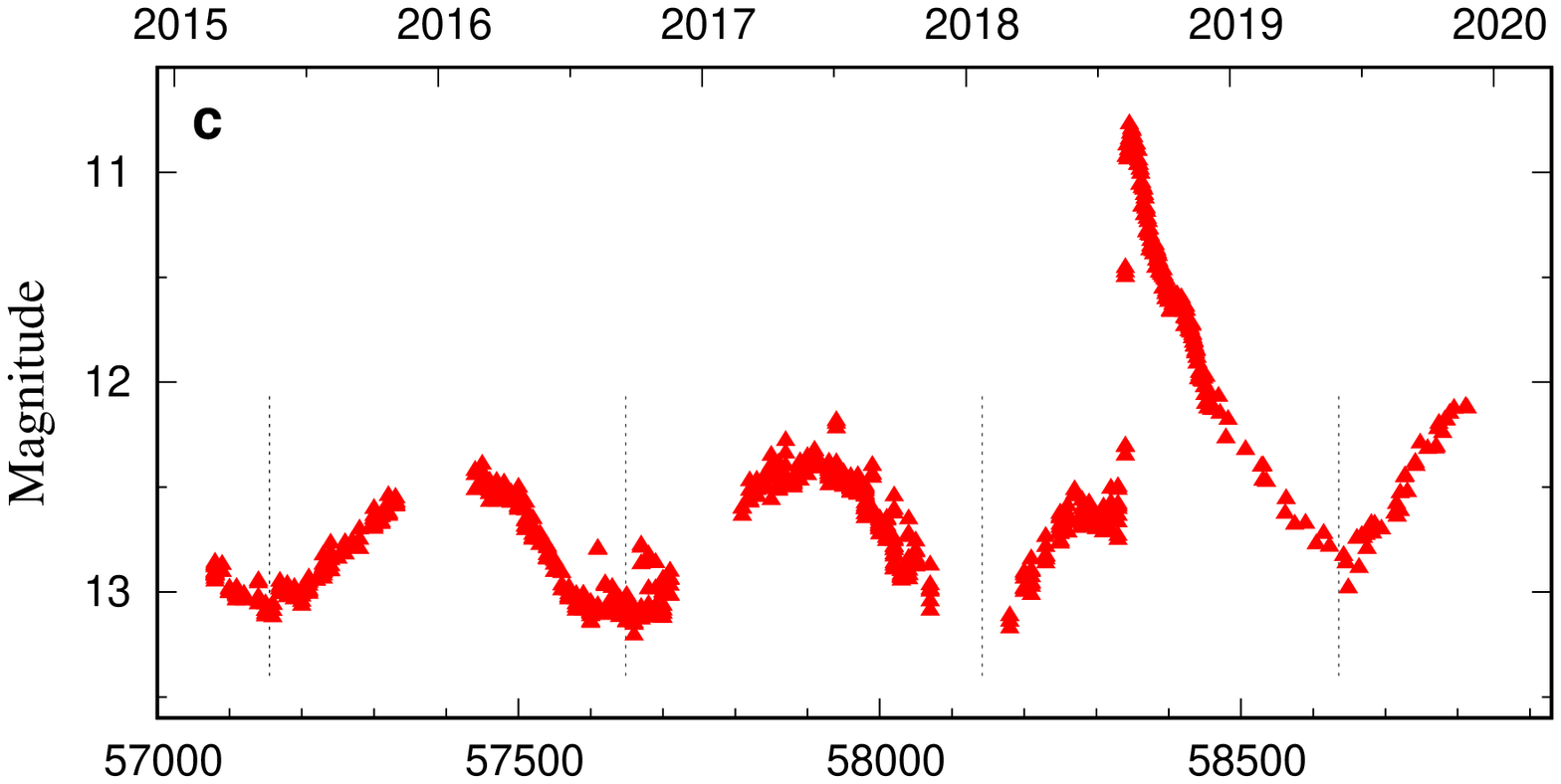}}
\resizebox{\hsize}{!}{\includegraphics[angle=-90]{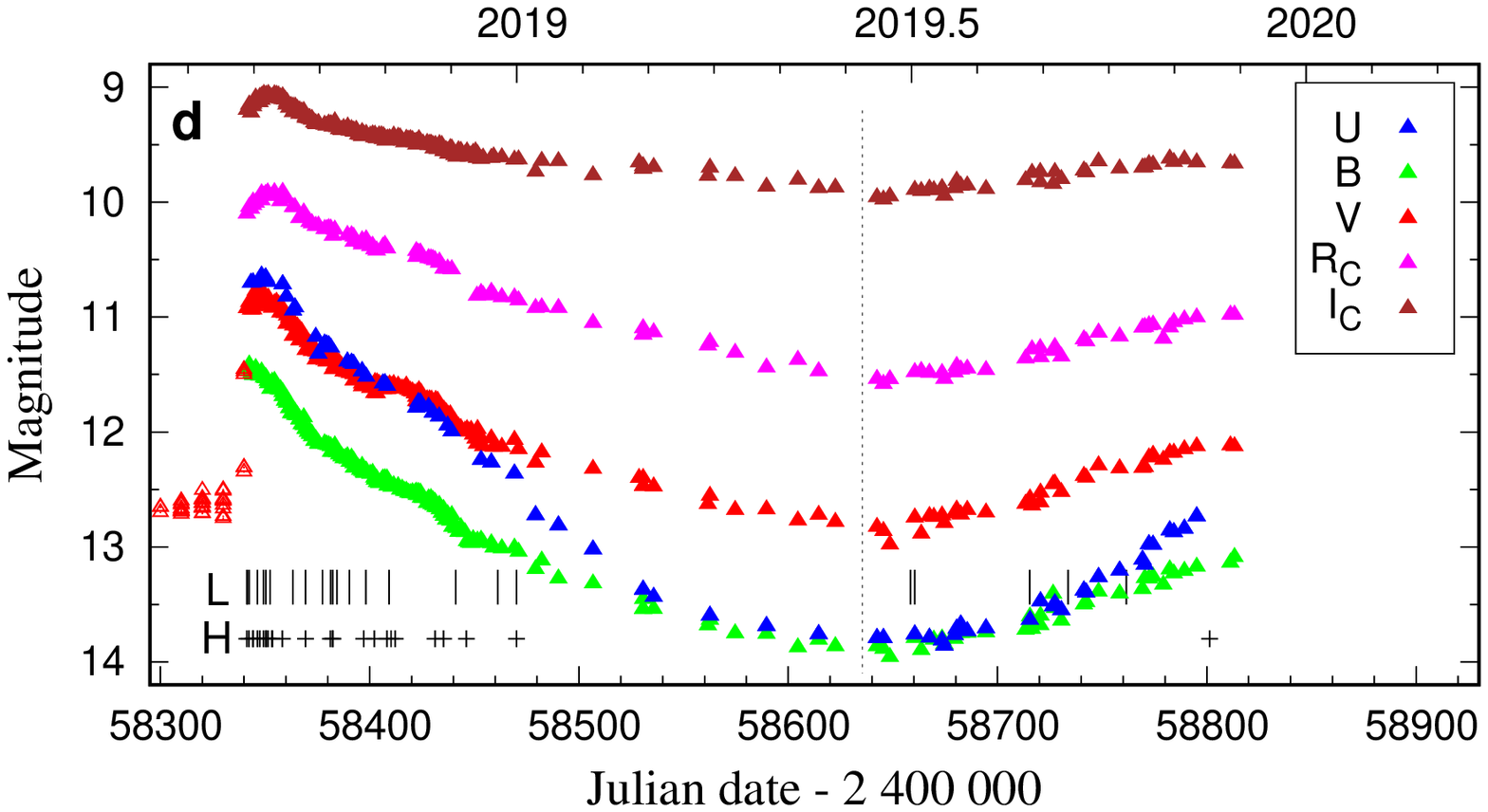}}
\end{center}
\caption[]{
{\bf a:} Historical LC of V426~Sge from 1900 to the present 
as given by our photographic m$_{\rm pg}$/m$_{\rm pv}$ 
magnitudes (blue/green triangles), ASASSN and our $V$ magnitudes 
(red triangles). Grey triangles with arrows denote a limiting 
magnitude on plates with unseen target. 
{\bf b:} Detail of the LC covering the 1968 symbiotic nova outburst. 
Decline from the maximum and wave-like periodic variation 
are denoted by the dotted and full line, respectively. 
{\bf c:} The $V$-LC from 2015 to the present showing the ASASSN 
and our data. 
{\bf d:} $UBVR_{\rm C}I_{\rm C}$ LCs covering the 2018 outburst 
and the present quiescence. Vertical bars and crosses denote 
the dates of our low (L) and high (H) resolution spectra. 
Dotted vertical lines in the panel {\bf c} and {\bf d} mark 
the light minima according to the ephemeris (\ref{eq:eph}). 
          }
\label{fig:hlc}
\end{figure}

Photographic measurements revealed that the first recorded 
outburst started around 1967.5 from $m_{\rm pg}\sim 13.5$ 
and peaked around 1968.5 at $m_{\rm pg}\sim 11.5$\,mag. 
After a 60-day decline by $\sim$0.5\,mag, the star kept 
its brightness at $m_{\rm pg}\sim 12.0$ to approximately 
mid-1971, before a gradual decline to approximately 1990. 
During the decline, a pronounced wave-like variation 
developed in the LC (see panel {\bf b}). We did not find 
this type of variability prior to the outburst, although 
the LC is well defined from 1928 to 1967 (Fig.~\ref{fig:hlc}). 
It is of interest to note that the same type of LC profile 
was observed for the symbiotic nova V1329~Cyg around its 1964 
outburst: (i) No symbiotic-like variability prior to the 
outburst, (ii) a two-year brightening by 2--3\,mag, followed 
by (iii) a gradual $\sim$20-year decline, during which periodic 
wave-like variations connected with the binary motion developed 
\citep[see Figs.~4 and 5 of][ and Fig.~\ref{fig:snovae} here]{sk98}. 
Similar evolution of the wave-like variability was also recorded 
in the LC of the symbiotic nova AG~Peg during its decline from 
the 1850 outburst (Fig.~\ref{fig:snovae}). 
Accordingly, we classify the 1968 outburst of V426~Sge as 
a symbiotic nova outburst. 
%
%
%

\subsubsection{Wave-like orbitally related variation}
\label{sss:wave}
\begin{figure}
\begin{center}
\resizebox{\hsize}{!}{\includegraphics[angle=-90]{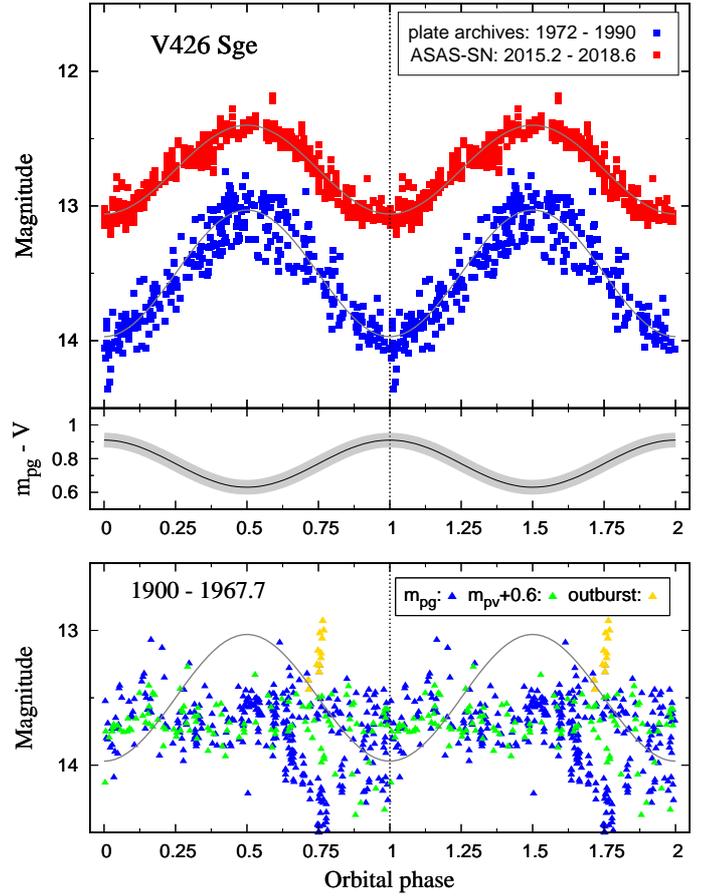}}
\end{center}
\caption[]{
Top: Phase diagram for periodic wave-like variations in the 
ASASSN $V$ and $m_{\rm pg}$ LCs of V426~Sge according to 
the ephemeris (\ref{eq:eph}). Grey lines represent their 
sinusoidal fits. 
Middle: Corresponding colour index $m_{\rm pg} - V$ 
 illustrated by the black line inside the grey belt 
(= its uncertainty) (Sect.~\ref{sss:wave}). 
Bottom: Photographic magnitudes prior to the 1968 outburst 
phased with the ephemeris (\ref{eq:eph}). The LCs do not show 
any sinusoidal light variation along the orbit 
(Sect.~\ref{sss:outobs}). 
          }
\label{fig:phase}
\end{figure}

The ASASSN $V$-LC, recorded from February 26, 2015, to July 30, 
2018 (671 measurements), shows a wave-like variability 
with a period of $\sim$500 days (see panel {\bf c} of 
Fig.~\ref{fig:hlc}). 
Using the Lafler-Kinnman algorithm \citep[see][]{lafler+65}, 
we determined the ephemeris of the light minima as, 
$JD_{\rm Min,\,V} = 2\,456\,660(\pm 7) + 494(\pm 5)\times E$. 
In the same way we analysed this type of variability in the 
$m_{\rm pg}$-LC during the transition from the 1968 outburst 
to quiescence. We obtained the ephemeris, 
$JD_{\rm Min,\,mpg} = 2\,440\,872(\pm 10) + 493.5(\pm 2.0)\times E$ 
using 309 magnitudes between July 29, 1971, and July 18, 1990, 
for which we subtracted their decline using a third-order 
polynomial (dotted line in panel {\bf b}). Within the uncertainties, 
both ephemerides are identical in zero phase and period, which 
suggests a common origin of the wave-like variability in both colours. 
Analysing both datasets together, we obtained the ephemeris 
of the light minima as,\footnote{We also verified this 
ephemeris using the {\em date-compensated discrete Fourier 
transform} \citep[][]{ferraz-mello81} and by the {\em CLEANest} 
algorithm \citep[][]{foster95} we confirmed unicity of 
the period. We used the softver of \cite{paunzen+16}.}
\begin{equation}
 JD_{\rm Min} = 
    2\,440\,873(\pm 7) + 493.4(\pm 0.7)\times E .
\label{eq:eph}
\end{equation}
Figure~\ref{fig:phase} shows the phase diagram of both the 
datasets plotted with the ephemeris (\ref{eq:eph}). Their 
fits by sinusoidal curve determine amplitudes of the wave-like 
variations, $\Delta$$V\sim 0.66$\,mag and 
$\Delta$$m_{\rm pg}\sim 0.94$\,mag, which is the difference 
between the maximum and minimum. 
The corresponding colour index $m_{\rm pg} - V$ also varies with 
the phase, with a maximum around the phase $\varphi = 0$ 
and a minimum around $\varphi = 0.5$. 
These characteristics are typical for the orbitally related 
wave-like variability that develops during quiescent phases 
of symbiotic stars 
\citep[e.g.][]{hoff68,mein79,sekeras+19}. 
This type of variability is caused by the optically thick part 
of the symbiotic nebula, whose contributions are thus different 
at different orbital phases \citep[][]{sk98,sk01}. 
Therefore, we suggest that the orbital period of V426~Sge 
is $493.4\pm 0.7$ days and the inferior conjunction of the cool 
component occurs around the orbital phase $\varphi = 0$, the 
timing of which is determined by Eq.~(\ref{eq:eph}). 
Measurements of radial velocities of the red giant orbital motion 
during the following quiescent phase should definitely confirm 
this suggestion. 

\subsubsection{Z~And-type outburst in 2018}
\label{sss:2018o}
\begin{figure}
\begin{center}
\resizebox{\hsize}{!}{\includegraphics[angle=-90]{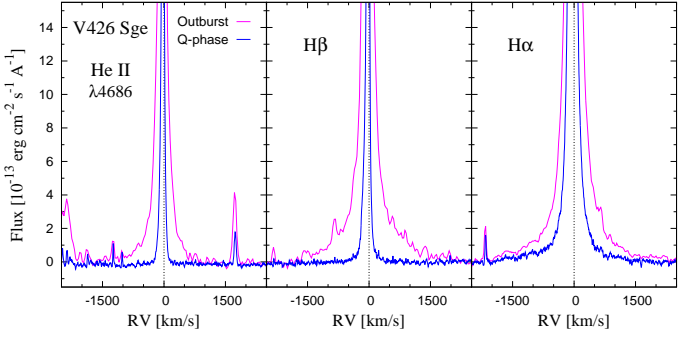}}
\end{center}
\caption[]{
Broadening of \ion{He}{ii}\,$\lambda$4686, \hb\ and \ha\ 
line profiles during the outburst (magenta, spectrum 
2018/08/13.814 in Table~\ref{tab:med}) with respect to quiescent 
phase (blue, spectrum 2019/11/13.760 in Table~\ref{tab:med}). 
Local continuum is subtracted. 
          }
\label{fig:broad}
\end{figure}

During the first decade of August 2018, V426~Sge commenced 
a new outburst (see Fig.~\ref{fig:hlc}). According to ASASSN 
$V$ magnitudes, the brightening started between July 30 
($V\sim 12.6$) and August 6, 2018 ($V\sim 11.5$). 
On August 10, our photometry confirmed the brightening indicating 
the object at $V\sim 10.9$. Following measurements revealed 
a maximum in $V\sim 10.80$, $R_{\rm C}\sim 9.95$, and 
$I_{\rm C}\sim 9.07$ on August 17$\pm 2$, 19$\pm 3$, and 20$\pm 3$, 
respectively. Then the star was gradually weakening and, between 
February 15 and March 19 2019, it reached its pre-outburst 
brightness in $V$ and $m_{\rm pg}$ (see Fig.~\ref{fig:hlc}). 
Assuming that the outburst ceased at this time, its maximum 
brightening, $\Delta U\sim 3.0$, $\Delta B\sim 2.1$, and 
$\Delta V\sim 1.7$\,mag, happened on a timescale of a few days 
and the outburst lasted for a few months. 
Figure~\ref{fig:broad} shows a significant increase of 
fluxes and broadening of \ion{He}{ii}\,$\lambda$4686, 
\hb,\ and \ha\ line profiles during the outburst relative 
to the quiescent phase. 
Such evolution in the multicolour LC and signatures of 
the enhanced mass outflow are classified as the Z~And-type 
outburst \citep[e.g.][]{kenyon86}. Following the outburst, and 
until the present, the LC represents a fraction of the wave-like 
variability around $\varphi = 0$ of the current quiescent 
phase (see Fig.~\ref{fig:hlc}). 
%
%
%
%
%
%
%
%
\subsection{Physical parameters during the 2018 outburst}
\label{ss:spec_evol}
In this section we determine physical parameters of V426~Sge 
and describe their temporal evolution throughout the 2018 outburst. 
Our analysis is similar to the one we used for the 2015 outburst 
of AG~Peg, because of their close similarity 
\citep[see][ and Sect.~\ref{ss:compagpeg} here]{skopal+17}. 
Therefore, our approach is explained here only briefly. 
\subsubsection{Parameters from SED models}
\label{sss:sed}
Modelling the near-UV (NUV) to NIR continuum allows us to 
determine parameters of the main components of radiation 
contributing to the observed spectrum. 
In the given spectral domain, we consider radiation emitted by 
the giant hot component 
($\equiv$ the WD pseudophotosphere\footnote{as the modelling 
does not allow us to discern the geometry of radiative sources, 
we ascribed the whole stellar component of radiation to the 
WD pseudophotosphere during the outburst.}) 
and nebula. Using the method of disentangling the composite 
continuum of symbiotic binaries \citep[see][]{sk05a}, we 
can determine the effective temperature of the giant, 
$T_{\rm eff}$, its radius, $R_{\rm G}$, the electron temperature 
of the nebula, $T_{\rm e}$, and its emission measure, \textsl{EM}. 
The luminosity of the WD pseudophotosphere, $L_{\rm WD}$, and 
its effective radius, $R_{\rm WD}^{\rm eff}$ (i.e. the radius 
of a sphere with the same luminosity), can be obtained for 
independently determined temperature, $T_{\rm WD}$ 
(Sect.~\ref{sss:twd}). 
We selected the model corresponding to a minimum of the reduced 
$\chi^2$ function as the best model of our SED-fitting analysis. 
Owing to systematic errors in the calibration of some 
parts of our spectra, we adopted relatively large uncertainties 
of 10\%\ for selected continuum fluxes in all our low-resolution 
spectra. Therefore, models of well-calibrated spectra are 
characterised with $\chi_{\rm red}^2 < 1$ (Table~\ref{tab:sed}). 
More details can be found in \cite{skopal+17}. 
Modelling observations within different spectral domains 
requires a specific approach: 

(i) 
In the case of fitting just the optical continuum, the method only  
makes it possible to disentangle contributions from the nebula and 
the cool giant, because of a very small contribution from 
the hot WD pseudophotosphere in the optical (see Sect.~\ref{sss:twd}). 
Therefore, here we only determined parameters, $T_{\rm e}$, 
\textsl{EM}, $R_{\rm G}$, and $T_{\rm eff}$, where $T_{\rm eff}$ 
corresponds to the spectral type (ST) of the giant according 
to the calibration of \cite{fluks+94}. 
The ST subclass was estimated by a linear interpolation of 
two neighbouring best-fitting STs. 

\begin{figure*}
\begin{center}
\resizebox{18cm}{!}{\includegraphics[angle=-90]{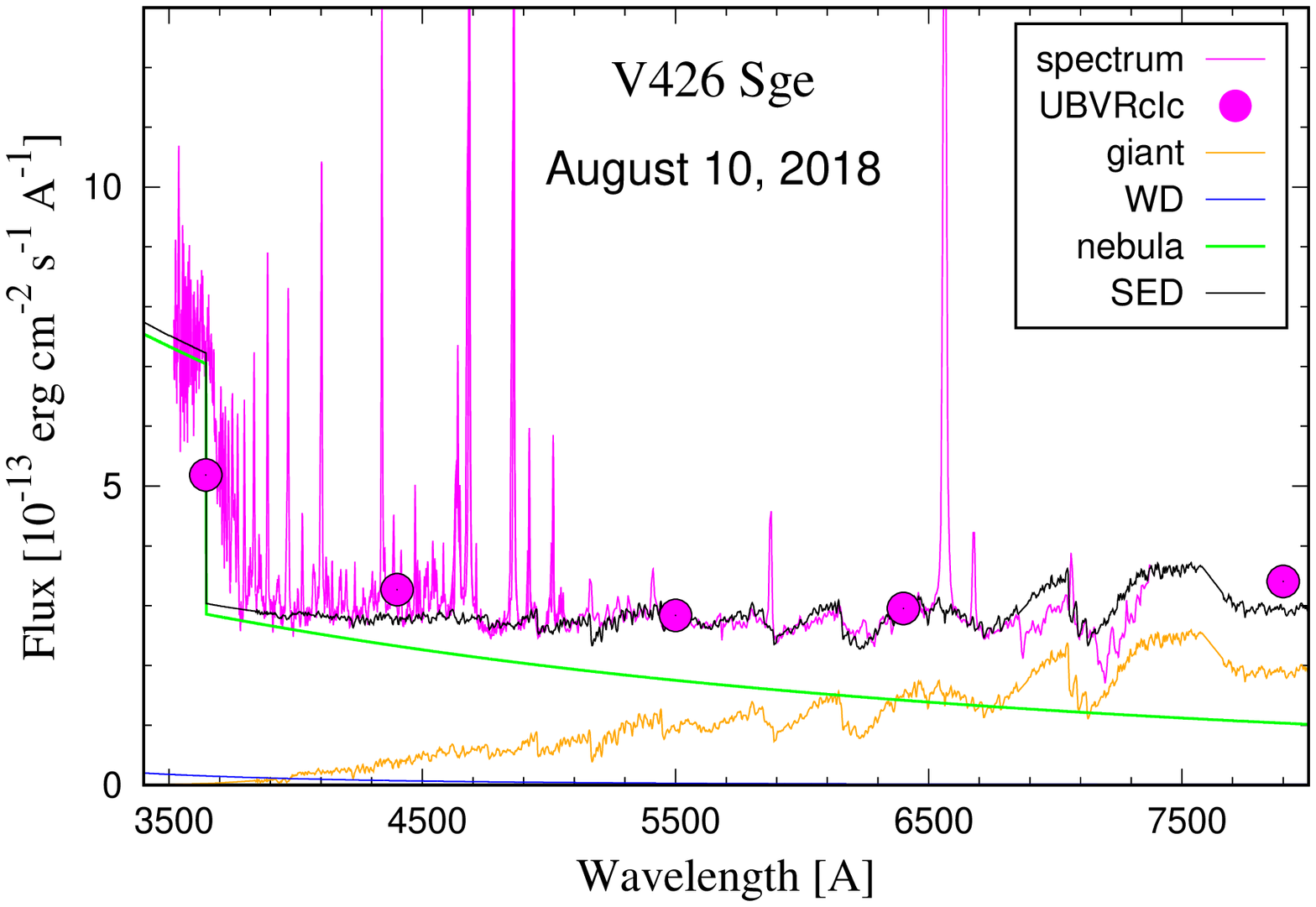}
                    \includegraphics[angle=-90]{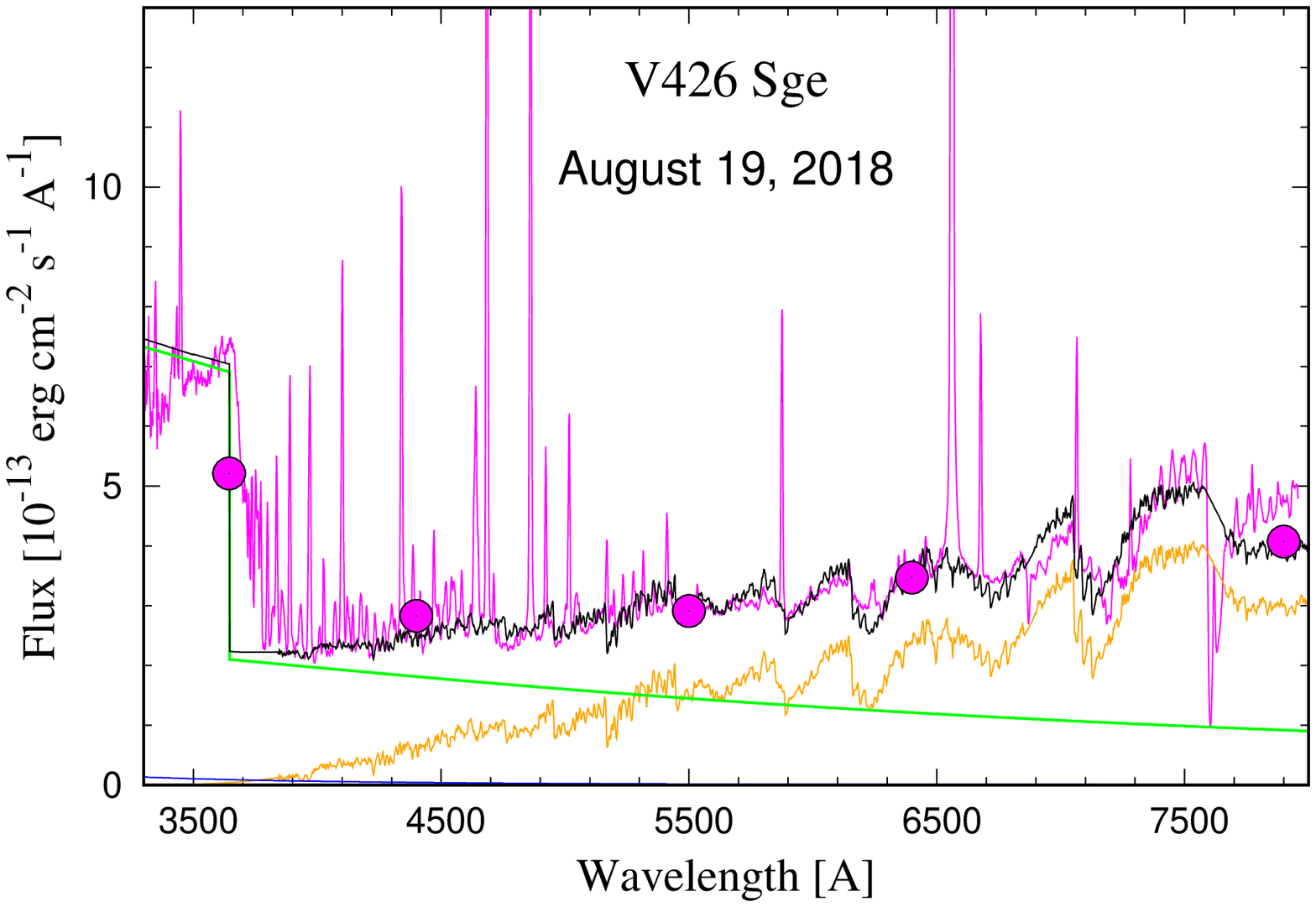}
                    \includegraphics[angle=-90]{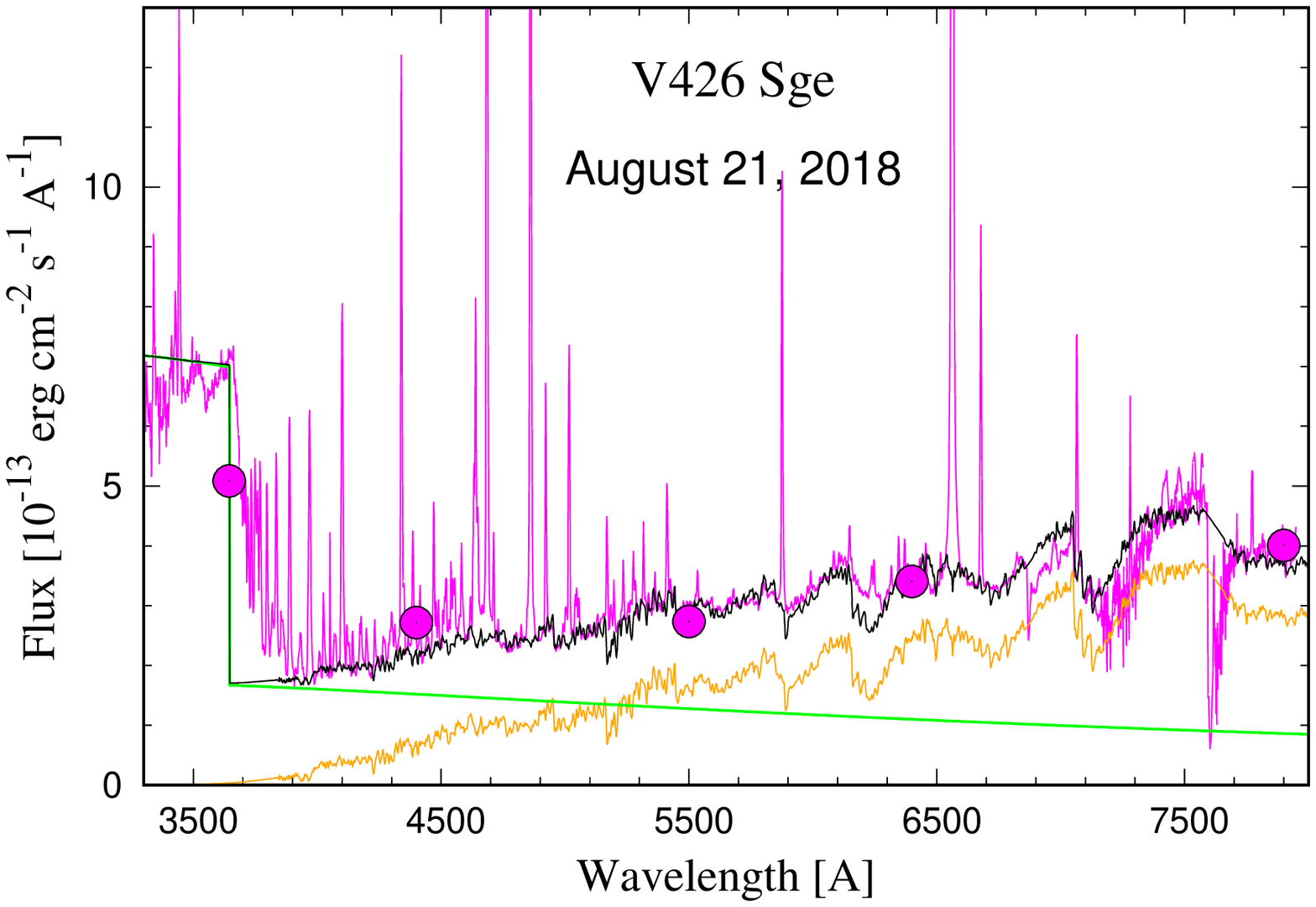}}
\resizebox{18cm}{!}{\includegraphics[angle=-90]{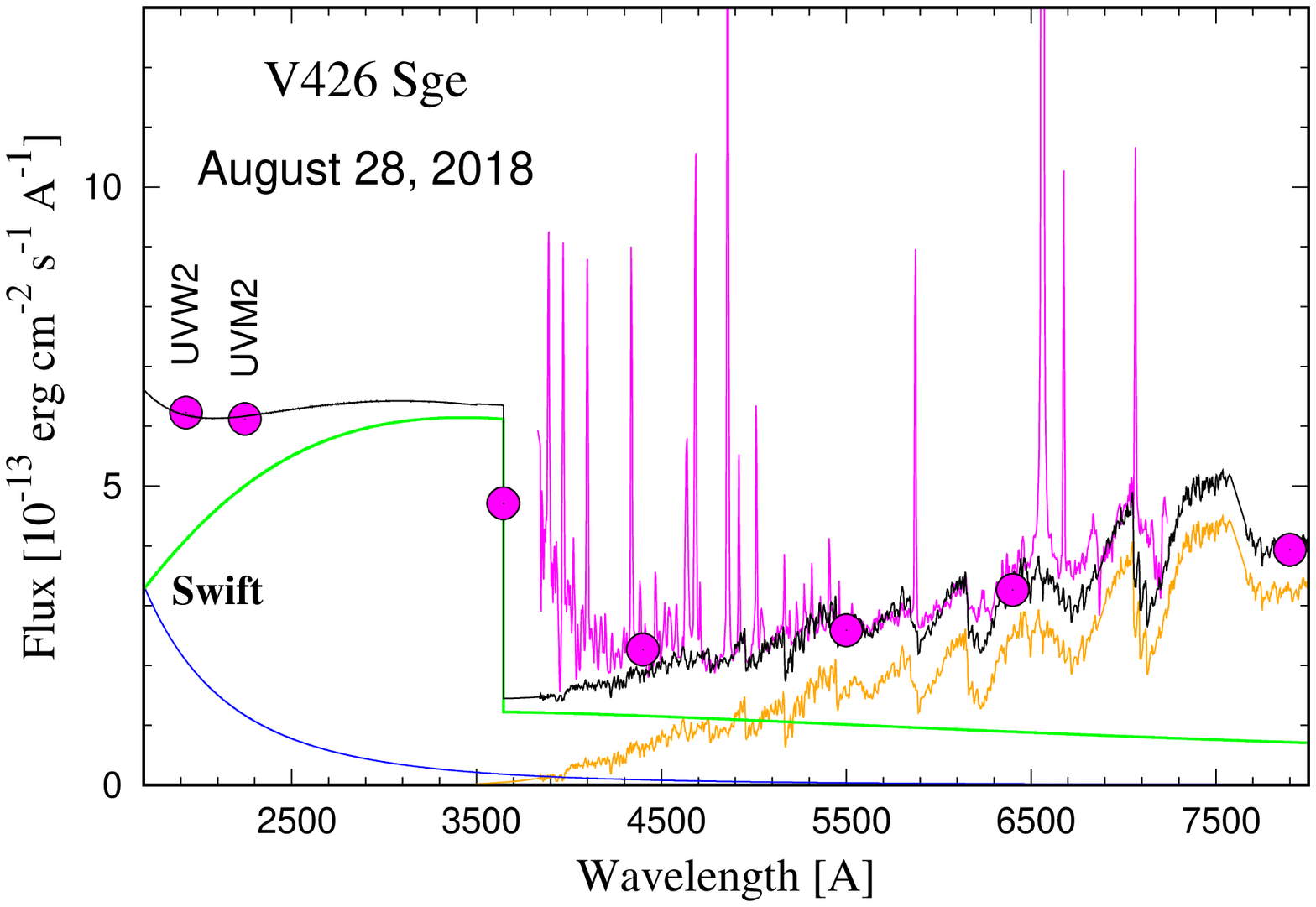}
                    \includegraphics[angle=-90]{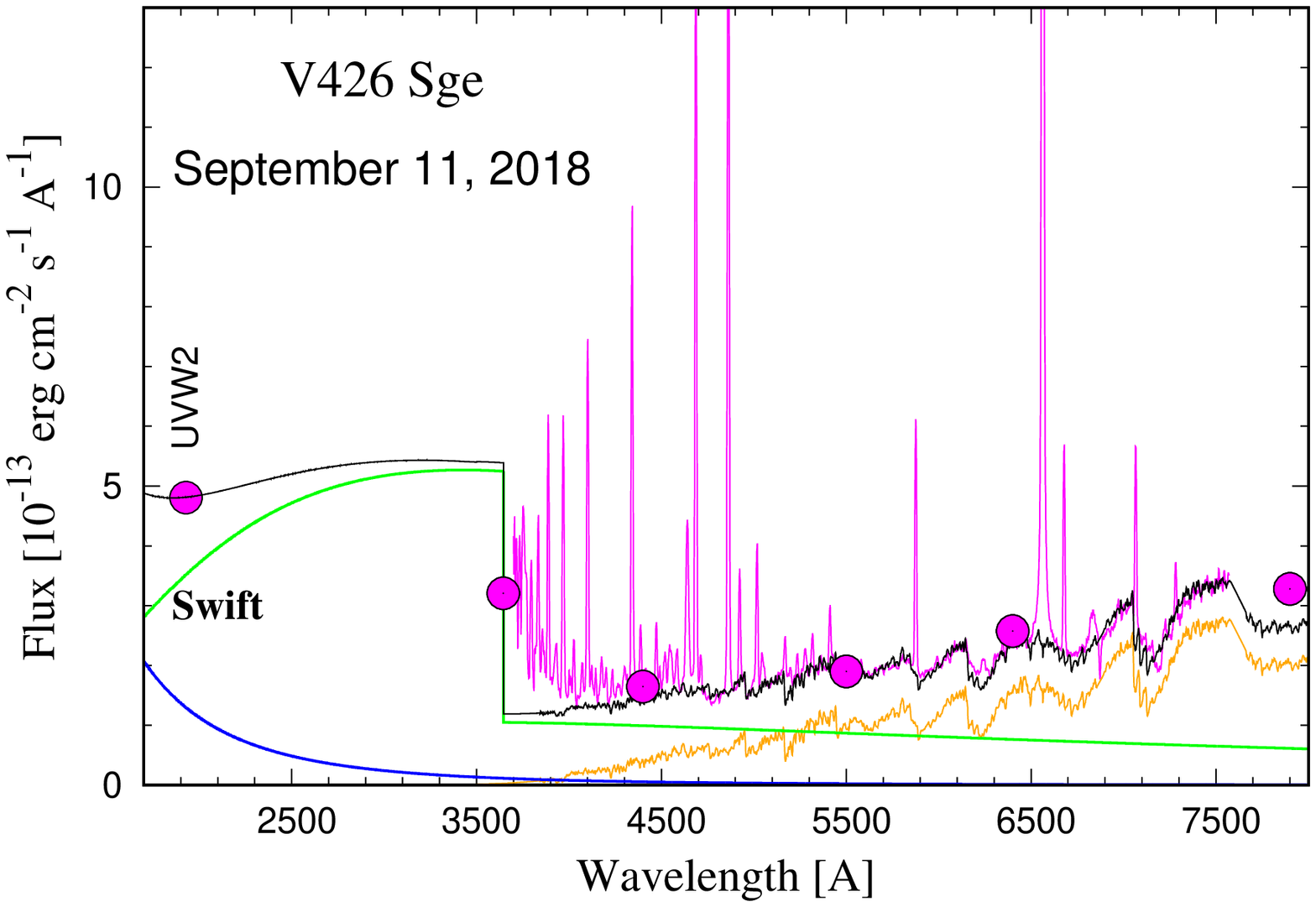}
                    \includegraphics[angle=-90]{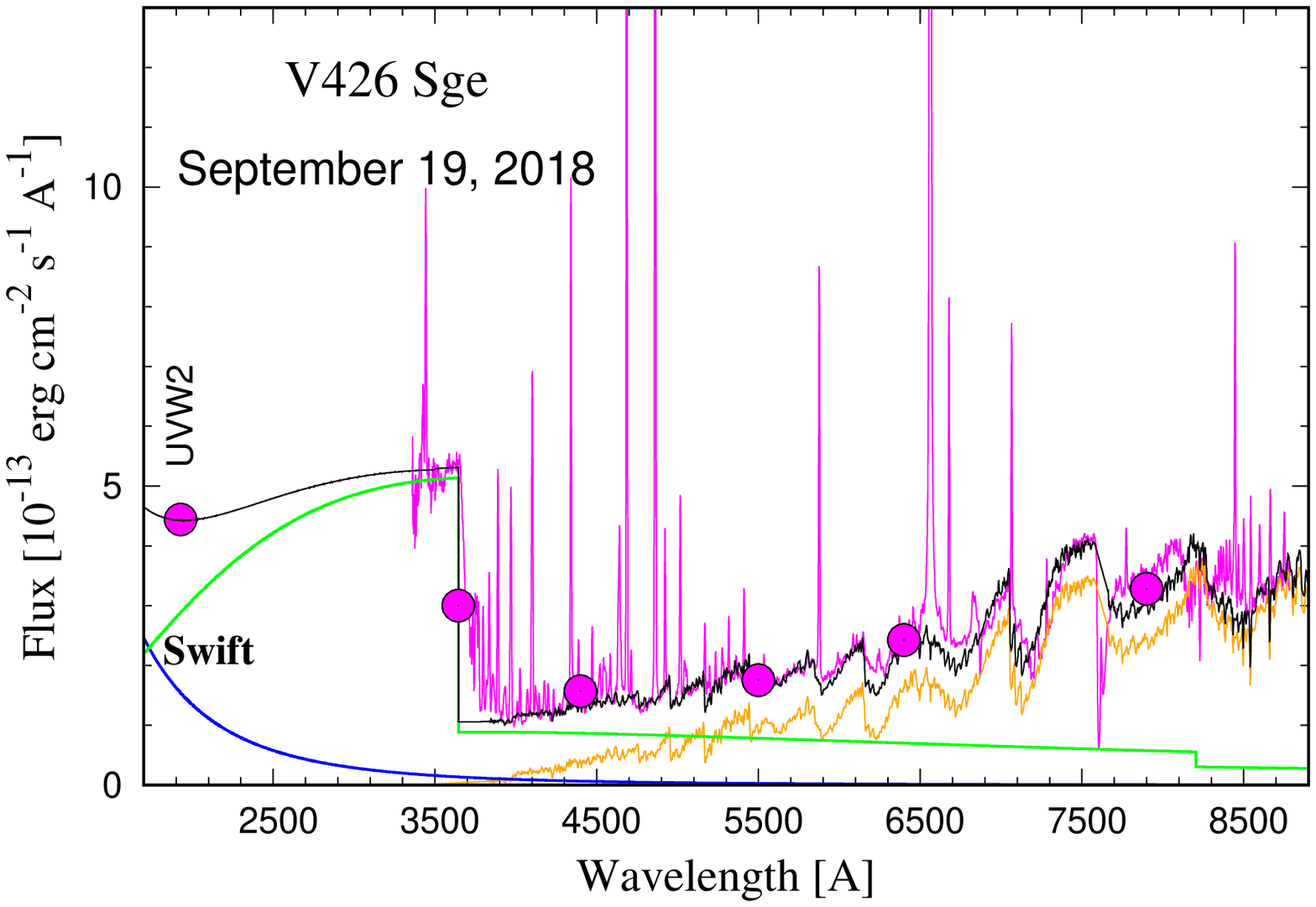}}
\resizebox{18cm}{!}{\includegraphics[angle=-90]{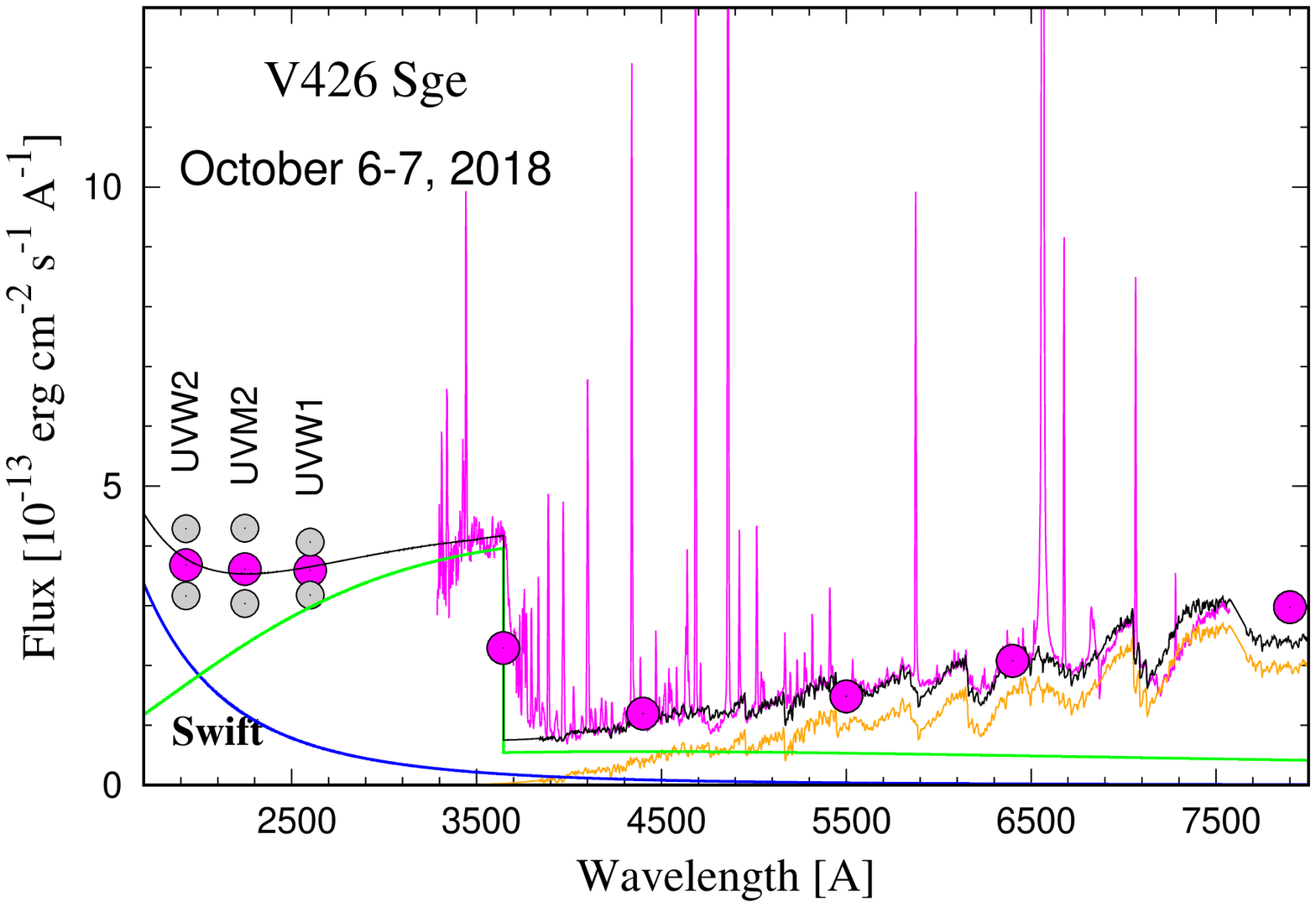}
                    \includegraphics[angle=-90]{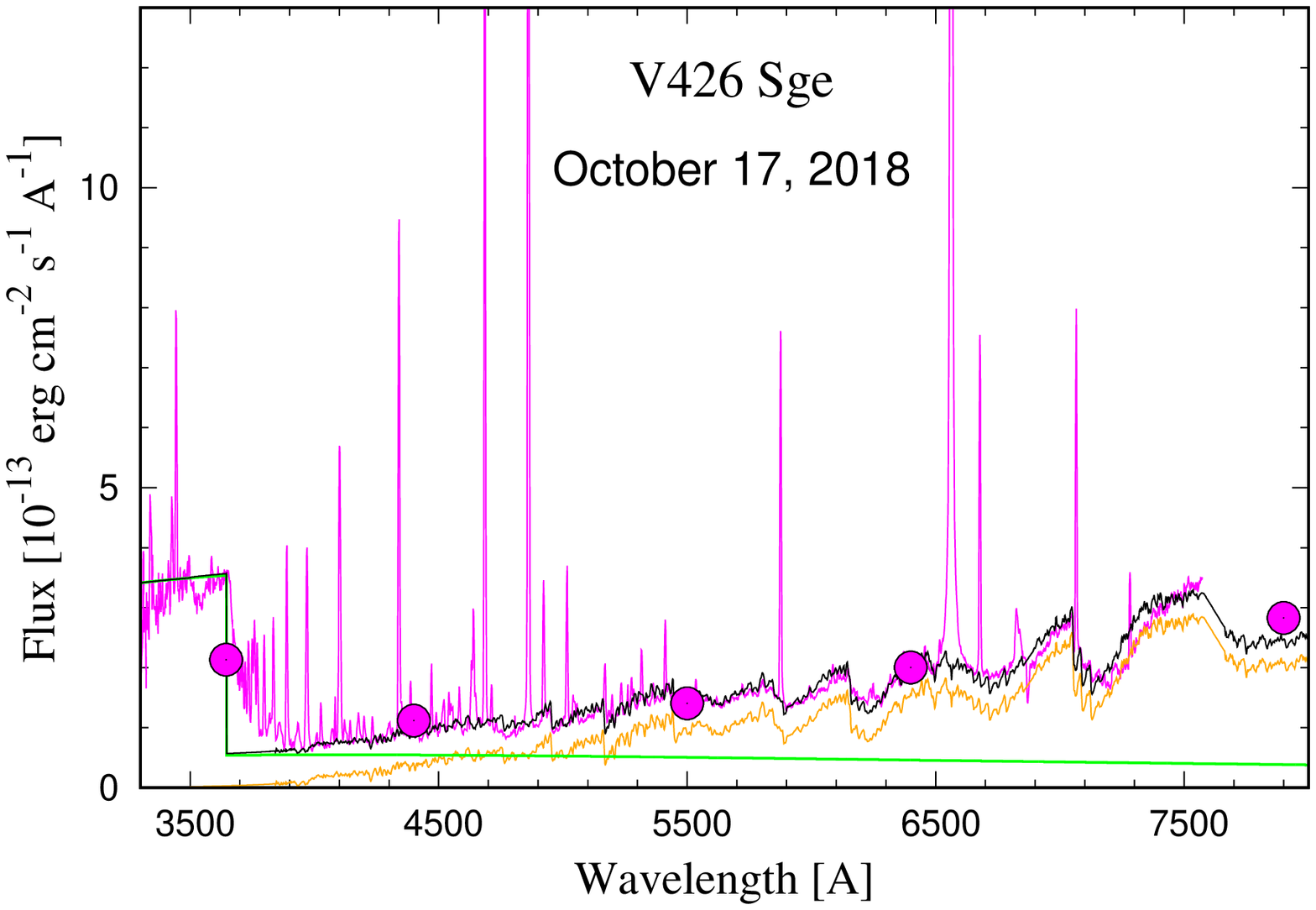}
                    \includegraphics[angle=-90]{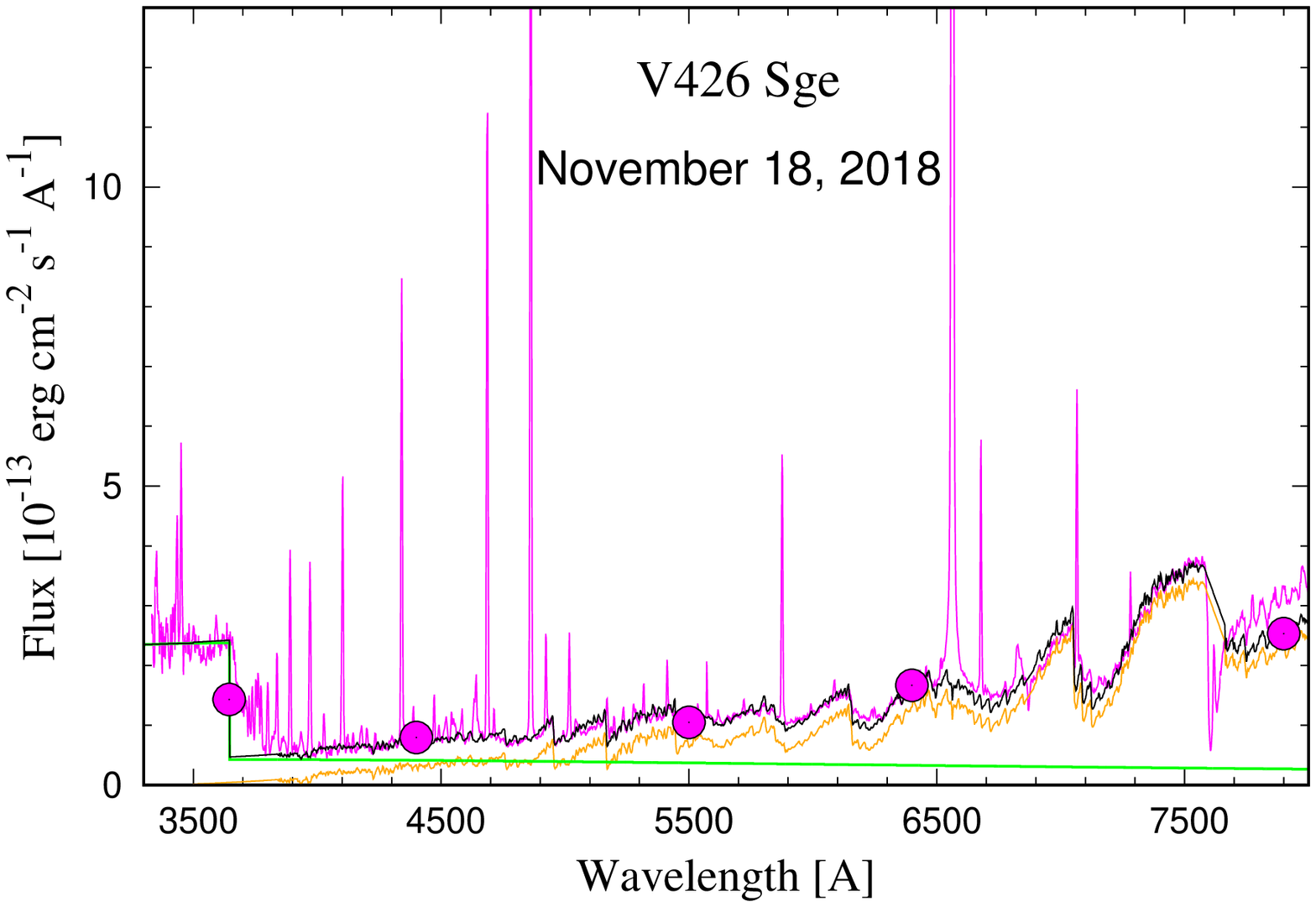}}
\resizebox{18cm}{!}{\includegraphics[angle=-90]{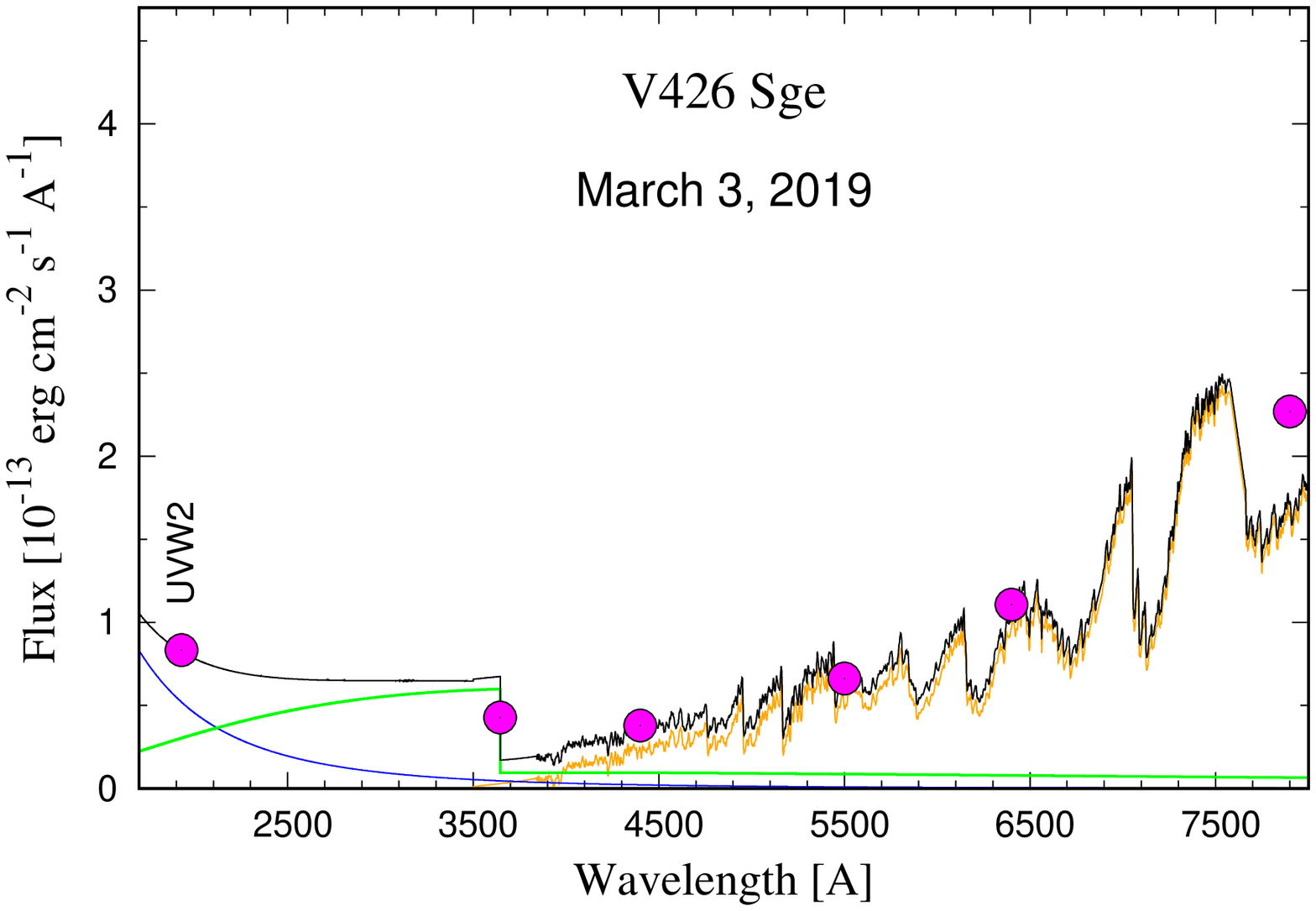}
                    \includegraphics[angle=-90]{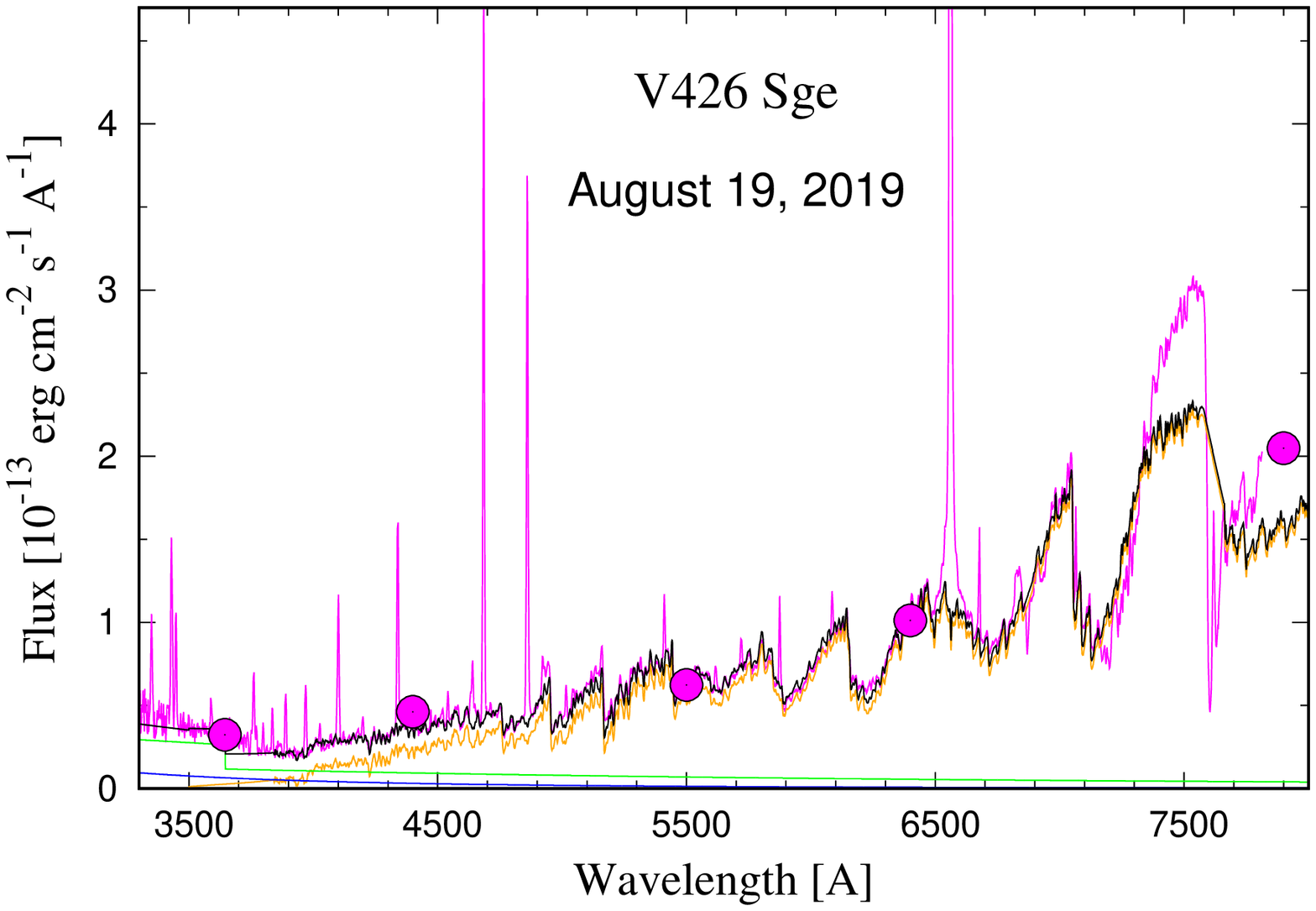}
                    \includegraphics[angle=-90]{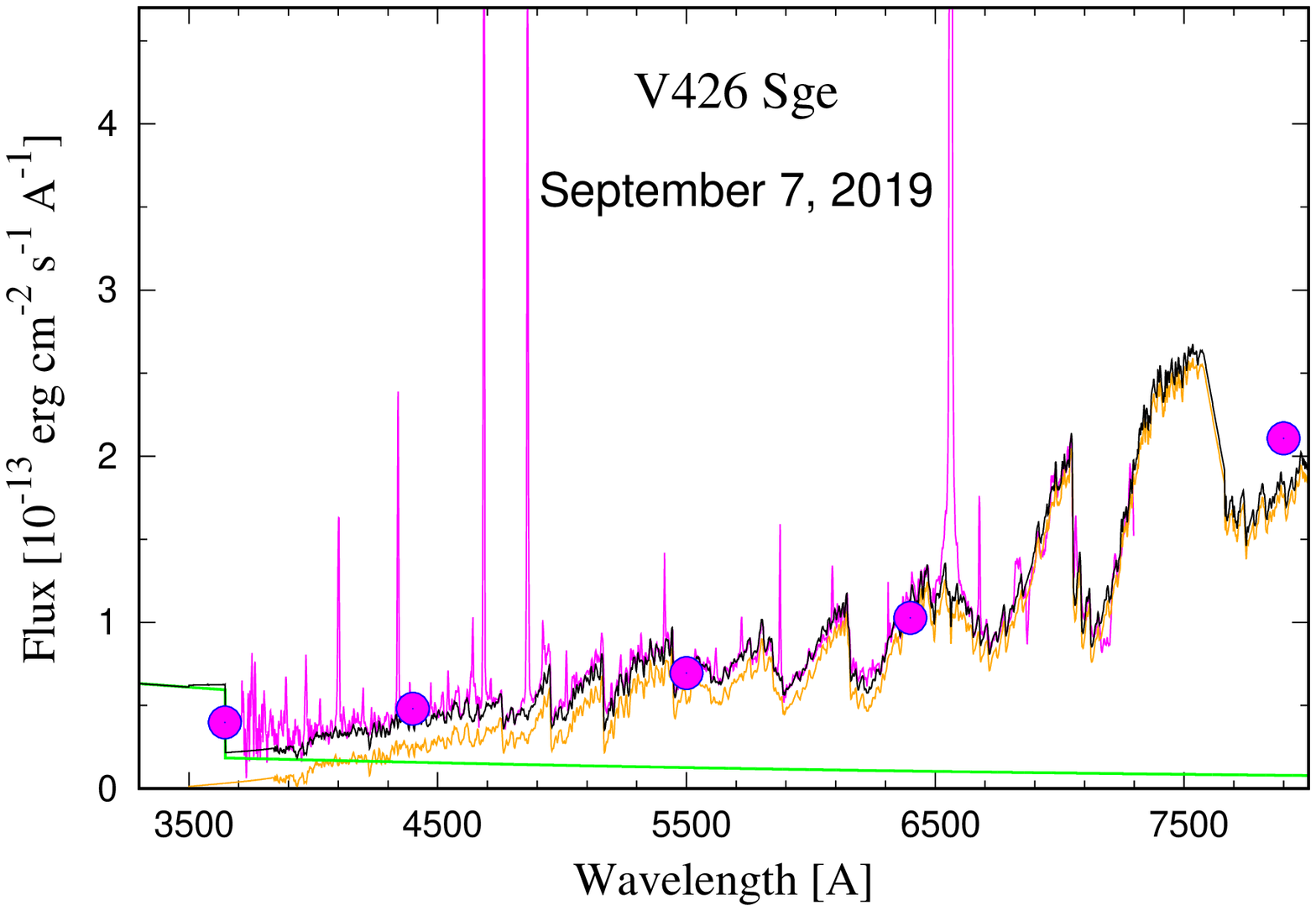}}
\end{center}
\caption[]{
Examples of the observed (in magenta) and modelled (black lines) 
SEDs of V426~Sge at selected dates during its 2018 outburst 
and 2019 quiescent phase (bottom row). 
Lines and symbols are as shown in the keys (top left panel). 
Grey circles in the panel (3,1) are {\em Swift}-UVOT 
fluxes dereddened with the upper and lower limit of the 
$E_{\rm B-V}$ excess, 0.22 and 0.18\,mag, respectively 
(see Sect.~\ref{sss:sed}). 
          }
\label{fig:sedopt}
\end{figure*}

(ii) 
In modelling the {\em Swift}-UVOT/optical continuum we first 
fitted its optical part following the method (i) above. 
We then added 
a blackbody radiation for $T_{\rm WD}$ interpolated to dates 
of {\em Swift}-UVOT observations (see Table~\ref{tab:lrt}) 
and tuned all other variables to obtain an appropriate 
SED model. 
In this way we estimated the parameters $R_{\rm WD}^{\rm eff}$, 
$T_{\rm e}$, \textsl{EM}, $R_{\rm G}$, $T_{\rm eff}$, and 
$L_{\rm WD}$ from the Stefan-Boltzman law. 
%
In addition, the extreme sensitivity of mainly the UVM2 flux 
($\lambda$ = 2246\,\AA) to the interstellar attenuation by dust 
and a strong dependence of both the height of the Balmer jump and 
the profile of the nebular continuum in the NUV on $T_{\rm e}$ 
allow us to verify the light reddening. 
This is because the lower values of the UV fluxes require a lower 
$T_{\rm e}$, which produces a higher Balmer jump, and vice versa. 
Accordingly, modelling the observations from October 6 and 7, 2018, 
when fluxes in all {\em Swift}-UVOT filters are available and 
the Balmer jump is well defined, allowed us to specify the colour 
excess $E_{\rm B-V}$ to 0.2$\pm 0.02$\,mag. 
The corresponding upper and lower limits of the {\em Swift}-UVOT 
fluxes are shown in the panel (3,1) of Fig~\ref{fig:sedopt}. 

(iii)
In modelling the optical/NIR continuum fluxes, we first 
fitted the optical part of the SED as in point (i), because 
the models of \cite{fluks+94} are calculated only for 
$\lambda < 10\,000$\,\AA. The optical SED was then connected 
with a synthetic spectrum of the same ST selected from a grid 
of models made by \cite{h+99} so as to match the $JHKL$ fluxes. 
Here, we obtained the same fitting parameters as in (i), but 
the well-determined 
bolometric flux of the giant provided trustworthy fundamental 
parameters, that is, $L_{\rm G}$, $R_{\rm G}$, and $T_{\rm eff}$ 
(see Sect.~\ref{ss:rg}, Fig.~\ref{fig:sedir}). 
%
%
%
\begin{table}
\caption[]{Parameters from SED models: ST of the M giant, $T_{\rm e}$, 
           $EM$, and a minimum of the $\chi_{\rm red}^2$ function 
           (see Sect.~\ref{sss:sed}).}
\begin{center}
\begin{tabular}{clccc}
\hline
\hline
\noalign{\smallskip}
Date           &  ST  & $T_{\rm e}$ & $EM$ & $\chi_{\rm red}^2$/d.o.f. \\
yyyy/mm/dd.ddd &      &     (K)     & (10$^{60}$\cmt) &                \\
\noalign{\smallskip}
\hline
\noalign{\smallskip}
\multicolumn{5}{c}{2018 outburst} \\
\noalign{\smallskip}
\hline
\noalign{\smallskip}
 2018/08/10.851       & 3.4 & 45000 & 6.61 & 0.47/1359 \\
 2018/08/10.905$^{a}$ & 3.3 & 40000 & 7.29 & 0.48/982  \\
 2018/08/11.944$^{b}$ & 2.5 & 43000 & 7.34 & 1.11/1330 \\
 2018/08/15.888       & 2.8 & 30000 & 6.54 & 0.34/935  \\
 2018/08/18.896       & 2.7 & 30000 & 6.86 & 0.43/849  \\
 2018/08/19.895       & 3.2 & 29500 & 5.22 & 0.63/1186 \\
 2018/08/21.899       & 2.6 & 24000 & 4.24 & 0.80/1557 \\
 2018/08/28.860$^{c}$ & 3.5 & 21000 & 3.19 & 1.51/1954 \\
 2018/09/01.828       & 3.5 & 21000 & 2.93 & 1.04/1614 \\
 2018/09/07.886       & 3.3 & 20000 & 2.78 & 0.72/894  \\
 2018/09/11.903$^{d}$ & 3.5 & 21000 & 2.74 & 0.65/1283 \\
 2018/09/15.911       & 3.7 & 19500 & 2.57 & 0.67/867  \\
 2018/09/20.888       & 3.7 & 22000 & 2.27 & 0.76/1279 \\
 2018/09/21.318$^{e}$ & 4.3 & 19000 & 2.39 & 1.57/5607 \\
 2018/09/28.911       & 4.2 & 20000 & 2.42 & 0.78/910  \\
 2018/10/07.712$^{f}$ & 3.3 & 16000 & 1.55 & 1.60/1628 \\
 2018/10/17.732       & 3.7 & 17600 & 1.50 & 0.67/1554 \\
 2018/11/18.708       & 4.9 & 19500 & 1.14 & 0.93/1388 \\
 2018/12/08.681       & 4.8 & 22000 & 1.08 & 1.37/1440 \\
 2018/12/17.704       & 4.5 & 14000 & 0.60 & 0.80/1330 \\
\noalign{\smallskip}
\hline
\noalign{\smallskip}
\multicolumn{5}{c}{Quiescent phase} \\
\noalign{\smallskip}
\hline
\noalign{\smallskip}
2019/03/03.179 & 4.8$^{g}$ & 18000$^{h}$ & 0.26  & $^{i}$    \\
2019/06/23.872 & 4.8       & 15000$^{h}$ & 0.18  & $^{i}$    \\
2019/08/19.832 & 4.6       & 45000       & 0.28  & 0.83/1418 \\
2019/09/07.191 & 4.8       & 30000$^{h}$ & 0.45  & 0.76/1018 \\
2019/10/05.112 & 4.6       & 40000$^{h}$ & 0.72  & 0.39/~935 \\
\noalign{\smallskip}
\hline
\end{tabular}
\end{center}
{\bf Notes.} 
$^{(a)}$\,spectra in Table~\ref{tab:low}, 2018/08/10.878 and 2018/08/10.932, 
$^{(b)}$\,2018/08/11.915 and 2018/08/11.972, 
$^{(c)}$\,UVW2, UVM2 from 2018/08/28.7 and spectrum 2018/08/28.860, 
$^{(d)}$\,UVW2 from 2018/09/11.45 and spectrum 2018/09/11.903, 
$^{(e)}$\,UVW2 from 2018/09/18.10 and spectra 2018/09/19.827 and 
         2018/09/22.808, 
$^{(f)}$\,UVW2, UVM2, UVW1 from 2018/10/06.44 and spectrum 
         2018/10/07.712, 
$^{(g)}$\,adapted value, 
$^{(h)}$\,given by the $U$ flux-point, 
$^{(i)}$\,a comparison only. 
\label{tab:sed}
\end{table}  
%
%

\subsubsection{Temperature of the WD pseudophotosphere}
\label{sss:twd}
According to the presence of strong \ion{He}{ii}\,$\lambda$4686 
and \hb\ emission lines in the spectrum of V426~Sge, we 
apply the \ion{He}{ii}($\lambda 4686$)/\hb\ method to estimate 
the temperature of the WD pseudophotosphere, $T_{\rm WD}$. 
We introduce our approach as follows. 

(i) The output energy via the \ion{He}{ii}\,$\lambda$4686 
transitions produced by the He$^{+2}$ zone is 
\begin{equation}
 4\pi d^2 F_{\rm 4686} = \alpha_{4686}(T_{\rm e})h\nu_{4686}\,
                         EM({\rm He}^{+2}), 
\label{eq:f4686}
\end{equation}
where $F_{\rm 4686}$ is the observed line flux, 
$EM({\rm He}^{+2}) = n_{\rm e}n({\rm He}^{+2})V$ is the emission 
measure of the He$^{+2}$ zone, $\alpha_{4686}(T_{\rm e})$ is 
the effective recombination coefficient for the transition, 
$n_{\rm e}$ and $n({\rm He}^{+2})$ are mean concentrations of 
electrons and He$^{+2}$ ions, and $V$ is the volume of the He$^{+2}$ 
zone. 

(ii) The input energy for that given by Eq.~(\ref{eq:f4686}) 
is represented by the flux of stellar photons capable of 
ionising He$^{+}$ ions, 
\begin{equation}
 Q(4\nu_0,\infty) = \alpha_{\rm B}({\rm He}^{+},T_{\rm e})
                    EM({\rm He}^{+2}), 
\label{eq:q4oo}
\end{equation}
where $\nu_0$ is the ionising frequency of hydrogen and 
$\alpha_{\rm B}({\rm He}^{+},T_{\rm e})$ is the total recombination 
coefficient of He$^{+}$ for Case B (i.e. the nebula is optically 
thick to ionising radiation). 
Accordingly, substituting $EM({\rm He}^{+2})$ from (\ref{eq:q4oo}) 
to (\ref{eq:f4686}), we can write Eq.~(\ref{eq:f4686}) in the form 
\begin{equation}
 4\pi d^2 F_{\rm 4686} = \alpha_{4686}(T_{\rm e})h\nu_{4686}\,
 \frac{Q(4\nu_0,\infty)}{\alpha_{\rm B}({\rm He}^{+},T_{\rm e})}. 
\label{eq:fhe2}
\end{equation}
Using the same approach for the \hb\ line created within the 
H$^{+}$ zone, we can write the flux ratio 
\citep[see also][]{gurz97}, 
\begin{equation}
 \frac{F_{4686}}{F_{{\rm H}\beta}} = 
 \frac{\alpha_{4686}(T_{\rm e})}{\alpha({\rm H}\beta,T_{\rm e})}\,
 \frac{\alpha_{\rm B}({\rm H}^0,T_{\rm e})}
      {\alpha_{\rm B}({\rm He}^{+},T_{\rm e})}\,
 \frac{\nu_{4686}}{\nu_{4861}}\,
 \frac{Q(4\nu_0,\infty)}{Q(\nu_0,4\nu_0)}, 
\label{eq:ratio}
\end{equation}
where the flux of hydrogen ionising photons, $Q(\nu_0,4\nu_0)$, 
is calculated only between $\nu_0$ and $4\nu_0$, because 
recombinations of He$^{+2}$ to He$^{+}$ within the innermost 
He$^{+2}$ zone ($\nu > 4\nu_0$) produce sufficient quanta to 
keep the hydrogen ionised \citep[see][]{hs64}. 
The principle of this method was first suggested by 
\cite{ambartsumyan}. The method was later elaborated by \cite{harman+66}
and further modified by other authors \citep[e.g.][]{iijima,kj89}. 

Optically thick conditions where the nebula is not transparent 
to ionising radiation are indicated by the presence of elements 
at very different ionisation states in the spectrum, such as during 
the outburst of AG~Peg 
\citep[see][ and references therein]{skopal+17} and the increase 
of particle density around the WD due to the increase of 
the mass-loss rate from the WD (see Sect.~\ref{sss:mdot}). 
Applying Eq.~(\ref{eq:ratio}) to \ion{He}{ii}($\lambda 4686$)/\hb\ 
flux ratios (Table~\ref{tab:fluxes}), 
we used recombination coefficients from \cite{hs87} for 
$T_{\rm e}$ = 30\,000 and 20\,000\,K for August 10 to 22 
and August 27 to December 17, 2018, respectively 
(see Table~\ref{tab:sed}), and for the electron concentration 
of $10^{10}$\cmt\ \citep[e.g.][]{skopal+11}. 
Equation~(\ref{eq:ratio}) can subsequently be expressed as 
\begin{equation}
  \frac{F_{4686}}{F_{{\rm H}\beta}} = 
  \xi\,\frac{Q(4\nu_0,\infty)}{Q(\nu_0,4\nu_0)}, 
\label{eq:ratio2}
\end{equation}
where $\xi$ = 1.30 and 1.33 for $T_{\rm e}$ = 30\,000 and 
20\,000\,K, respectively. The number of quanta $Q$ was calculated for 
Planck's function. Resulting $T_{\rm WD}$ are listed in 
Table~\ref{tab:lrt} and plotted in Fig.~\ref{fig:lrt}. 
During the quiescent phase, the method is not applicable 
because the nebula is only partly ionisation-bounded, that is, 
a fraction of ionising photons escape the nebula without 
being converted to diffuse radiation. 

Uncertainties in $T_{\rm WD}$ are around 5\%. They are given by 
errors in the used \ion{He}{ii}($\lambda 4686$) and \hb\ line 
fluxes (see Table~\ref{tab:fluxes}), the values of which 
depend mostly on the determination of the true continuum. 
Here, the main sources of uncertainty are the noise in the local 
continuum and its calibration to photometric flux points. 

\subsubsection{Luminosity and radius of the ionising source}
\label{sss:lrwd}
Having only the optical spectra, we can only indirectly estimate 
the luminosity $L_{\rm WD}$, under the assumption that 
the total flux of hydrogen ionising photons produced by 
the WD pseudophotosphere, $Q(\nu_0,\infty)$, is balanced by 
the total rate of recombinations within the ionised volume, 
that is, 
\begin{equation}
  Q(\nu_0,\infty) = 
      \alpha_{\rm B}({\rm H^0},T_{\rm e})\,\textsl{EM}, 
\label{eq:QH}
\end{equation}
where $\alpha_{\rm B}({\rm H^0},T_{\rm e})$ is the recombination 
coefficient to all but the ground state of hydrogen 
(i.e. Case $B$). 
Having the quantity of \textsl{EM} from SED models we can then 
estimate the corresponding $L_{\rm WD}$ for the given temperature 
of the ionising source, $T_{\rm WD}$ (Sect.~\ref{sss:twd}), 
according to expression, 
\begin{equation}
  L_{\rm WD} = \alpha_{\rm B}({\rm H^0},T_{\rm e})\,\textsl{EM}
               \frac{\sigma T_{\rm WD}^{4}}{f(T_{\rm WD})},   
\label{eq:lwdem}
\end{equation}
where the function $f(T_{\rm WD})$ determines the flux of 
ionising photons emitted by the 1\,cm$^2$ area of the ionising 
source \citep[see][ in detail]{skopal+17}. The effective 
radius of the WD pseudophotosphere, $R_{\rm WD}^{\rm eff}$, 
is then given by the Stefan-Boltzmann law. 

\begin{figure}
\begin{center}
\resizebox{\hsize}{!}{\includegraphics[angle=-90]{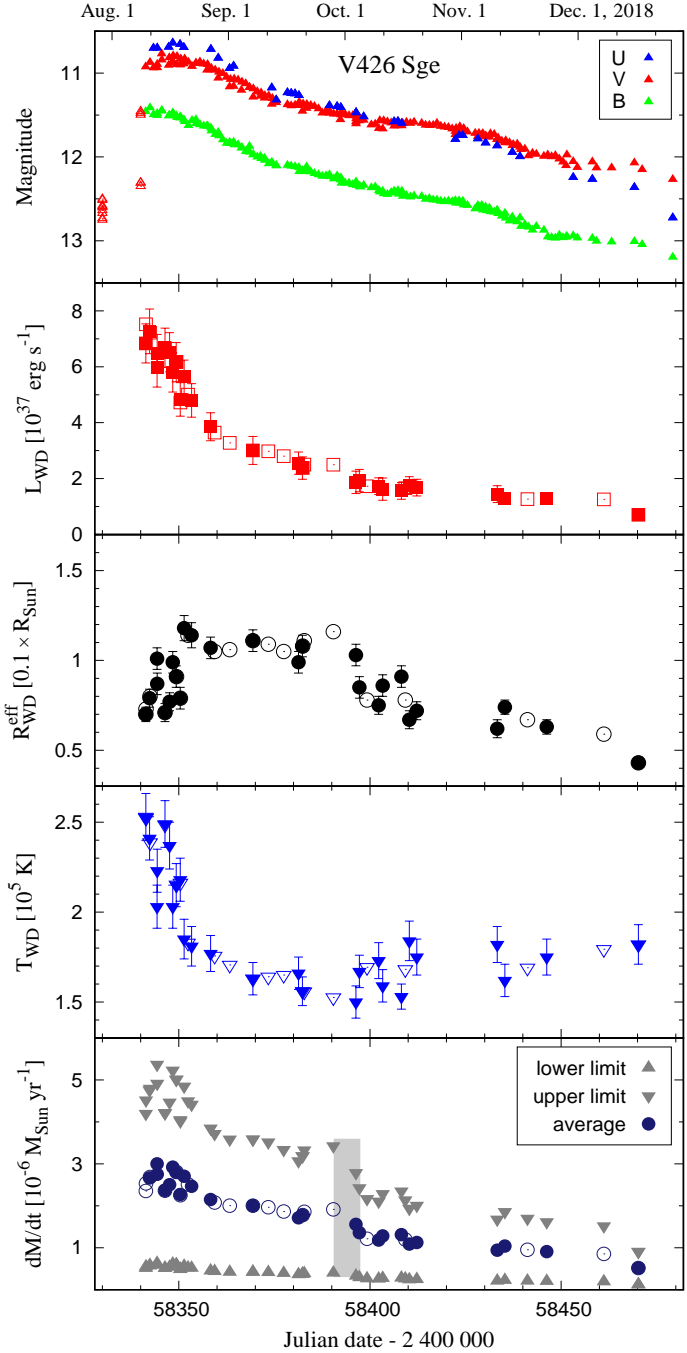}}
\end{center}
\caption[]{
Evolution of the parameters $L_{\rm WD}$, $R_{\rm WD}^{\rm eff}$, 
$T_{\rm WD}$ and $\dot M_{\rm WD}$ throughout the 2018 outburst of 
V426~Sge (data from Table~\ref{tab:lrt}). 
The meaning of filled and open symbols is explained in 
Sect.~\ref{sss:lrwd}. 
The grey belt in the bottom panel denotes the time of a drop 
in $\dot M_{\rm WD}$ (see Sect.~\ref{ss:h-r}). 
          }
\label{fig:lrt}
\end{figure}

Having $T_{\rm WD}$ from high-resolution spectra 
(Table~\ref{tab:med}) and \textsl{EM} from low-resolution 
spectra (Table~\ref{tab:low}) we interpolated \textsl{EM} 
values to dates of high-resolution spectra. Corresponding 
$L_{\rm WD}$ from Eq.~(\ref{eq:lwdem}) with other parameters 
are listed in Table~\ref{tab:lrt} and their temporal evolution 
along the outburst is shown in Fig.~\ref{fig:lrt} with filled 
symbols. We also interpolated $T_{\rm WD}$ values to dates 
of low-resolution spectra and plotted the corresponding 
parameters in Fig.~\ref{fig:lrt} with open symbols. 
Parameters $L_{\rm WD}$ and $R_{\rm WD}^{\rm eff}$ 
determined directly from the radiation of the WD pseudophotosphere 
with the aid of modelling the {\em Swift}-UVOT/optical 
continuum (Sect.~\ref{sss:sed}) are shown in Table~\ref{tab:lrt}. 
Their quantities are comparable to those obtained indirectly 
from the \textsl{EM}. However, their large scatter is a result 
of less accurate scaling and the temperature 
estimated from line ratios. 
%
%
%
\begin{table}
\caption[]{Parameters of the WD pseudophotosphere, 
           $L_{\rm WD}$ (10$^{37}$\es), 
           $R_{\rm WD}^{\rm eff}$ (\ro), 
           $T_{\rm WD}$ (10$^{5}$\,K) and 
           $\dot{M}_{\rm WD}$ (10$^{-6}$\myr) 
along the 2018 outburst of V426~Sge 
(see Sects.~\ref{sss:twd}, \ref{sss:lrwd} and \ref{sss:mdot}).}
\begin{center}
\begin{tabular}{ccccc}
\hline
\hline
\noalign{\smallskip}
Date                       & 
$L_{\rm WD}$               & 
$R_{\rm WD}^{\rm eff}$     &
$T_{\rm WD}$               &
$\dot{M}_{\rm WD}$         \\
\noalign{\smallskip}
\hline
\noalign{\smallskip}
 2018/08/10.845&  6.8$\pm$0.7&  0.070$\pm$0.004&  2.53$\pm$0.13&   2.36 \\
 2018/08/11.826&  7.3$\pm$0.8&  0.079$\pm$0.005&  2.41$\pm$0.12&   2.67 \\
 2018/08/13.814&  6.0$\pm$0.7&  0.101$\pm$0.006&  2.03$\pm$0.12&   3.00 \\
 2018/08/13.827&  6.5$\pm$0.7&  0.087$\pm$0.006&  2.23$\pm$0.12&   2.75 \\
 2018/08/15.853&  6.7$\pm$0.7&  0.071$\pm$0.005&  2.49$\pm$0.13&   2.37 \\
 2018/08/17.070&  6.5$\pm$0.7&  0.077$\pm$0.005&  2.37$\pm$0.13&   2.50 \\
 2018/08/17.896&  5.8$\pm$0.7&  0.099$\pm$0.006&  2.03$\pm$0.12&   2.92 \\
 2018/08/18.823&  6.2$\pm$0.7&  0.091$\pm$0.006&  2.15$\pm$0.12&   2.80 \\
 2018/08/19.842&  4.8$\pm$0.6&  0.079$\pm$0.006&  2.18$\pm$0.12&   2.27 \\
 2018/08/20.869&  5.6$\pm$0.6&  0.118$\pm$0.007&  1.85$\pm$0.11&   2.71 \\
 2018/08/22.840&  4.8$\pm$0.6&  0.114$\pm$0.007&  1.81$\pm$0.11&   2.47 \\
 2018/08/27.796&  3.9$\pm$0.5&  0.107$\pm$0.006&  1.77$\pm$0.10&   2.15 \\
 2018/09/07.867&  3.0$\pm$0.5&  0.111$\pm$0.006&  1.63$\pm$0.09&   2.00 \\
 2018/09/19.821&  2.6$\pm$0.4&  0.099$\pm$0.006&  1.66$\pm$0.09&   1.71 \\
 2018/09/20.837&  2.4$\pm$0.4&  0.108$\pm$0.006&  1.56$\pm$0.08&   1.78 \\
 2018/10/04.840&  1.9$\pm$0.4&  0.103$\pm$0.006&  1.50$\pm$0.09&   1.56 \\
 2018/10/05.724&  1.9$\pm$0.4&  0.085$\pm$0.006&  1.67$\pm$0.09&   1.36 \\
 2018/10/10.788&  1.7$\pm$0.3&  0.075$\pm$0.005&  1.73$\pm$0.10&   1.18 \\
 2018/10/11.835&  1.6$\pm$0.4&  0.086$\pm$0.006&  1.59$\pm$0.09&   1.28 \\
 2018/10/16.713&  1.6$\pm$0.3&  0.091$\pm$0.006&  1.53$\pm$0.07&   1.31 \\
 2018/10/18.797&  1.7$\pm$0.3&  0.067$\pm$0.005&  1.84$\pm$0.11&   1.09 \\
 2018/10/20.757&  1.7$\pm$0.3&  0.072$\pm$0.005&  1.75$\pm$0.10&   1.13 \\
 2018/11/10.773&  1.4$\pm$0.3&  0.062$\pm$0.005&  1.82$\pm$0.10&   0.95 \\
 2018/11/12.803&  1.3$\pm$0.2&  0.074$\pm$0.004&  1.62$\pm$0.09&   1.04 \\
 2018/11/23.729&  1.3$\pm$0.2&  0.063$\pm$0.004&  1.75$\pm$0.10&   0.91 \\
 2018/12/17.701&  0.7$\pm$0.1&  0.043$\pm$0.003&  1.82$\pm$0.11&   0.52 \\
\noalign{\smallskip}
\hline
\noalign{\smallskip}
    \multicolumn{5}{c}{From {\em Swift}-UVOT/optical SED models} \\
\noalign{\smallskip}
\hline
\noalign{\smallskip}
 2018/08/28.722&  5.4$\pm$0.6& 0.128$\pm$0.008& 1.76$^{a}$& -- \\
 2018/09/11.451&  2.9$\pm$0.4& 0.107$\pm$0.007& 1.64$^{a}$& -- \\
 2018/09/18.101&  3.4$\pm$0.5& 0.116$\pm$0.007& 1.65$^{a}$& -- \\
 2018/10/06.440&  4.7$\pm$0.6& 0.133$\pm$0.008& 1.67$^{a}$& -- \\
 2019/03/03.179& 0.89$\pm$0.13&0.071$\pm$0.005& 1.50$^{b}$& -- \\
\noalign{\smallskip}
\hline
\end{tabular}
\end{center}
{\bf Notes.} 
$^{(a)}$~Interpolated value to dates of {\em Swift}-UVOT 
        observations, $^{(b)}$~adopted value
\label{tab:lrt}
\end{table}  
%
%
%

\subsubsection{Mass-loss rate from the burning WD}
\label{sss:mdot}
A strong nebular emission with \textsl{EM} of a few times 
10$^{60}$\cmt\ develops during active phases of symbiotic 
stars \citep[][]{sk05a}. It is produced by the ionised wind 
from the burning WD, the emissivity of which is given by 
the mass-loss rate \citep[][]{sk06}. 
Therefore, having \textsl{EM} from SED models and adopting 
an appropriate model for the stellar wind, we can determine 
the corresponding mass-loss rate, $\dot M_{\rm WD}$. 

As in the case of AG~Peg, we consider the $\beta$-law wind 
according to \cite{lamcass99}. The wind begins at the radial 
distance from the WD centre $r = R_0$ with the initial velocity 
$a$ and becomes optically thin at $r = R_{\rm in}$. 
Further, the velocity profile of the wind is characterised by 
the acceleration factor $\beta$ and the terminal velocity 
$v_{\infty}$. Assuming a spherically symmetric wind around 
the WD, we can express a relationship between \textsl{EM} and 
$\dot M_{\rm WD}$ of the ionised wind as \citep[][]{skopal+17}, 
\begin{equation}
\textsl{EM} = \xi\,
              \Big(\frac{\dot M_{\rm WD}}{v_{\infty}}\Big)^{2}
              \frac{1}{b R_0 (1-2\beta)}
              \Big[1-\Big(1-\frac{b R_0}{R_{\rm in}}
              \Big)^{1-2\beta}\Big]~,
\label{eq:em}
\end{equation}
where $\xi = 1.45\times 10^{46}$\,g$^{-1}$ and the parameter 
$b = 1-(a/v_{\infty})^{1/\beta}$. 
Using Eq.~(\ref{eq:em}) we determined $\dot M_{\rm WD}$ 
for \textsl{EM} (Table~\ref{tab:sed}), $v_{\infty} = 2000$\kms\ 
(from \ha\ wings, Fig.~\ref{fig:broad}), $\beta \sim 1.7$, 
$a$ = 50\kms\ and assuming $R_{\rm in} = R_{\rm WD}^{\rm eff}$ 
\citep[see Sect.~3.2.6. of][ and references therein]{skopal+17}. 
Because a significant fraction of the wind emission is produced 
within its densest innermost part, the corresponding 
$\dot M_{\rm WD}$ strongly depends on the wind origin, $R_{\rm 0}$, 
which, however, cannot be derived from observations. 
Therefore, according to the theory of the optically thick wind 
in nova outbursts \citep[e.g.][]{k+h94}, we consider two 
limiting values of $R_{\rm 0}$ within the shell around 
the WD: 
\begin{enumerate}
\item
$R_{\rm 0} = R_{\rm WD}^{\rm eff}$, which corresponds to 
the lower limit of $\dot M_{\rm WD}$. Our values of \textsl{EM} 
and $R_{\rm WD}^{\rm eff}$ imply 
$\dot M_{\rm WD} = 2 - 5\times 10^{-7}$\myr. 
\item
$R_{\rm 0} = R_{\rm WD}$, which provides the upper limit of 
$\dot M_{\rm WD}$. Our measurements and $R_{\rm WD} \equiv 0.01$\ro\ 
correspond to $\dot M_{\rm WD} = 1 - 5\times 10^{-6}$\myr. 
\end{enumerate}
Table~\ref{tab:lrt} presents the average of these limiting 
values that are between $\sim$1 and $\sim$3$\times 10^{-6}$\myr, 
while Fig.~\ref{fig:lrt} shows all quantities of $\dot M_{\rm WD}$. 

\subsection{Raman-scattered \ion{O}{vi} lines}
\label{ss:raman}
\begin{figure}
\begin{center}
\resizebox{\hsize}{!}{\includegraphics[angle=-90]{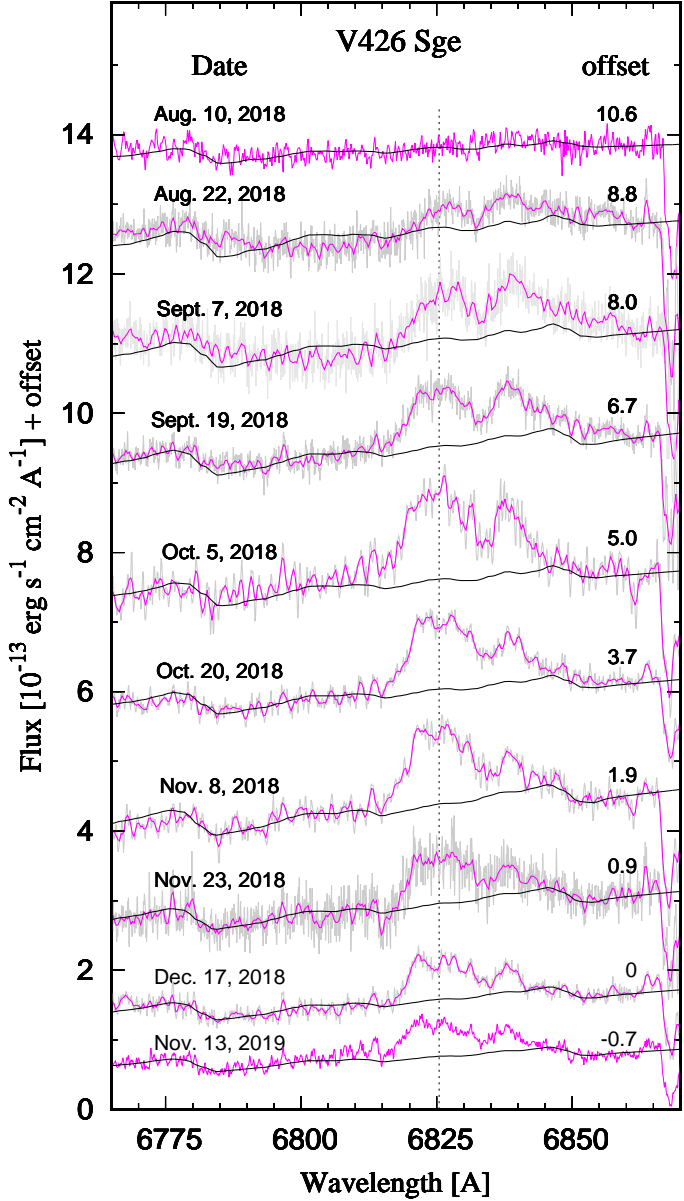}}
\end{center}   
\caption[]{
Evolution of the Raman-scattered \ion{O}{vi}\,6825\,\AA\ line 
along the 2018 outburst of V426~Sge. The grey and magenta lines 
show original and smoothed spectra, respectively, the vertical 
dotted line represents the wavelength of the Raman transition 
at 6825.44\,\AA, the black lines represent the continuum, and 
the numbers on the right mark the spectra offset. 
          }
\label{fig:ramevol}
\end{figure}

The evolution of the Raman-scattered \ion{O}{vi}\,6825\,\AA\ line 
profile and its flux along the outburst of V426~Sge are very 
similar to those observed during the 2015 outburst of AG~Peg 
\citep[see Figs.~5 and 9 of][ and Figs.~\ref{fig:ramevol} and 
\ref{fig:ramflux} here]{skopal+17}. 
Around the optical maximum, during August 10-12, 2018, the 
Raman line was not detectable, and hardly recognisable 
until August 19. 
From August 20, the Raman line was clearly seen above 
the continuum showing two emission bumps in its profile 
at $\sim$6827 and $\sim$6838\,\AA. Both components were 
comparable in the profile and flux until about September 7. 
The blueshifted component then became stronger and broader 
being located around the Raman transition at 6825.44\,\AA\ 
during the whole active phase. 
The redshifted component was stable at its position with 
no significant variation in the profile and flux 
(see Fig.~\ref{fig:ramevol}). 
Finally, it is of interest to note that maximum fluxes of the 
Raman line observed during the 2015 outburst of AG~Peg 
($\sim$5$\times 10^{-11}$\ecs) and during the 2018 outburst 
of V426~Sge ($\sim$2.2$\times 10^{-12}$\ecs) correspond to 
similar luminosities of 
$\sim$3.8$\times 10^{33}\,(d/0.8\,{\rm kpc})^2$\es\
and 
$\sim$2.9$\times 10^{33}\,(d/3.3\,{\rm kpc})^2$\es\ 
for AG~Peg and V426~Sge, respectively. 
%
During the following quiescent phase, our only high-resolution 
spectrum from November 2019 indicates a similar profile to that 
from the end of the outburst, but fainter by a factor of 
about 0.5 (Fig.~\ref{fig:ramevol}, Table~\ref{tab:fluxes}). 
Similar evolution of the Raman line profile and flux during 
recent outbursts of AG~Peg and V426~Sge suggests that their 
ionisation structure was similar (see Sect.~\ref{ss:compagpeg}). 

\begin{figure}
\begin{center}
\resizebox{\hsize}{!}{\includegraphics[angle=-90]{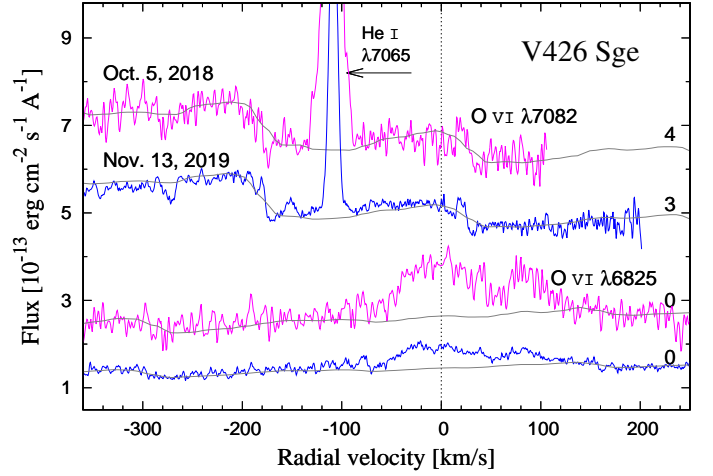}}
\end{center}   
\caption[]{Example of the Raman scattered \ion{O}{vi}\,7082 
and 6825\,\AA\ lines measured during the outburst (in magenta) 
and quiescence (in blue). Other denotations as in 
Fig.~\ref{fig:ramevol}. 
          }
\label{fig:7082}
\end{figure}

\begin{figure}
\begin{center}
\resizebox{\hsize}{!}{\includegraphics[angle=-90]{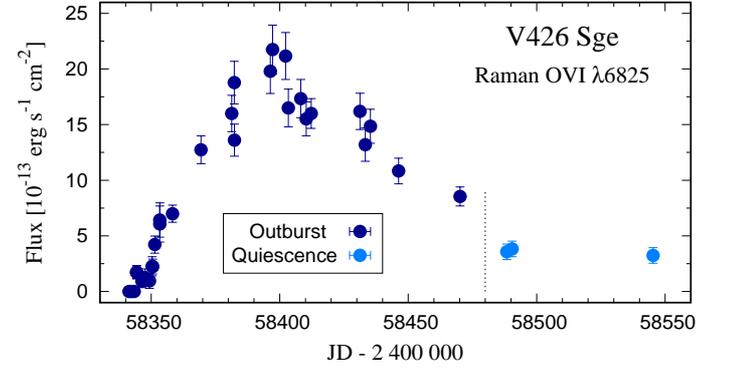}}
\end{center}   
\caption[]{
Fluxes of the Raman-scattered \ion{O}{vi}\,6825\,\AA\ line 
during the outburst (dark blue) and following quiescence 
(light blue). Values from quiescence were shifted by -170 days 
for better comparison. 
          }
\label{fig:ramflux}
\end{figure}

In regards to the Raman \ion{O}{vi}\,7082\,\AA\ line, its 
profile was difficult to indicate in the spectrum. If we can 
associate the small emission at $\sim$+20\kms\ to the 
\ion{O}{vi}\,7082\,\AA\ line (see Fig.~\ref{fig:7082}), then 
the flux ratio $F(6825)/F(7082)$ is between 10 and 30, which 
is anomalously high. 
For the ratio of the parent \ion{O}{vi} lines 
$F(1032)/F(1038) \sim 1$, \cite{lee+16} found the flux 
ratio $F(6825)/F(7082)$ to be between 1 and 3, depending on 
the H$^0$ column density. 
However, in some cases we observe this ratio to be around 6 
\citep[][]{birriel+00} and probably even larger for AG~Peg 
\citep[see Fig.~10 of][]{skopal+17}. 
A very high ratio in the range of $F(1032)/F(1038) \sim 7 - 20$ 
was reported by \cite{birriel+98} and \cite{schmid+99}, who 
explained this anomaly by a significant attenuation of the 
\ion{O}{vi}\,1038\,\AA\ line by the interstellar H$_2$ absorption. 
If we can assume the presence of the H$_2$ molecules also within 
the circumstellar H$^0$ region, then H$^0$ column densities of 
10$^{23-24}$\cmd\ with the abundance of H/H$_2 = 10^{2-3}$ 
give a flux ratio of $F(6825)/F(7082) = 10-20$ or more 
\citep[see Fig.~8 of][]{schmid+99}. 
However, this suggestion requires theoretical verification, which 
is beyond the scope of this paper. 
%
%
%
%
%
%
%
%

\subsection{Swift-XRT emission from colliding winds}
\label{ss:xrt}
The X-ray emission from V426~Sge during its 2018 outburst is 
very soft, as more than 92\% of X-ray photons are detected 
below 2\,keV (Sect.~\ref{sss:xphot}). 
On the other hand, the super-soft X-rays from a $\sim$180\,kK 
hot WD pseudophotosphere producing a luminosity of 
$\sim 3\times 10^{-8}$\ecs\ at the Earth (see values in 
Table~\ref{tab:lrt} after August 28, 2018 -- the first XRT 
observations) cannot be detected within the {\em Swift}-XRT 
energy range \citep[see Fig.~13 of][ for a comparison]{sk19}. 
Such a location of the X-ray emission from V426~Sge is most 
consistent with the so-called $\beta$-type 
\citep[see][]{murset+97,luna+13}. 
In addition, the X-ray fluxes correspond to luminosities of 
a few times $10^{32}\,(d/3.3\,{\rm kpc})^2$\es\ 
(Table~\ref{tab:resx}) which are typical for symbiotic binaries 
producing this type of X-ray \citep[see Table~2 of][]{mukai17}. 
According to \cite{murset+95}, this type of X-ray spectrum 
can be reproduced by shock-heated plasma as a result of colliding 
winds. Also, \cite{mukai17} connected these X-ray properties with 
the energy source resulting from colliding winds in symbiotic 
binaries. 

The same origin of the {\em Swift}-XRT emission from the outburst 
of V426~Sge is strongly supported by the enhanced wind from the WD 
(Fig.~\ref{fig:lrt}). According to the ionisation structure 
during the outburst (Sect.~\ref{ss:compagpeg}), a biconical wind 
from the WD can collide with the wind from the cool giant in the 
interaction zone well above the orbital plane and at distances greater 
than the size of the binary \citep[e.g. Fig.~4 of][]{murset+95}. 
Therefore, such emission is well observable as it is only 
slightly attenuated by the interstellar absorption corresponding 
to $E_{\rm B-V} = 0.2$\,mag. 

Accordingly, during the quiescent phase, the indication of 
the X-ray emission was negative 
(Table~\ref{tab:resx}, {\em Swift}-XRT pointing on March 3, 2019), 
because the wind from the WD terminated (see Sect.~\ref{ss:trans}). 

\subsection{Transition to quiescent phase}
\label{ss:trans}
\begin{figure}
\begin{center}
\resizebox{\hsize}{!}{\includegraphics[angle=-90]{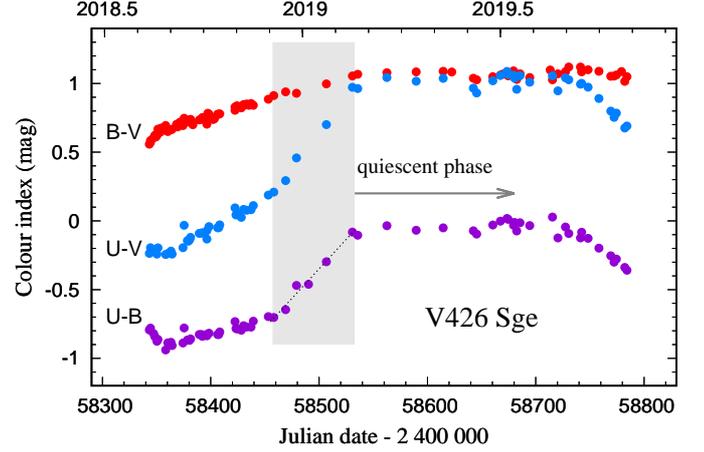}}
\end{center}   
\caption[]{
Transition of V426~Sge from its 2018 outburst to quiescent 
phase as indicated by our $UBV$ photometry (Table~\ref{tab:ubvri}). 
The grey belt indicates a rapid decrease of $\dot M_{\rm WD}$ to 
around the middle of February 2019, when V426~Sge moved to 
quiescent phase (see Sect.~\ref{ss:trans}). 
          }
\label{fig:trans}
\end{figure}

The optical brightening during the outburst was governed 
exclusively by the nebular component of radiation, whose 
contribution dominates the $U$ band (Fig.~\ref{fig:sedopt}). 
Therefore, the $U$ magnitude is also proportional to 
$\dot M_{\rm WD}$ during the outburst, because 
$EM \propto (\dot M_{\rm WD})^{1/2}$ (Eq.~(\ref{eq:em})). 
As a result, a rapid decline in the brightness of the star 
in $U$ from December 5-7, 2018, to February 15, 2019 
(see Fig.~\ref{fig:trans}), reflects a rapid decrease in 
$\dot M_{\rm WD}$. 
After this period, $UBV$ colour indices came to a constant state 
and the LCs indicated development of the wave-like orbitally 
related variation, a typical feature of the quiescent phase 
(see panels {\bf c} and {\bf d} of Figs.~\ref{fig:hlc} and 
\ref{fig:trans}). This means that the wind from the WD, 
and thus the outburst, ceased around the middle of February, 
2019, and V426~Sge moved into a quiescent phase. 
A new source of a relative faint nebular emission 
($EM \sim 2-7\times 10^{59}$\cmt, Table~\ref{tab:sed}) is 
now represented by the ionised fraction of the wind from 
the giant (Sect.~\ref{s:intro}). 

\subsection{Parameters of the red giant}
\label{ss:rg}
\begin{figure}
\begin{center}
\resizebox{\hsize}{!}{\includegraphics[angle=-90]{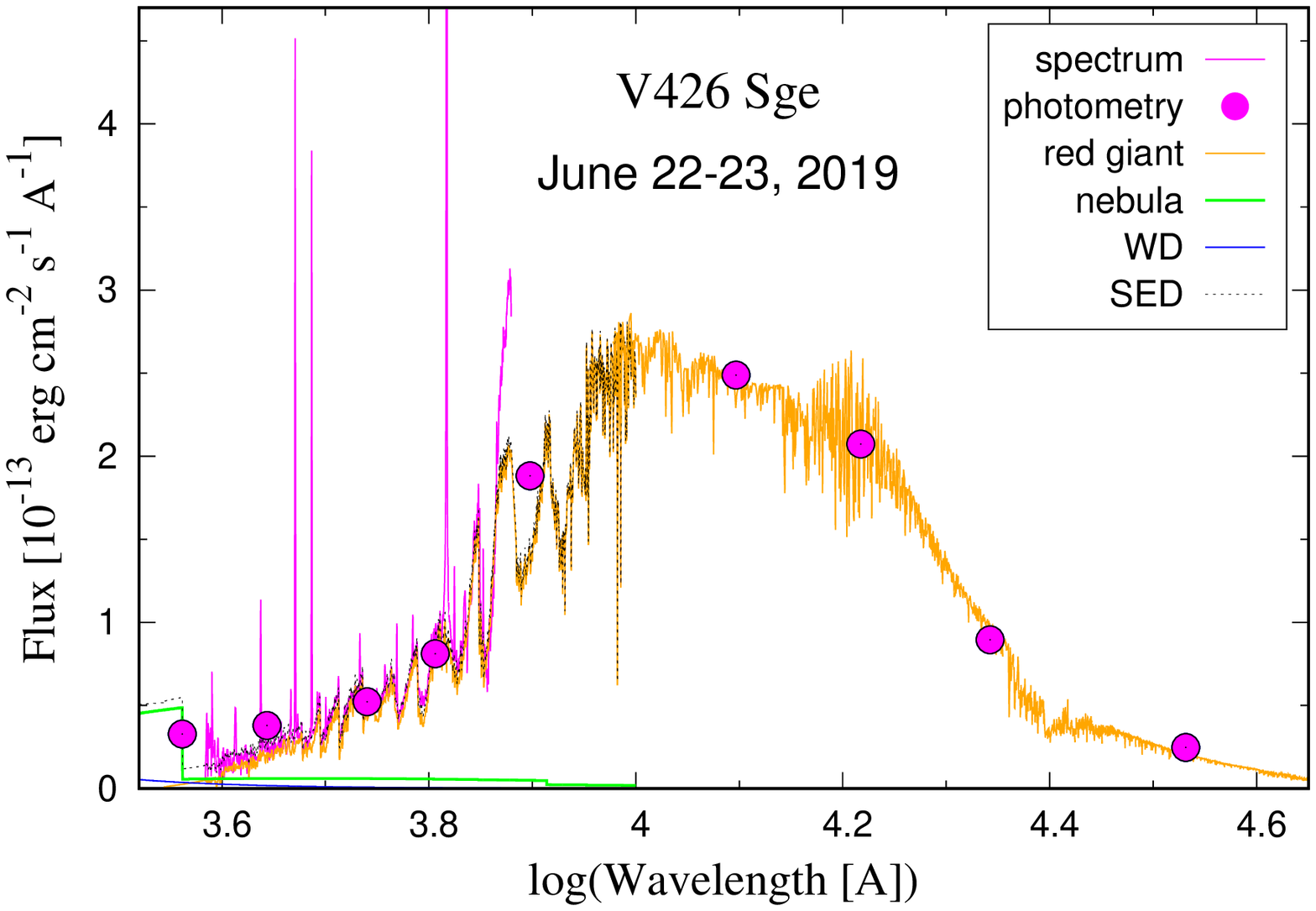}}
\resizebox{\hsize}{!}{\includegraphics[angle=-90]{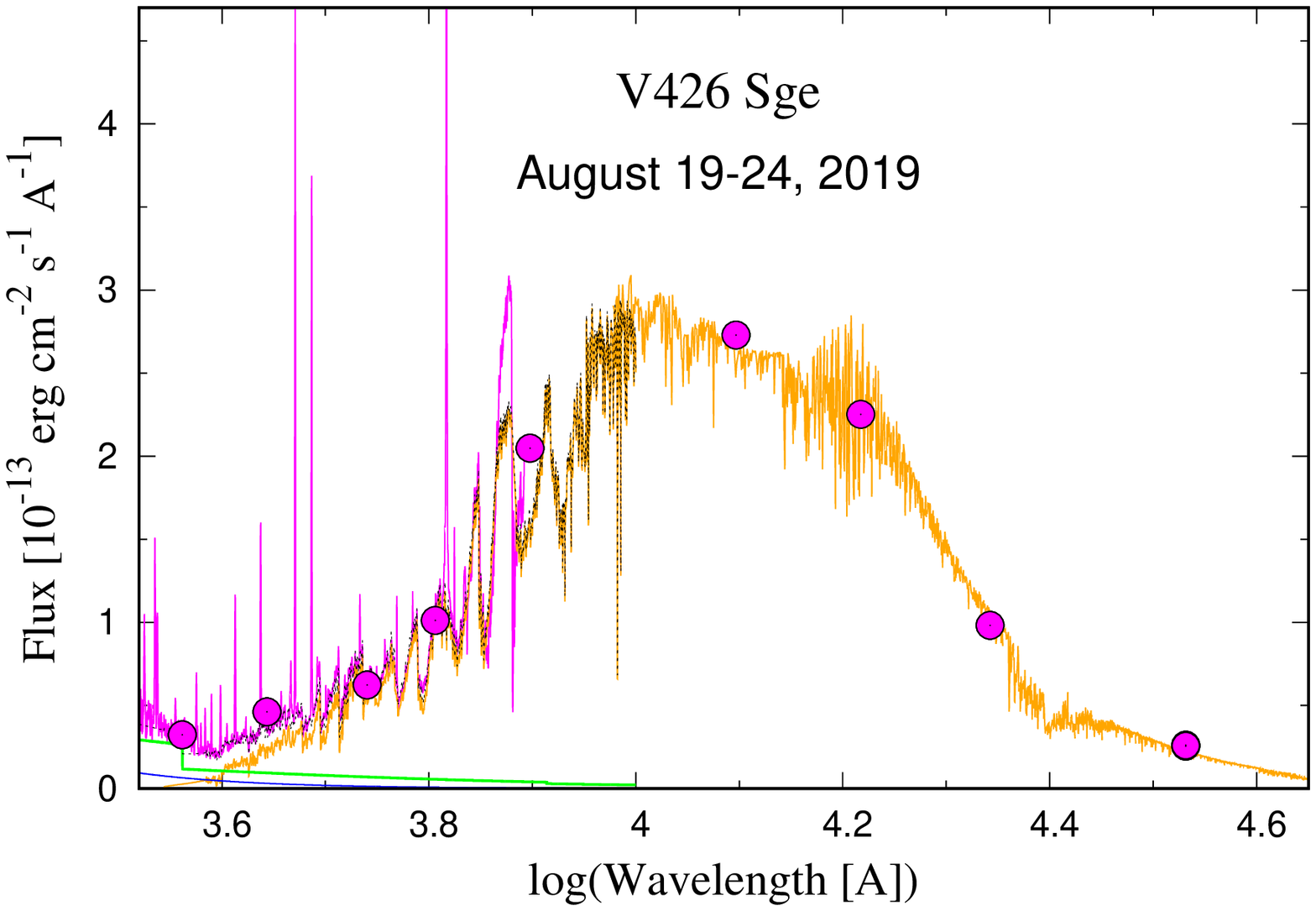}}
\end{center}
\caption[]{
Optical/NIR SED of V426~Sge during quiescent phase. 
The observed SED (low-resolution spectrum and 
$UBVR_{\rm C}I_{\rm C}JHKL$ photometry) is matched by a synthetic 
spectrum, corresponding to spectral type M4.8 (top) and M4.6 (bottom) 
or $T_{\rm eff} \sim 3400$\,K (see Sect.~\ref{ss:rg}). 
          }
\label{fig:sedir}
\end{figure}

To obtain parameters of the red giant in V426~Sge, we 
modelled two SEDs determined by the optical spectrum and 
$UBVR_{\rm C}I_{\rm C}JHKL$ photometry, observed 
almost simultaneously during quiescent phase on June 22 and 23 
and August 19 and 24, 2019 (see Tables~\ref{tab:jhkl} and 
\ref{tab:low} for precise timing). Observations and models 
are depicted in Fig.~\ref{fig:sedir}. 

Fitting the optical spectrum, we determined the ST of the giant 
as M4.7($\pm$0.2)\,{\small III} which corresponds to 
$T_{\rm eff} \sim 3\,400$\,K according to the calibration of 
\cite{fluks+94}. We then matched both the optical and the 
NIR fluxes by a synthetic spectrum calculated for 
$T_{\rm eff} = 3\,400$\,K. 
The model corresponds to the observed bolometric flux 
 $F_{\rm G}^{\rm obs} = 3.85$ and $4.15\times 10^{-9}$\ecs\ 
and the angular radius of the giant, 
 $\theta_{\rm G} = 7.2$ and $7.4\times 10^{-10}$ 
\citep[see Eqs.~(3) and (4) of][]{sk05a}, 
for the model from June and August, 2019, respectively. 
Our values of $\theta_{\rm G}$ are in good agreement with those 
derived from the surface brightness relation for M-giants 
\citep[see][]{ds98}, $\theta_{\rm G} = 7.3$ and 
$7.6\times 10^{-10}$, given by the reddening-free magnitudes 
$J = 7.77$, $K = 6.62$, and $J = 7.67$, $K = 6.52$ for the two 
SED models above, respectively. 
Finally, our values of $\theta_{\rm G}$ give the radius 
  $R_{\rm G} = 104$ and $108\,(d/3.3\,\kpc)$\ro\ 
and the luminosity 
  $L_{\rm G} = 1\,300$ and $1400\,(d/3.3\,\kpc)^2$\lo\ 
for the first and second model SED, respectively. 
Uncertainties for such determined parameters are of $\sim$5\% 
only, but for fixed $T_{\rm eff}$. 

Variability in the giant is suggested by the large scatter in the 
model ST (M2.5 -- M4.9\,{\small III}, see Table~\ref{tab:sed}) 
and in its scaling $\theta_{\rm G}$ (see Fig.~\ref{fig:sedopt}). 
It is of interest to note that the earlier ST is limited to 
the early outburst evolution, while the changes in the angular 
radius seem to be random, probably connected with the intrinsic 
variability of the giant. The former effect was observed also 
during the 2015 outburst of AG~Peg 
\citep[see Table~4 of][]{skopal+17} 
and during the 2015 super-active state of recurrent nova T~CrB 
\citep[][]{munari+16}. 
This effect requires further investigation, which  is beyond 
the scope of this paper. 
%
%
%
%
%

\section{Discussion}
\label{s:dis}
\subsection{Comparison with AG~Peg}
\label{ss:compagpeg}
The SED models during the 2015 and 2018 outburst of AG~Peg and 
V426~Sge are essentially the same: a strong nebular 
component of radiation, which is responsible for the optical 
brightening, a more or less constant contribution from the cool 
giant, and a very hot stellar component from the WD whose 
contribution is negligible in the optical 
\citep[see Fig.~4 of][ and Fig.~\ref{fig:sedopt} here]{skopal+17}. 
Thus, the brightness increase and its evolution along both the 
outbursts have the same origin: injection of particles into the 
particle-bounded nebula around the WD, which as a result becomes 
 ionisation-bounded. This change leads to production of an extra 
nebular emission due to the increase of the rate of recombinations 
that balance the increased flux of ionising photons. Under 
the optically thick conditions we can then derive fundamental 
parameters of the source of the original stellar radiation (which 
is the WD pseudophotosphere) from its fraction converted to the 
nebular radiation (see Sects.~\ref{sss:twd} and \ref{sss:lrwd}). 
In both cases, we derived a more than one order of magnitude 
increase in the \textsl{EM} relative to values from the quiescent 
phase, which corresponds to a mass-loss rate from the WD of a few 
times 10$^{-6}$\myr\ (see Sect.~\ref{sss:mdot}). 

Both stars show also the same type of the X-ray emission during 
their recent outbursts. The spectrum is of $\beta$-type, 
concentrated below 2\,keV and with a luminosity of a few 
times $10^{32}$\es\ \citep[see][ for AG~Peg and Sect.~\ref{ss:xrt} 
for V426~Sge]{ramsay+16}. 

The similar quantities found for the parameters $L_{\rm WD}$, 
$R_{\rm WD}^{\rm eff}$, and $T_{\rm WD}$ of the WD pseudophotosphere 
during the 2015 outburst of AG~Peg and the 2018 outburst of 
V426~Sge 
\citep[see Fig.~7 of][ and Fig.~\ref{fig:lrt} here]{skopal+17} 
suggest that these outbursts are of the same nature. 
The high value of $L_{\rm WD} \sim 6\times10^{37}$\es\ observed 
around the maximum of the 2018 outburst can only be generated by 
nuclear hydrogen burning on the surface of the WD. 
The corresponding accretion rate of $\sim 2\times 10^{-7}$\myr\ 
exceeds the upper limit of the stable-burning regime for 
a low-mass WD \citep[see e.g.][]{shen+bild07}, which leads to 
the production of optically thick wind 
\citep[see e.g.][]{hachisu+96} 
whose particles convert the ionising photons to the nebular 
emission that dominates the NUV/optical domain; we observe 
the Z~And-type outburst of the 2nd-type 
(i.e. the `hot-type' outburst; see Sect.~\ref{ss:h-r}). 

Also during quiescent phase, fundamental parameters of the 
burning WD are comparable for both stars (here, 
Table~\ref{tab:lrt}, the {\em Swift}-UVOT/optical SED model 
on March 3, 2019). 
The luminosity of $\sim$2300\lo\ emitted by the WD photosphere 
at a high temperature of $\approx$150\,kK corresponds to 
a $\sim$0.5\mo\ WD that burns hydrogen on its surface 
at the rate of its accretion 
\citep[see e.g. Fig.~1 of][]{nomoto+07}. 
Thus, both symbiotic binaries also harbour a low-mass WD 
\citep[see][ for AG~Peg]{kenyon+93}. 

Finally, a striking similarity in the evolution of the 
Raman-scattered \ion{O}{vi}\,6825\,\AA\ line profile and 
the increase of its flux during the recent outbursts of AG~Peg 
and V426~Sge (see Sect.~\ref{ss:raman}) suggest similar 
ionisation structure in these two systems: the presence of a neutral 
disc-like zone at the equatorial plane expanding from the WD 
and the ionised region located above and below the disc \citep[see 
Fig.~12 and Sect.~4.7.3. of][ and references therein]{skopal+17}. 
Accordingly, (i) the significant increase of the Raman line flux 
during both outbursts could be caused by an increase of the Raman 
scattering efficiency, because a larger fraction of the O$^{+5}$ 
sky is covered by the neutral hydrogen during outbursts than 
in quiescence. 
During outbursts, the initial O$^{+5}$ photons arising around 
the WD are located just above and below the scattering region 
(= the neutral disc), and can therefore `see' the scattering 
region 
under a large solid angle, while during quiescence the \ion{H}{i} 
zone around the giant (= the neutral fraction of the giant's wind) 
is located far from the O$^{+5}$ zone, and thus occupies a much 
smaller fraction of the O$^{+5}$ sky. 
%
(ii) Broadening of the Raman line and development of a strong 
redshifted component at $\sim$6838\,\AA\ during outbursts 
can arise from such a disc-like zone of scatterers expanding 
from the source of the initial O$^{+5}$ photons. 
This interpretation is independently supported by the presence 
of the Raman-scattered \ion{O}{vi} lines in the spectrum of the 
luminous B[e] star LHA~115-S~18, where the Raman emission 
arises from a dense circumstellar disc of neutral hydrogen 
illuminated by the radiation from the central hot star 
\citep[see][]{torres+12}. 
%
%
\begin{figure}
\begin{center}
\resizebox{\hsize}{!}{\includegraphics[angle=-90]{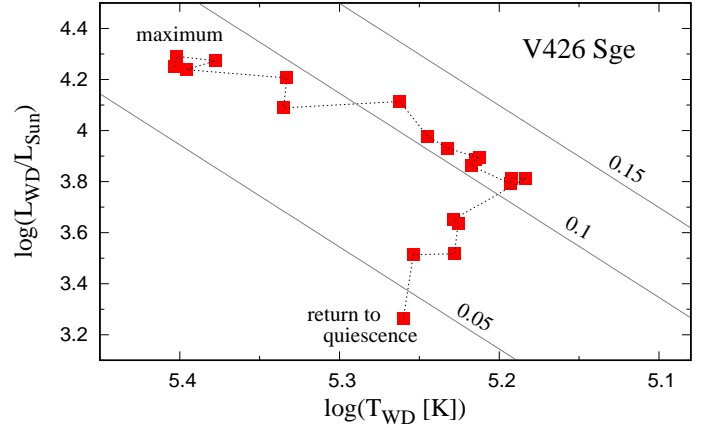}}
\end{center}   
\caption[]{
H-R diagram for the 2018 outburst of V426~Sge from around 
its optical maximum to the return to quiescence as given by 
our ground-based observations (Table~\ref{tab:lrt}). 
The dotted line follows the temporal evolution of the given 
parameters. Full lines represent loci with constant effective 
radii. Labels are in solar units (see Sect.~\ref{ss:h-r}). 
          }
\label{fig:h-r}
\end{figure}

\subsection{The outburst in the H-R diagram}
\label{ss:h-r}
Figure~\ref{fig:h-r} shows the evolution of the WD parameters, 
$L_{\rm WD}$, $R_{\rm WD}^{\rm eff}$, and $T_{\rm WD}$ in the 
H-R diagram from around the maximum of the outburst 
on August 10, 2018, to almost its end on December 17, 2018 
(see Sects.~\ref{sss:2018o} and \ref{ss:trans} for the outburst 
timing and Table~\ref{tab:lrt} for parameters). 
The highest values of $L_{\rm WD}$ and $T_{\rm WD}$ were measured 
around the maximum of the outburst, while the lowest $L_{\rm WD}$ 
and $R_{\rm WD}^{\rm eff}$ were observed close to its end. 
The most interesting feature in the diagram is the `knee', 
when the effective radius starts to shrink more rapidly with a small 
increase of $T_{\rm WD}$ and a gradual decline of $L_{\rm WD}$. 
This change happened between September 28 and October 7, when 
$\dot M_{\rm WD}$ dropped by a factor of $\sim$1.5, from 
$\sim 2\times10^{-6}$ to $\sim 1.3\times 10^{-6}$\myr\ (see grey 
vertical band in the bottom panel of Fig.~\ref{fig:lrt}). 
The following gradual decrease in $\dot M_{\rm WD}$ caused 
a relevant shrinkage of the optically thick/thin interface 
of the wind, i.e., the WD's pseudophotosphere. 

The position of fundamental parameters in the H-R diagram and 
the model SED suggest that the 2018 outburst of V426~Sge is 
of the 2nd-type; see Sect.~\ref{s:intro}. 
These outbursts are characterised by the immediate occurrence 
of a strong nebular emission and of a hot 
($\approx 2\times 10^{5}$\,K) ionising source from the very 
beginning of the outburst. Therefore, it is suitable to refer to 
these outbursts as `hot-type' outbursts. On the other hand, it 
is appropriate to refer to the 1st-type outbursts 
as `warm-type', because they are characterised by a warm 
(1--2$\times10^{4}$\,K) pseudophotosphere dominating 
the optical during outburst. 
According to \cite{sk05a}, this classification implies that 
the V426~Sge binary has a low orbital inclination, which 
allows us to directly see the hot WD and its product during 
outbursts, that is, the ionised high-velocity wind as indicated 
by a remarkable increase of \textsl{EM}. 
Similar, well-observed systems showing hot-type outbursts are: 
AG~Dra \citep[e.g.][]{mika+95,greiner+97,skopal+09,sion+12}, 
AG~Peg \citep[e.g.][]{ramsay+16,skopal+17,sion+19}, and 
LT~Del \citep[see][]{arkhipova+95,ikonnikova+19}. 

\subsection{The path to Z~And-type outbursts}
\label{ss:zandout}

\subsubsection{Observational constraints}
\label{sss:outobs}
\begin{figure}
\begin{center}
\resizebox{\hsize}{!}{\includegraphics[angle=-90]{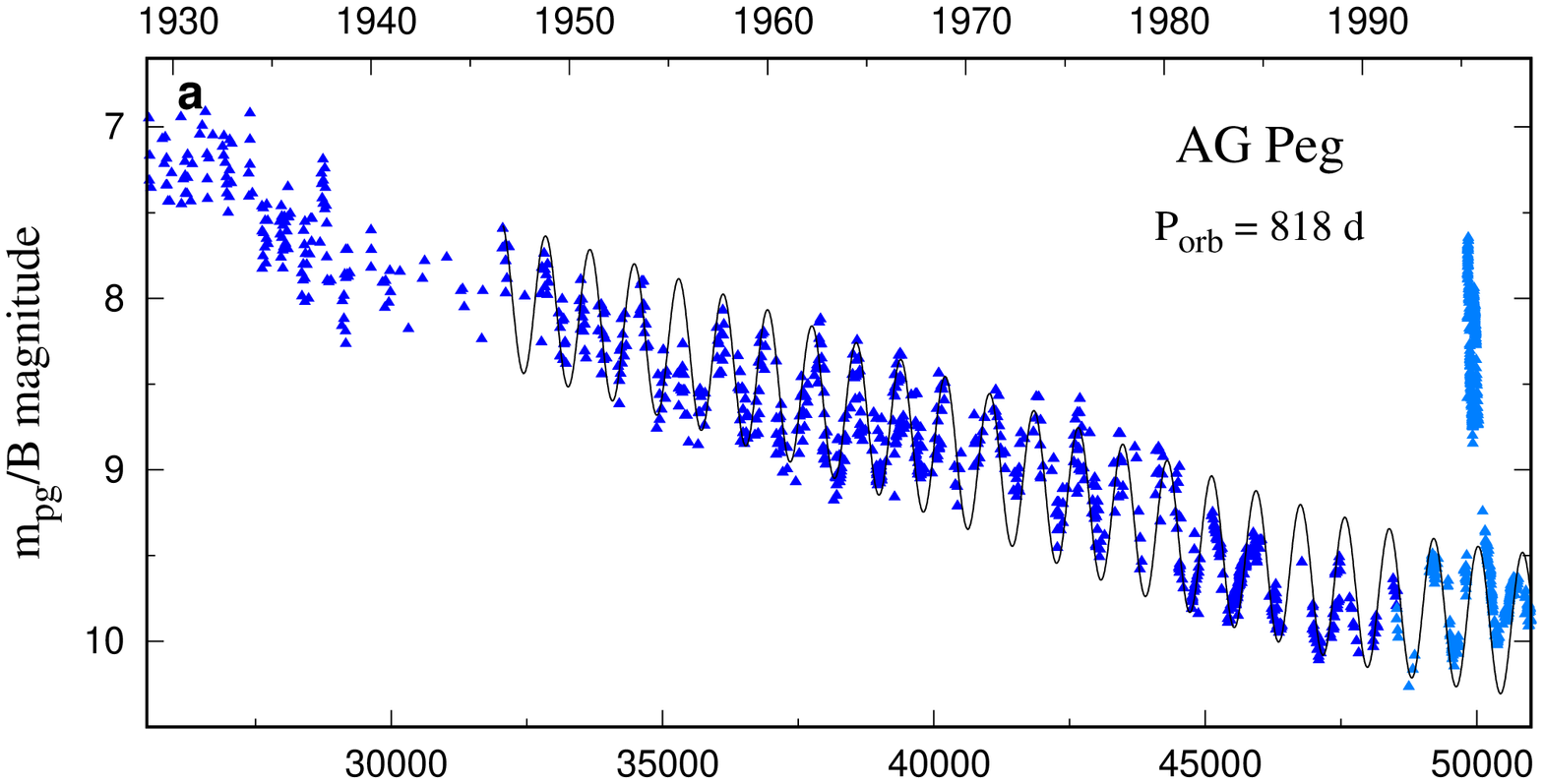}}
\resizebox{\hsize}{!}{\includegraphics[angle=-90]{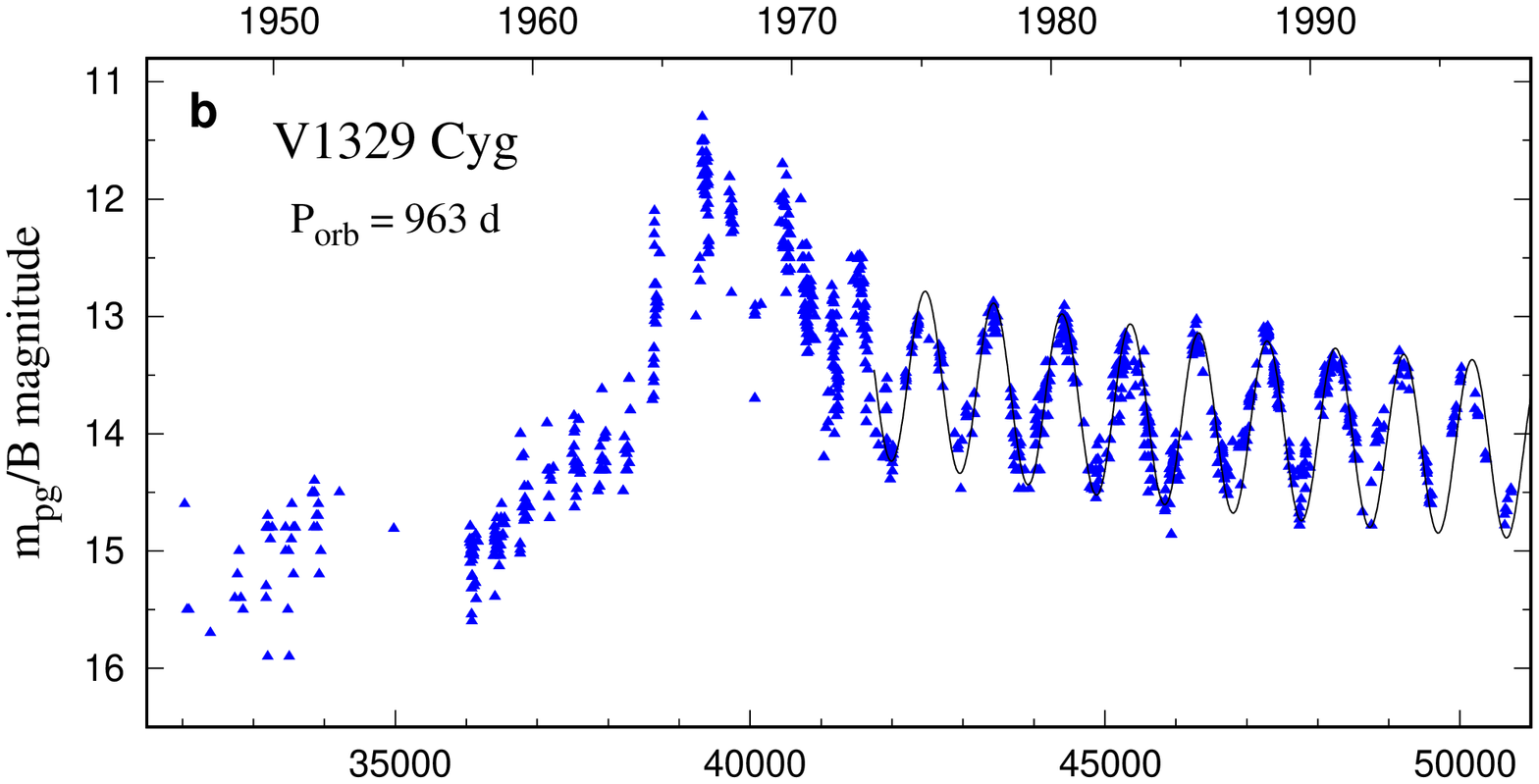}}
\resizebox{\hsize}{!}{\includegraphics[angle=-90]{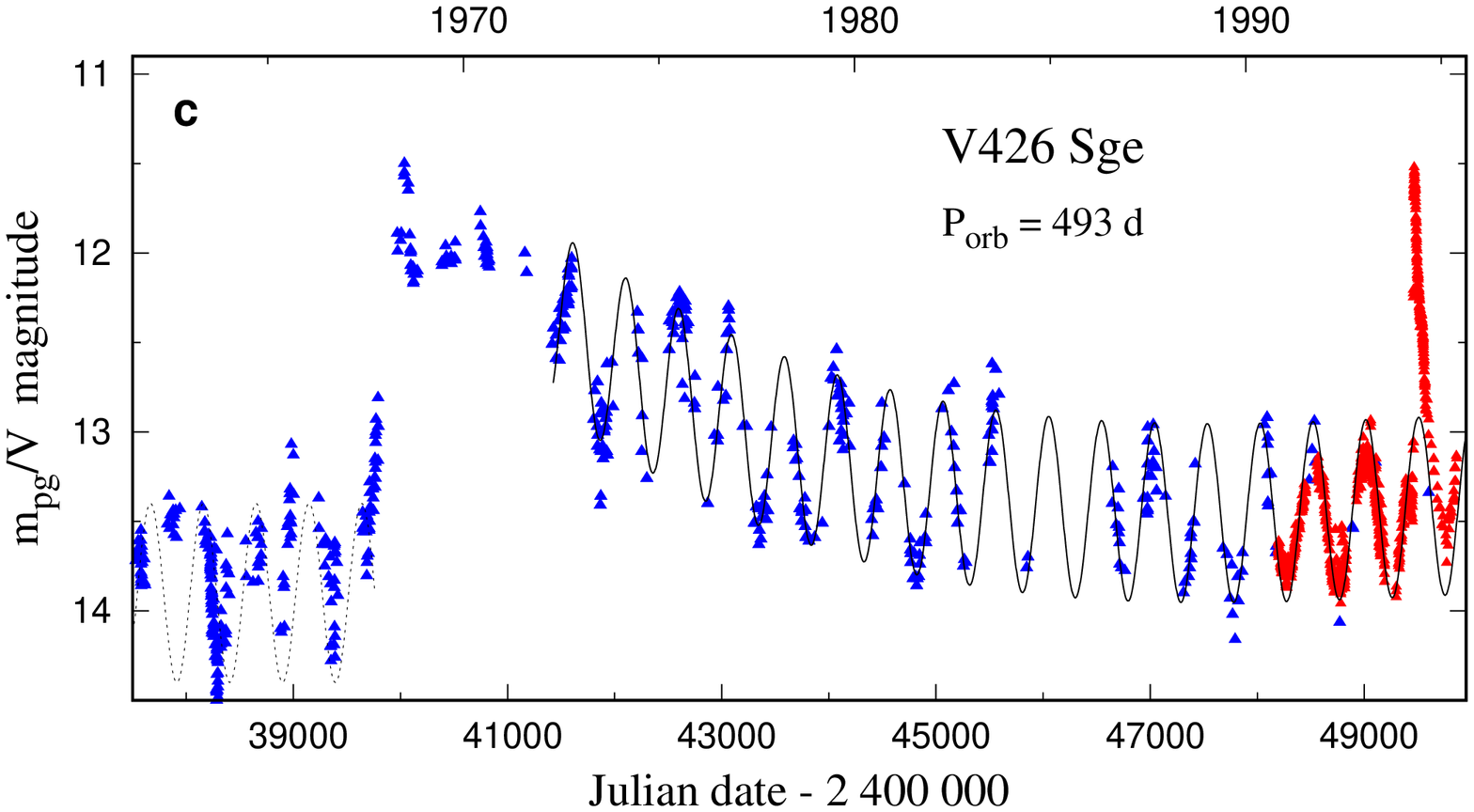}}
\end{center}
\caption[]{
Transition from the symbiotic nova outburst to quiescent phase 
and eventually to Z~And-type outburst for AG~Peg, V1329~Cyg, and 
V426~Sge. The sine curve fits the wave-like orbitally 
related variation. 
{\bf a:} 
Descending part of the historical LC of AG~Peg from its 1850 
symbiotic nova outburst. The 2015 Z~And-type outburst is shifted 
by $-9\times P_{\rm orb}$ for a better comparison. Data are 
from \cite{meinunger83} and \cite{sekeras+19}. 
{\bf b:} 
Historical LC of V1329~Cyg covering its symbiotic nova outburst 
around 1965. Data were summarised by \cite{chochol+99}. 
{\bf c:} 
The 1968 symbiotic nova outburst ($m_{\rm pg}$ in blue) and the 2018 
Z~And-type outburst ($V$ in red) of V426~Sge. Data for the latter 
event are shifted by $-18\times P_{\rm orb}$ and +0.75\,mag. 
          }
\label{fig:snovae}
\end{figure}

Figure~\ref{fig:snovae} shows development of the wave-like 
orbitally related variation in the LCs of symbiotic novae 
AG~Peg, V1329~Cyg, and V426~Sge during the transition from 
their symbiotic nova outburst to quiescent phase. This 
type of the variability also developed  in the LC of RT~Ser 
after its symbiotic nova outburst in 1909 
\citep[see][]{shugarov+03} and in the LC of the symbiotic 
nova PU~Vul during its return from the nova outburst in 1979 
to quiescence \citep[see][]{cuneo+18}. 

After around 20 and 28 years of quiescence, when the star's 
brightness varied around a constant level, AG~Peg and V426~Sge 
underwent an Z~And-type outburst and became classical symbiotic 
stars \citep[see][ for AG~Peg and this paper for 
V426~Sge]{tomov+16,ramsay+16,skopal+17}. 
For V1329~Cyg and V426~Sge, there is around 70 years of 
photometric observations prior to their symbiotic nova outbursts. 
In both cases, the pre-outburst LC does not show the sinusoidal 
light variation along the orbit or any direct evidence of 
symbiotic-like activity; some activity of this kind was recorded, 
but only approximately 5-7 years before the nova outburst. 
In the case of V1329~Cyg, oscillations within $\sim$0.8\,mag 
with a gradual increase by $\sim$1\,mag were indicated, while 
for V426~Sge, three $\sim$1\,mag brightness decreases were 
measured, but these were out of the phase with the post-outburst 
variation 
(see Figs.~\ref{fig:hlc}, \ref{fig:phase} and \ref{fig:snovae}). 
%
%
%

\subsubsection{Accretion-powered symbiotics: Progenitors 
               to symbiotic nova outbursts}
\label{sss:progen}
The event of symbiotic nova outburst requires a long-lasting 
preceding accretion by the WD from the red giant wind to 
accumulate a sufficient amount of hydrogen-rich material 
to ignite a thermonuclear outburst on the surface of the WD 
(see \cite{mn94} and \cite{bode08} for a review). Therefore, 
WDs in symbiotic binaries accreting in the nova regime 
\citep[i.e. below the stable-burning limit; see e.g.][]{nomoto+07} 
are accretion-powered only, and thus represent progenitors 
to symbiotic nova outbursts; they are far less abundant 
among all the known symbiotics\footnote{in part due to 
a selection effect, because their symbiotic activity is 
indicated mostly in the UV and X-rays 
\citep[see][]{luna+13,sokoloski+17}.} 
with only a few well-observed objects in the optical 
\citep[e.g. EG~And, 4~Dra, SU~Lyn, see][]{sk05b,mukai+16}. 
The accretion-powered symbiotics generate a low luminosity of 
$\sim 10^1 - 10^2$\lo, and thus also a low flux of ionising 
photons. As a result, a faint nebular continuum with only a few 
emission lines (mostly \ha, \hb\ and some forbidden lines) are 
superposed on the red giant continuum in the optical. 
Examples for EG~And were illustrated by 
\cite{smith80}, \cite{munari93}, and \cite{kenyon+16}, 
for CQ~Dra by \cite{wheatley+03} and \cite{sk05a} and 
for SU~Lyn by \cite{mukai+16} and \cite{teyssier+19}. 

Their photometric variability in the optical is also low. 
Observed are 0.1--0.2\,mag variations, which can be modulated 
with the orbit (EG~And), and $\sim$0.1\,mag oscillations 
probably caused by a variability of the giant \citep[see the 
long-term photometry of EG~And and CQ~Dra by][]{sekeras+19,
hric+91,hric+94}. However, changes in the ultraviolet and X-rays 
can be significant, as recently reported for SU~Lyn by 
\cite{lopes+18}, who measured a dramatic drop in the 
UV-to-X-ray luminosity ratio on a timescale of 9 months. 
The authors interpreted this event as being a consequence of 
the almost 90\,\% decrease in the accretion rate. This suggests 
that the activity of red giants in symbiotic stars is important 
for fueling their WDs, and can therefore be used to determine 
their evolutionary path. 

\subsubsection{Evolution from the symbiotic nova 
               to the first Z~And-type outburst}
\label{sss:aftersno}
At a certain point during the decline from the symbiotic nova 
outburst, the sinusoidal variation develops along the orbit. 
This signals a gradual transition to quiescent phase, where 
a dense symbiotic nebula is represented by the ionised fraction 
of a slow massive wind from the cool giant (Sect.~\ref{s:intro}). 
The luminosity of the burning WD decreases to a few times 
$10^3$\lo\ at a high temperature of $1-2\times 10^5$\,K 
\citep[e.g.][]{mn94}, which generates a large flux of 
ionising photons ($\sim 10^{47}$\,s$^{-1}$) 
giving rise to an extended nebula. The nebula occupies 
a significant part of the circumbinary environment, whose 
optically thick part is responsible for the pronounced 
sinusoidal variation along the orbit \citep[][]{sk01}. 

In the case of AG~Peg, the nova outburst ceased around 1995, 
when the accretion process switched on and the decline in the 
brightness of the star levelled out showing 
the pronounced sinusoidal variation until the Z~And-type 
outburst in June 2015 \citep[see][]{skopal+17}. 
%
During this $\sim$20 years of quiescence, the high $L_{\rm WD}$ 
of $2-3\times 10^3$\lo\ was sustained by the accretion in the 
stable-burning regime. Subsequently, an increase in the accretion 
rate above the upper limit of the stable-burning ignited the 
Z~And-type outburst. 

The same evolutionary path from symbiotic nova to Z~And-type 
outburst is also indicated for V426~Sge by the same type of 
LC evolution (Fig.~\ref{fig:snovae}) and the parameters of 
the burning WD during its 2018 Z~And-type outburst, which 
are similar to those determined for AG~Peg 
(Sect.~\ref{ss:compagpeg}). 
Thus, V426~Sge also came to be a classical symbiotic star. 

\subsubsection{Connections between basic types of symbiotic stars}
\label{sss:stages}
%
%
The above described evolution of V426~Sge and AG~Peg from their 
symbiotic nova to Z~And-type outbursts suggests that all classical 
symbiotic stars showing the well pronounced sinusoidal variation 
along the orbit have experienced a symbiotic nova outburst in 
the past. 
If we assume that the pronounced wave-like orbitally related 
variability, which develops after a symbiotic nova outburst, 
results from a high $L_{\rm WD}$ generated by the stable 
hydrogen burning on the WD surface, then the following 
connections between basic types of symbiotic stars can be 
considered. 
\begin{enumerate}
\item
If the giant is not capable of sustaining the stable burning, 
the shell burning from prior nova will gradually expire, 
resulting in a gradual fading of the symbiotic activity with 
significantly damped sinusoidal variation. EG~And is an example 
here. The WD will accrete in the nova regime and the system 
will become accetion powered, giving rise to the next symbiotic 
nova outburst in future (Sect.~\ref{sss:progen}). 
As this happens on very different timescales, 
from a few times $10^5$ years to that of human life 
%
\citep[see, e.g.][]{yaron+05}, only the recurrent symbiotic 
novae are regularly observed as a result of this path. 
\item
As long as the giant is capable of fueling the WD at the rate 
required by the stable burning, the pronounced wave-like 
variation will continue at a constant level. Here, all quiet 
symbiotic stars showing this type of variability 
(e.g. SY~Mus, RW~Hya, V443~Her, BD--21$^{\circ}$3873) 
are found at this stage. 
\item
If the accretion rate temporarily exceeds the upper limit of 
the stable burning, the Z~And-type outburst occurs 
\citep[see][ for details]{skopal+17}. 
Here, all the best-studied symbiotic stars whose LCs show 
the pronounced wave-like orbitally related variation during 
the quiescent phase and indicate at least one Z~And-type 
outburst belong to this category. We call them 
{\em classical symbiotic stars}, and Z~And is their prototype. 
\end{enumerate}

\section{Summary}
\label{s:sum}
In this paper we primarily analyse the newly discovered 2018 
outburst of V426~Sge using our high-cadence optical spectroscopy 
and multicolour photometry complemented with {\em Swift}-XRT 
and UVOT observations and NIR $JHKL$ photometry. 
Using the Moscow and Sonneberg photographic plate archives, we 
re-constructed the historical LC of V426~Sge from approximately 
the year 1900. The main results of our analysis can be summarised 
as follows. 
\begin{enumerate}
\item
From around 1900 to 1967, V426~Sge did not show any typical 
symbiotic-like activity. In 1968, V426~Sge experienced the 
symbiotic nova outburst that ceased around 1990. Similarly 
to other symbiotic novae (e.g. AG~Peg and V1329~Cyg), 
a pronounced wave-like orbitally related variation developed 
in its LC, when returning from the nova outburst to quiescent 
phase (Figs.~\ref{fig:hlc}, \ref{fig:phase} and \ref{fig:snovae}). 
The corresponding period of $493.4\pm 0.7$ days is the orbital 
period of the binary (Eq.~(\ref{eq:eph})). 
\item
After around 28 years of quiescence, at the beginning of August 
2018, V426~Sge erupted again showing characteristics of a 
Z~And-type outburst. Around maximum, the WD increased its 
temperature to $\ga 2\times 10^5$\,K, generated luminosity 
of $(7\pm 1)\times 10^{37}\,(d/3.3\,{\rm kpc})^2$\es\ and blew 
a wind at the rate of $\sim 3\times 10^{-6}$\myr, whose 
emission measure of 
$6-7\times 10^{60}\,(d/3.3\,{\rm kpc})^2$\cmt\ was indicated 
by SED models (Figs.~\ref{fig:sedopt} and \ref{fig:lrt}, 
Table~\ref{tab:sed}). 
Quantities of these parameters are very close to those observed 
during the 2015 Z~And-type outburst of the symbiotic nova AG~Peg 
(see Sect.~\ref{ss:compagpeg}). 
\item
Around the middle of February 2019, V426~Sge came to quiescent 
phase, and began to develop wave-like variation along 
its orbit (Sect.~\ref{ss:trans}, Figs.~\ref{fig:trans} and 
\ref{fig:hlc}). 
\item
Our {\em Swift}-UVOT/optical SED model from the quiescent phase 
(on March 3, 2019) revealed a high $L_{\rm WD}$ of $\sim$2300\lo\ 
emitted at a high $T_{\rm WD}$ of $\approx$150\,kK. This suggests 
that the accretor is a low-mass $\sim$0.5\mo\ WD burning hydrogen 
on its surface at steady state conditions 
(Table~\ref{tab:lrt}, Sect.~\ref{ss:compagpeg}). 
\item
Our optical/NIR SED models from the quiescent phase 
revealed that the donor is a normal (probably variable) 
M4.7($\pm$0.2)\,{\small III} red giant characterised with 
$T_{\rm eff}\sim 3\,400$\,K, 
$R_{\rm G}\sim 106$\ro\, and 
$L_{\rm G}\sim 1350$\lo\ 
(Sect.~\ref{ss:rg}, Fig.~\ref{fig:sedir}). 
\item
The cases of V426~Sge and AG~Peg, where the symbiotic nova 
outburst was followed by the Z~And-type outburst and similar 
evolution in their LCs, lead us to sketch out some possible 
connections between the basic types of symbiotic stars, as follows: 

Symbiotic nova outbursts result from prolonged accretion by 
the WD in an accretion-powered symbiotic star 
(Sect.~\ref{sss:progen}). 
During the transition from the nova outburst to quiescent phase, 
a pronounced sinusoidal variation along the orbit develops in 
the LC (Sect.~\ref{sss:outobs}, Fig.~\ref{fig:snovae}). 
The following evolution depends on the accretion rate by the WD 
(Sect.~\ref{sss:stages}). 

(i) 
If the giant is capable of sustaining stable hydrogen burning 
on the WD surface, the wave-like variability continues at 
a constant level: we observe quiet symbiotic stars. 

(ii) 
If the accretion exceeds the upper limit for the stable 
burning, a Z~And-type outburst occurs and the system 
becomes a classical symbiotic star. 

(iii) 
If the giant is not capable of sustaining the stable burning, 
the system becomes accretion-powered, and the progenitor 
of a future symbiotic nova outburst. 
\end{enumerate}

\begin{acknowledgements}
We thank the anonymous referee for constructive comments. 
Hee-Won Lee is thanked for a discussion on the anomalous 
flux ratio $F(6825)/F(7082)$ in Sect.~\ref{ss:raman}. 
Matej Seker\'a\v{s}, Theodor Pribulla, Zolt\'an Garai, 
Peter Sivani\v{c} and Alisa Shchurova are thanked for their 
assistance in acquiring some spectra at the Skalnat\'e Pleso 
and Star\'a Lesn\'a observatories. 
Some optical spectra presented in this paper were obtained 
within the {\it Astronomical Ring for Access to Spectroscopy 
(ARAS)}, an initiative promoting cooperation between professional 
and amateur astronomers in the field of spectroscopy, coordinated 
by Francois Teyssier. Here, we thank contributions by 
Paolo Berardi, Stephane Charbonnel, Lorenzo Franco, Tim Lester, 
Forrest Sims and Umberto Sollecchia. 
This work was supported by the Slovak Research and Development 
Agency under the contract No. APVV-15-0458 and by the Slovak 
Academy of Sciences grant VEGA No. 2/0008/17. 
This work was partially supported by the Program of Development 
of M. V. Lomonosov Moscow State University `Leading Scientific 
Schools', project `Physics of Stars, Relativistic Objects and Galaxies'. 
The spectrograph used at the Astronomical Observatory on the Kolonica 
Saddle was purchased from the Polish NCN grant 2015/18/A/ST9/00578. 
NM acknowledges financial support from ASI-INAF contract 
No. 2017-14-H.0. 
\end{acknowledgements}
%
%

%

\appendix

\section{Tables with optical photometry of V426~Sge}
\label{app:tab}
%
%
\begin{table*}
\caption[]{$UBVR_{\rm C}I_{\rm C}$ photometry of V426~Sge 
  obtained at the Star\'a Lesn\'a and Kolonica Saddle observatories. 
  The full table is available at the CDS.}
\begin{center}
\begin{tabular}{ccccccccccccc}
\hline
\hline
\noalign{\smallskip}
HJD        &       UT           & 
\multicolumn{2}{c}{$U$}         &
\multicolumn{2}{c}{$B$}         &
\multicolumn{2}{c}{$V$}         & 
\multicolumn{2}{c}{$R_{\rm C}$} &
\multicolumn{2}{c}{$I_{\rm C}$} & 
Tel.~ID \\
+2\,458\,000 & yyyy mm dd.ddd   &  mag & error & 
      mag & error & mag & error &  mag & error & mag & error &  \\
\noalign{\smallskip}
\hline
\noalign{\smallskip}
 343.357& 2018 08 12.857& 10.702& 0.055&  11.496& 0.030&  10.938&   0.014&
                          10.060& 0.036&   9.220& 0.052&  G2 \\
 344.342& 2018 08 13.842& 10.701& 0.038&  11.481& 0.025&  10.896&   0.007&
                           9.997& 0.029&   9.145& 0.046&  G2 \\
 344.348& 2018 08 13.848&       &      &  11.506& 0.025&  10.935&   0.007&
                          10.028& 0.028&   9.167& 0.038&  G1 \\
 347.337& 2018 08 16.837& 10.691& 0.039&  11.511& 0.026&  10.906&   0.009&
                           9.973& 0.030&   9.132& 0.047&  G2 \\
 348.396& 2018 08 17.896&       &      &  11.487& 0.026&  10.872&   0.008&
                           9.939& 0.029&        &      &  G1 \\
 348.406& 2018 08 17.906& 10.641& 0.039&  11.483& 0.027&  10.861&   0.012&
                           9.926& 0.031&   9.095& 0.047&  G2 \\
\noalign{\smallskip}
\hline
\end{tabular}
\end{center}
\label{tab:ubvri}
\end{table*}
%
%
\begin{table*}
\caption[]{$BVR_{\rm C}I_{\rm C}$ photometry of V426~Sge 
 obtained within the ANS collaboration. The full table is 
 available at the CDS.}
\begin{center}
\begin{tabular}{ccccccccccc}
\hline
\hline
\noalign{\smallskip}
HJD        &       UT          & \multicolumn{2}{c}{$B$}  &
\multicolumn{2}{c}{$V$} & \multicolumn{2}{c}{$R_{\rm C}$} &
\multicolumn{2}{c}{$I_{\rm C}$} & Tel.~ID \\
+2\,458\,000& yyyy mm dd.ddd  &  mag &  error & mag &  error &  
                                 mag &  error & mag &  error &  \\
\noalign{\smallskip}
\hline
\noalign{\smallskip}
341.33743 & 2018 08 10.837 & 11.459 & 0.010 & 10.926 & 0.009 &
                             10.101 & 0.013 &  9.199 & 0.010 & 0310 \\
342.46219 & 2018 08 11.962 & 11.414 & 0.010 & 10.872 & 0.010 &
                             10.042 & 0.009 &  9.153 & 0.011 & 0310 \\
345.52123 & 2018 08 15.021 & 11.453 & 0.007 & 10.771 & 0.004 &
                                    &       &  9.092 & 0.009 & 2100 \\
345.55826 & 2018 08 15.058 & 11.444 & 0.008 & 10.843 & 0.007 &
                             10.007 & 0.009 &  9.124 & 0.011 & 0310 \\
347.51701 & 2018 08 17.017 & 11.504 & 0.005 & 10.816 & 0.004 &
                                    &       &  9.085 & 0.009 & 2100 \\
348.40872 & 2018 08 17.909 & 11.476 & 0.011 & 10.873 & 0.010 &
                              9.985 & 0.013 &  9.112 & 0.012 & 0310 \\
\noalign{\smallskip}
\hline
\end{tabular}
\end{center}
\label{tab:asn}
\end{table*}
%
%
%
\begin{table*}
\caption[]{Photographic magnitudes of V426~Sge 
           obtained from archives. The full table is 
           available at the CDS.}
\begin{center}
\begin{tabular}{ccccc}
\hline
\hline
\noalign{\smallskip}
    HJD       &   UT  & \multicolumn{2}{c}{$m_{\rm pg}$} & Archive  \\
+2\,400\,000  & yyyy mm dd.ddd  &   mag &  error         &         \\
\noalign{\smallskip}
\hline
\noalign{\smallskip}
14578.264 & 1898 10 15.764 & $>$13.8  & not visible &  M \\
14908.317 & 1899 09 10.817 & $>$12.6  & not visible &  M \\
14909.375 & 1899 09 11.875 &    13.62 & 0.16        &  M \\
15614.344 & 1901 08 17.844 &    13.45 & 0.18        &  M \\
18950.246 & 1910 10 05.746 &    13.86 & 0.18        &  M \\
18973.220 & 1910 10 28.720 &    13.09 & 0.16        &  M \\
28751.397 & 1937 08 05.897 &    13.68 & 0.17        &  M \\
\noalign{\smallskip}
\hline
\end{tabular}
\end{center}
\label{tab:mpg}
\end{table*}

\section{Table of the used line fluxes}
\label{app:fluxes}
%
%
\begin{table*}
\caption[]{
Dereddened fluxes for the \ion{He}{ii}\,4686\,\AA, \hb\ and Raman 
scattered \ion{O}{vi}\,6825\,\AA\ lines in units of 10$^{-13}$\ecs. 
}
\begin{center}
\begin{tabular}{cccccccc}
\hline
\hline
\noalign{\smallskip}
UT                        &
HJD                       & 
$F_{4686}$                & 
error                     &
$F_{{\rm H}\beta}$        &
error                     &
$F_{\rm Raman}$           &
error                     \\
yyyy/mm/dd.ddd  &
+2\,400\,000    &
                &
                &
                &
                &
                &
                \\
\noalign{\smallskip}
\hline
\noalign{\smallskip}
 20180810.845&  58341.345& 162.5& 12.2&  125.4&  9.4&  0.00&  --  \\ 
 20180811.826&  58342.326& 165.6& 14.5&  142.3& 12.5&  0.00&  --  \\  
 20180812.872&  58343.372&   -- &  -- &    -- &  -- &  0.00&  --  \\ 
 20180813.814&  58344.314& 210.6& 20.0&  271.2& 25.8&  1.72& 0.58 \\  
 20180813.827&  58344.327& 239.4& 22.0&  245.8& 22.5&  1.72& 0.58 \\  
 20180815.853&  58346.353& 183.4& 19.0&  146.0& 15.2&  0.92& 0.58 \\  
 20180817.070&  58347.570& 144.2& 18.0&  128.2& 16.0&  1.25& 0.77 \\  
 20180817.896&  58348.396& 135.0& 16.0&  174.4& 17.0&   -- &  --  \\
 20180818.823&  58349.323& 212.2& 20.0&  237.4& 22.4&  0.95& 0.67 \\  
 20180819.842&  58350.342& 228.1& 23.0&  246.8& 24.9&  2.26& 0.86 \\  
 20180820.060&  58350.560& 192.6& 20.2&    -- &  -- &  2.23& 0.67 \\  
 20180820.868&  58351.369& 205.0& 27.0&  334.5& 44.1&  4.21& 0.77 \\  
 20180822.833&  58353.333&   -- &  -- &    -- &  -- &  6.42& 1.54 \\  
 20180822.840&  58353.340& 221.0& 16.0&  379.2& 24.4&  6.06& 1.63 \\  
 20180827.796&  58358.296& 206.0& 22.5&  369.5& 40.3&  6.99& 0.77 \\  
 20180907.867&  58369.367& 187.5& 20.0&  419.1& 44.8& 12.74& 1.25 \\  
 20180919.821&  58381.321& 238.6& 17.5&  511.6& 37.6& 16.00& 1.63 \\  
 20180920.837&  58382.337& 196.4& 18.5&  498.2& 47.0& 18.78& 1.92 \\  
 20180920.882&  58382.382&   -- &  -- &    -- &  -- & 13.60& 1.44 \\  
 20181004.840&  58396.340& 136.0& 14.0&  386.2& 40.0& 19.80& 2.00 \\
 20181005.724&  58397.224& 166.1& 18.0&  346.8& 37.5& 21.75& 2.20 \\  
 20181010.788&  58402.288& 190.1& 19.0&  364.1& 36.0& 21.17& 2.11 \\  
 20181011.835&  58403.335& 108.2& 22.0&  261.5& 50.0& 16.50& 1.70 \\
 20181016.713&  58408.213& 156.5& 12.0&  422.1& 32.5& 17.34& 1.72 \\  
 20181018.797&  58410.297& 122.8& 13.0&  200.3& 21.0& 15.52& 1.53 \\  
 20181020.757&  58412.257&  91.9& 11.0&  171.1& 20.5& 16.00& 1.34 \\  
 20181108.741&  58431.241& 124.6& 19.0&    -- &  -- & 16.19& 1.63 \\  
 20181110.773&  58433.273&  71.3& 10.0&  118.6& 20.0& 13.20& 1.50 \\
 20181112.803&  58435.303&  58.2&  4.8&  132.5& 11.0& 14.85& 1.53 \\  
 20181123.729&  58446.229&  77.5&  6.5&  142.7& 12.0& 10.83& 1.15 \\  
 20181217.701&  58470.201&  46.9&  5.0&   78.4&  8.3&  8.54& 0.86 \\  
\noalign{\smallskip}
\hline
\noalign{\smallskip}
         \multicolumn{8}{c}{Quiescent phase} \\
\noalign{\smallskip}
\hline
\noalign{\smallskip}
 20190623.872$^{a}$&  58658.372&  29.2&  3.5&   24.7&  3.0&  3.6& 0.7 \\
 20190625.931$^{a}$&  58660.431&  28.8&  3.5&   23.3&  3.0&  3.8& 0.7 \\
 20190819.832$^{a}$&  58715.332&  32.2&  4.0&   25.2&  3.0&  3.2& 0.6 \\
 20190907.191$^{a}$&  58733.691&  35.8&  4.0&   29.6&  3.0&  3.6& 0.4 \\
 20191005.112$^{a}$&  58761.612&  36.5&  4.0&   38.1&  4.0&  3.7& 0.5 \\
 20191113.760&  58801.260&  48.7&  2.0&   59.3&  3.0&  5.4& 0.5 \\
\noalign{\smallskip}
\hline
\end{tabular}
\end{center}
{\bf Notes.} 
$^{(a)}$~low-resolution spectrum 
\label{tab:fluxes}
\end{table*}

\begin{thebibliography}{}
\bibliographystyle{aa}
%
\bibitem[Allen(1980)]{allen80}
         Allen, D. A. 1980, MNRAS, 192, 521
\bibitem[Ambartsumyan(1932)]{ambartsumyan}
         Ambartsumyan, V. A. 1932, Pulkovo Obs. Circ., 4, 8
\bibitem[Arkhipova et al.(1995)]{arkhipova+95}
         Arkhipova, V. P., Esipov, V. F., \& Ikonnikova, N. P. 
         1995, Astronomy Letters, 21, 391
\bibitem[Bacher et al.(2005)]{bacher+05}
         Bacher, A., Kimeswenger, S., \& Teutsch, P. 
         2005, MNRAS, 362, 542
\bibitem[Bailer-Jones et al.(2018)]{bailer+18}
         Bailer-Jones, C. A. L., Rybizki, J., Fouesneau, M., 
         Mantelet, G., \& Andrae, R. 
         2018, AJ, 156, 58
\bibitem[Baudrand \& Bohm(1992)]{baud+92}
         Baudrand, J., \& Bohm, T. 
         1992, A\&A, 259, 711 
\bibitem[Belczy\'nski et al.(2000)]{belczynski+00}
         Belczy\'nski, K., Miko\l ajewska, J., Munari, U., 
         Ivison, R., \& J.; Friedjung, M. 2000
         A\&AS, 146, 407
\bibitem[Bessel(1979)]{bessel79}
         Bessel, M. S. 1979, PASP, 91, 589
\bibitem[Birriel et al.(1998)]{birriel+98}
         Birriel, J. J., Espey, B. R., \& Schulte-Ladbeck, R. E. 
         1998, ApJ, 507, L75
\bibitem[Birriel et al.(2000)]{birriel+00}
         Birriel, J. J., Espey, B. R., \& Schulte-Ladbeck, R. E. 
         2000, ApJ, 545, 1020
\bibitem[Blackburn(1995)]{1995ASPC...77..367B}
         Blackburn, J. K. 1995, in: Astronomical Data Analysis 
         Software and Systems IV, eds. R. A. Shaw, H. E. Payne, 
         \& J. J. E. Hayes, ASP Conf. Ser. (San Francisco: ASP), 
         77, 367
\bibitem[Bode \& Evans(2008)]{bode08}
         Bode, M. F., \& Evans, A., 2008, Classical Novae, 
         second edition, (Cambridge: Cambridge University Press)
\bibitem[Boyarchuk et al.(1966)]{boy+66}
         Boyarchuk, A.~A., Esipov, V.~F., \& Moroz, V.~I. 
         1966, Soviet Astronomy, 10, 331
\bibitem[Boyarchuk (1967)]{boy67}
         Boyarchuk, A. A. 1967, Soviet Astronomy, 11, 8
\bibitem[Burrows et al.(2005)]{burrows+05}
         Burrows, D. N., Hill, J. E., Nousek, J. A., et al. 
         2005, Space Sci. Rev., 120, 165
\bibitem[Cardelli et al.(1989)]{c+89}
         Cardelli, J. A., Clayton, G. C., \& Mathis, J. S. 
         1989, ApJ, 345, 245
\bibitem[Carikov\'a \& Skopal(2012)]{cs12}
         Carikov\'a, Z., \& Skopal, A. 2012, A\&A, 548, A21
\bibitem[Ciroi et al.(2014)]{ciori+14}
         Ciroi, S., Di Mille, F., Rafanelli, P., 
         Cracco, V. \& La Mura, G. 2014, CoSka, 43, 362
\bibitem[C\'uneo et al.(2018)]{cuneo+18}
         C\'uneo, V. A., Kenyon, S. J., G\'omez, M. N., et al.  
         2018, MNRAS, 479, 2728
\bibitem[Chochol et al.(1999)]{chochol+99}
         Chochol, D., Andronov, I. L., Arkhipova, V. P., et al. 
         1999, CoSka, 29, 31
\bibitem[D\"ohring et al.(2019)]{dohring+19}
         D\"ohring, T., Pribulla, T., Kom\v{z}\'{\i}k, R., 
         Mann, M., Sivani\v{c}, P., \& Stollenwerk, M. 
         2019, CoSka, 49, 154 
\bibitem[Dumm \& Schild(1998)]{ds98}
         Dumm, T., \& Schild, H. 1998, \na, 3, 137
\bibitem[Ferraz-Mello(1981)]{ferraz-mello81}
         Ferraz-Mello, S. 1981, AJ, 86, 619
\bibitem[Fern\'andez-Castro et al.(1995)]{f-c+95}
         Fern\'andez-Castro, T., Gonz\'alez-Riestra, R., 
         Cassatella, A., Taylor, A. R., Seaquist, E. R. 
         1995, ApJ, 442, 366
\bibitem[Fluks et al.(1994)]{fluks+94}
         Fluks, M. A., Plez, B., The, P. S., de Winter, D.,
         Westerlund, B. E., \& Steenman, H. C.
         1994, A\&AS, 105, 311
\bibitem[Foster(1995)]{foster95}
         Foster, G. 1995, AJ, 109, 1889
\bibitem[Fujimoto(1982a)]{fujimoto82a}
         Fujimoto, M. Y. 1982a, ApJ, 257, 752
\bibitem[Fujimoto(1982b)]{fujimoto82b}
         Fujimoto, M. Y. 1982b, ApJ, 257, 767
\bibitem[Gaia Collaboration(2018)]{gaia+18}
         Gaia Collaboration, Brown, A. G. A., Vallenari, A., et al.  
         2018, A\&A, 616, A1
\bibitem[Gehrels et al.(2004)]{gehrels+04}
         Gehrels, N., Chincarini, G., Giommi, P., et al. 
         2004, ApJ, 611, 1005
\bibitem[Green et al.(2018)]{green+18}
         Green, G. M., Schlafly, E. F., Finkbeiner, D., et al. 
         2018, MNRAS, 478, 651
\bibitem[Greiner et al.(1997)]{greiner+97}
         Greiner, J. Bickert, K., Luthardt, R., Viotti, R., 
         Altamore, A. Gonzalez-Riestra, R., Stencel, R. E. 
         1997, A\&A, 322, 576
\bibitem[Gurzadyan(1997)]{gurz97}
         Gurzadyan, G. A., 1997, The Physics and Dynamics
         of Planetary Nebulae. Springer-Verlag, Berlin, p. 105
\bibitem[Hachisu et al.(1996)]{hachisu+96}
         Hachisu, I., Kato, M., \& Nomoto, K. 
         1996, ApJL, 470, L97
\bibitem[Harman \& Seaton(1966)]{harman+66}
         Harman, R. J., \& Seaton, M. J. 
         1966, MNRAS, 132, 15
\bibitem[Hauschildt et al.(1999)]{h+99}
         Hauschildt, P. H., Allard, F., Ferguson, J., 
         Baron, E., \& Alexander, D. R. 1999, ApJ, 525, 871
\bibitem[Henden \& Kaitchuck(1982)]{hk82}
         Henden, A. A., \& Kaitchuck, R. H. 1982,
         Astronomical Photometry, 
         (New York: Van Nostrand Reinhold Company), 50
\bibitem[Henden et al.(2016)]{henden+16}
         Henden, A. A., Templeton, M., Terrell, D., 
         Smith, T. C., Levine, S. \& Welch, D. L. 
         2016, VizieR On-line Data Catalog: II/336. 
\bibitem[Hill et al.(2004)]{hill+04}
         Hill, J. E., Burrows, D. N., Nousek, J. A., et al. 
         2004, Proc. SPIE, 5165, 217
\bibitem[Hoffleit(1968)]{hoff68}
         Hoffleit, D. 1968, Irish Astr. J., 8, 149
\bibitem[Hric \& Urban(1991)]{hric+91}
         Hric, L., \& Urban, Z. 1991, IBVS No. 3683, 1
\bibitem[Hric et al.(1994)]{hric+94}
         Hric, L., Skopal, A., Chochol, D., at al. 
         1994, CoSka, 24, 31
\bibitem[Hummer \& Seaton(1964)]{hs64}
         Hummer, D. G., \& Seaton, M. J. 1964, MNRAS, 127, 217
\bibitem[Hummer \& Storey(1987)]{hs87}
         Hummer, D. G. \& Storey, P. J. 1987, MNRAS, 224, 801
\bibitem[Iijima(1981)]{iijima}
         Iijima, T. 1981, in: Photometric and Spectroscopic 
         Binary Systems, Proceedings of the NATO Advanced Study 
         Institute, E. B. Carling and Z. Kopal eds. 
         Dordrecht: D. Reidel Publishing Co., p. 517
\bibitem[Ikonnikova et al.(2019)]{ikonnikova+19}
         Ikonnikova, N. P., Burlak, M. A., Arkhipova, V. P., 
         \& Esipov, V. F. 2019, Astronomy Letters, 45, 217
\bibitem[Jayasinghe et al.(2018)]{2018MNRAS.477.3145J}
         Jayasinghe, T., Kochanek, C. S., Stanek, K. Z., et al. 
         2018, MNRAS, 477, 3145
\bibitem[Kaler \& Jacoby(1989)]{kj89}
         Kaler, J. B., \& Jacoby, G. H. 1989, ApJ, 345, 871
\bibitem[Kato \& Hachisu(1994)]{k+h94}
         Kato, M., \& Hachisu, I. 1994, ApJ, 437, 802
\bibitem[Kazarovets et al.(2019)]{kazarovets+19}
         Kazarovets, E. V., Samus, N. N., Durlevich, O. V., 
         Khruslov, A. V., Kireeva, N. N., \& Pastukhova, E. N. 
         2019, Peremennye Zvezdy, 39, no. 3 
\bibitem[Kenyon(1986)]{kenyon86}
         Kenyon, S. J. 1986, The symbiotic stars, 
         (Cambridge: Cambridge University Press)
\bibitem[Kenyon \& Webbink (1984)]{kw84}
         Kenyon, S. J., \& Webbink, R. F. 1984, ApJ, 279, 252
\bibitem[Kenyon et al.(1993)]{kenyon+93}
         Kenyon, S. J., Miko\l ajewska, J., Miko\l ajewski, M., 
         Polidan, R. S., Slovak, M. H. 
         1993, AJ, 106, 1573
\bibitem[Kenyon \& Garcia(2016)]{kenyon+16}
         Kenyon, S. J., \& Garcia, M. R. 2016, AJ, 152, 1 
\bibitem[Kohoutek \& Wehmeyer(1999)]{1999A&AS..134..255K}
         Kohoutek, L., \& Wehmeyer, R. 
         1999, A\&ASS, 134, 255
\bibitem[Kudzej \& Dubovsk\'y(2014)]{kudub14}
         Kudzej, I., \& Dubovsk\'y, P. 2014, CoSka, 43, 429 
\bibitem[Kudzej et al.(2019)]{kudzej+19}
         Kudzej, I., Savanevych, V. E., Briukhovetskyi, O. B., et al. 
         2019, AN, 340, 68 
\bibitem[Lafler \& Kinman(1965)]{lafler+65}
         Lafler, J., \& Kinman, T. D. 
         1965, ApJS, 11, 216
\bibitem[Lamers \& Cassinelli(1999)]{lamcass99}
         Lamers, H. J. G. L. M., \& Cassinelli, J. P. 
         1999, Introduction to stellar winds,
         Cambridge University Press
\bibitem[Lee et al.(2016)]{lee+16}
         Lee, Y.-M., Lee, D.-S., Chang, S.-J., et al. 
         2016, ApJ, 833:75 
\bibitem[Lopes de Oliveira et al.(2018)]{lopes+18}
         Lopes de Oliveira, R., Sokoloski, J. L., 
         Luna, G. J. M., Mukai, K., \& Nelson, T. 
         2018, ApJ, 864, 46
\bibitem[Luna et al.(2013)]{luna+13}
         Luna, G. J. M., Sokoloski, J. L., Mukai, K., \& Nelson, T. 
         2013, A\&A, 559, A6
\bibitem[Matthews \& Karovska(2006)]{matt+karov06}
         Matthews, L.~D., \& Karovska, M. 
         2006, ApJ, 637, L49
\bibitem[Meinunger(1983)]{meinunger83}
         Meinunger, L. 
         1983, Mitt. Ver\"anderliche Sterne, 9, 92 
\bibitem[Meinunger(1979)]{mein79}
         Meinunger, L. 1979, IBVS No. 1611
\bibitem[Miko\l ajewska et al.(1995)]{mika+95}
         Miko\l ajewska, J., Kenyon, S. J., Miko\l ajewski, M, 
         Garcia, M. R., \& Polidan, R. S. 1995, AJ, 109, 1289
\bibitem[Mukai(2017)]{mukai17}
         Mukai, K. 2017, PASP 129, 062001 
\bibitem[Mukai et al.(2016)]{mukai+16}
         Mukai, K., Luna, G. J. M. Cusumano, G., et al. 
         2016, MNRAS, 461, L1
\bibitem[Munari(1993)]{munari93}
         Munari, U. 1993, A\&A, 273, 425
\bibitem[Munari(1997)]{munari97}
         Munari, U. 1997, in Physical Processes in Symbiotic
         Binaries, ed. J. Miko\l ajewska (Warsaw: Copernicus
         Foundation for Polish Astronomy), 37
\bibitem[Munari(2019)]{munari19}
         Munari, U. 2019, in The Impact of Binary Stars 
         on Stellar Evolution, eds. G. Beccari and M.J. Boffin, 
         Cambridge Astrophysical Series vol. 54 
         (Cambridge: CUP), 77 
\bibitem[Munari \& Lattanzi(1992)]{1992PASP..104..121M}
         Munari U., \& Lattanzi, M. G. 1992, PASP, 104, 121
\bibitem[Munari et al.(2012a)]{2012BaltA..21...13M}
         Munari U., et al., 2012a, BaltA, 21, 13 
\bibitem[Munari \& Moretti(2012b)]{2012BaltA..21...22M}
         Munari U., \& Moretti S. 2012b, BaltA, 21, 22 
\bibitem[Munari et al.(2016)]{munari+16}
         Munari, U. Dallaporta, S., Cherini, G. 
         2016, New Astron., 47, 7
\bibitem[Munari et al.(2018)]{2018ATel11937....1M}
         Munari, U., Dallaporta, S., Valisa, P., et al. 
         2018, ATel, 11937
\bibitem[M\"urset et al. (1991)]{mu+91}
         M\"urset, U., Nussbaumer, H., Schmid, H. M., \& Vogel, M.
         1991, A\&A, 248, 458
\bibitem[M\"urset et al.(1995)]{murset+95}
         M\"urset, U., Jordan, S., Walder, R. 
         1995, A\&A, 297, L87
\bibitem[M\"urset et al.(1997)]{murset+97}
         M\"urset, U., Wolff, B., Jordan, S. 
         1997, A\&A, 319, 201
\bibitem[M\"urset \& Nussbaumer(1994)]{mn94}
         M\"urset, U. \& Nussbaumer, H. 
         1994, A\&A, 282, 586
\bibitem[M\"urset \& Schmid(1999)]{ms99}
         M\"urset, U., \& Schmid, H. M.
         1999, A\&AS, 137, 473
\bibitem[Nomoto et al.(2007)]{nomoto+07}
         Nomoto, K., Saio, H., Kato, M., \& Hachisu, I. 
         2007, ApJ, 663, 1269
\bibitem[Nussbaumer \& Vogel (1987)]{nussvog87}
         Nussbaumer, H., Vogel, M. 1987, A\&A, 182, 51
\bibitem[Paczy\'nski \& \.{Z}ytkow(1978)]{paczyt78}
         Pac\'zynski, B., \& \.{Z}ytkow, A. N.
         1978, ApJ, 222, 604
\bibitem[Paczy\'nski \& Rudak(1980)]{pacrud80}
         Paczy\'nski, B., \& Rudak, R. 1980, A\&A, 82, 349
\bibitem[Parimucha et al.(2002)]{parimucha+02}
         Parimucha, \v{S}., Chochol, D., Pribulla, T., 
         Buson, L.~M., \& Vittone, A.~A. 2002, 
         A\&A, 391, 999
\bibitem[Paunzen \& Vanmunster(2016)]{paunzen+16}
         Paunzen, E., \& Vanmunster, T. 
         2016, AN, 337, 239
\bibitem[Poole et al.(2008)]{poole+08}
         Poole, T. S., Breeveld, A. A., Page, M. J., et al. 
         2008, MNRAS, 383, 627
\bibitem[Predehl \& Schmitt(1995)]{predehl+95}
         Predehl, P., \& Schmitt, J. H. M. M. 1995, A\&A, 293, 889 
\bibitem[Pribulla et al.(2015)]{pribulla+15}
         Pribulla, T., Garai, Z., Hamb\'alek, \v{L}, et al. 
         2015, AN, 336, 682
\bibitem[Pringle(1981)[]{pringle81}
         Pringle, J. E. 
         1981, Ann. Rev. Astron. Astrophys., 19, 137
\bibitem[Ramsay et al.(2016)]{ramsay+16}
         Ramsay, G., Sokoloski, J. L., Luna, G. J. M., 
         Nu\~{n}ez, N. E. 2016, MNRAS, 461, 3599
\bibitem[Roming et al.(2005)]{roming+05}
         Roming, P. W. A., Kennedy, T. E., Mason, K. O., et al. 
         2005, Space Sci. Rev., 120, 95
\bibitem[Schmid \& Schild(2002)]{schmid+schild02}
        Schmid, H. M., \& Schild, H. 
        2002, A\&A, 395, 117
\bibitem[Schmid et al.(1999)]{schmid+99}
         Schmid, H. M., Krautter, J., Appenzeller, I., et al. 
         1999, A\&A, 348,950
\bibitem[Seaquist et al. (1984)]{stb}
         Seaquist, E. R., Taylor, A. R., \& Button, S.
         1984, ApJ, 284, 202
\bibitem[Seker\'a\v{s} et al.(2019)]{sekeras+19}
         Seker\'a\v{s}, M., Skopal, A., Shugarov, S. Yu., et al. 
         2019, CoSka, 49, 19
\bibitem[Shappee et al.(2014)]{2014ApJ...788...48S}
         Shappee, B. J., Prieto, J. L., Grupe, D., et al. 
         2014, ApJ, 788, 48
\bibitem[Shen \& Bildsten(2007)]{shen+bild07}
         Shen, K. J., \& Bildsten, L. 
         2007, ApJ, 660, 1444
\bibitem[Shenavrin et al.(2011)]{shenavrin+11}
         Shenavrin, V. I., Taranova, O. G., \& Nadzhip, A. E.  
         2011, Astron. Rep. 55, 31
\bibitem[Shugarov et al.(2003)]{shugarov+03}
         Shugarov, S., Pavlenko, E., \& Malanushenko, V. 2003, 
         in: Symbiotic Star Probing Stellar Evolution, 
         R. L. M. Corradi, J. Miko\l ajewska, \& T. J. Mahoney, eds., 
         ASP Conf. Ser. 303, San Francisco: ASP, p. 87
\bibitem[Sion et al.(2012)]{sion+12}
         Sion, E. M., Moreno, J., Godon, P., Sabra, B., 
         \& Miko\l ajewska, J. 2012, AJ, 144, 171
\bibitem[Sion et al.(2019)]{sion+19}
         Sion, E. M., Godon, P., Miko\l ajewska, J., \& Katynski, M. 
         2019, ApJ, 874, id. 178 
\bibitem[Siviero(2014)]{siviero14}
         Siviero, A. 2014, CoSka, 43, 301
\bibitem[Skopal(1998)]{sk98}
         Skopal, A. 1998, A\&A, 338, 599
\bibitem[Skopal(2001)]{sk01}
         Skopal, A. 2001, A\&A, 366, 157
\bibitem[Skopal(2005a)]{sk05a}
         Skopal, A. 2005a, A\&A, 440, 995
\bibitem[Skopal(2005b)]{sk05b}
         Skopal, A. 2005b, in: The Astrophysics of Cataclysmic 
         Variables and Related Objects, J.-M. Hameury, 
         \& J.-P. Lasota, eds., ASP Conf. Ser. 330, 
         San Francisco: ASP, p. 463
\bibitem[Skopal(2006)]{sk06}
         Skopal, A. 2006, A\&A, 457, 1003
\bibitem[Skopal(2007)]{sk07}
         Skopal, A. 2007, New Astron., 12, 597
\bibitem[Skopal(2015)]{sk15}
         Skopal, A. 2015, New Astronomy, 36, 116
\bibitem[Skopal(2019)]{sk19}
         Skopal, A. 2019, ApJ, 878, 28
\bibitem[Skopal et al.(2009)]{skopal+09}
         Skopal, A., Seker\'a\v{s}, M., Gonz\'alez-Riestra, R.,
        \& Viotti, R. F. 2009, A\&A, 507, 1531
\bibitem[Skopal et al.(2011)]{skopal+11}
         Skopal, A., Tarasova, T. N., Carikov\'a, Z., et al. 
         2011, A\&A, 536, A27
\bibitem[Skopal et al.(2014)]{skopal+14}
         Skopal, A., Drechsel, D., Tarasova, T., et al. 
         2014, A\&A, 569, A112
\bibitem[Skopal et al.(2017)]{skopal+17}
         Skopal, A., Shugarov, S. Yu., Seker{\'a}{\v s}, M., et al. 
         2017, A\&A, 604, A48
\bibitem[Skopal et al.(2018)]{skopal+18}
         Skopal, A., Tarasova, T. N., Wolf, M., 
         Dubovsk\'y, P. A., \& Kudzej, I. 
         2018, ApJ, 858:120 
\bibitem[Smith(1980)]{smith80}
         Smith, S. E. 1980, ApJ, 237, 831
\bibitem[Sokoloski et al.(2017)]{sokoloski+17}
         Sokoloski, J. L., Lawrence, S., Crotts, A. P. S. \& Mukai, K. 
         2017, arXiv170205898
\bibitem[Sokolovsky et al.(2016)]{sokolovsky+16}
         Sokolovsky, K.~V., Kolesnikova, D.~M., Zubareva, A.~M., 
         Samus, N.~N., \& Antipin, S.~V. 
         2016, arXiv e-prints no. 1605.03571 
\bibitem[Starrfield et al.(2016)]{starrfield+16}
         Starrfield, S., Iliadis, Ch., \& Hix, W. R. 
         2016, PASP, 128, 051001
\bibitem[Teyssier(2019)]{teyssier19}
         Teyssier, F. 2019, CoSka, 49, 217
\bibitem[Teyssier et al.(2019)]{teyssier+19}
         Teyssier, F., Boyd, D., Guarro, J., et al. 
         2019, Eruptive Stars Information Letter, 41, 2
\bibitem[Tomov et al.(2016)]{tomov+16}
         Tomov, T. V., Stoyanov, K. A., \& Zamanov, R. K. 
         2016, MNRAS, 462, 4435
\bibitem[Torres et al.(2012)]{torres+12}
         Torres, A. F., Kraus, M., Cidale, L. S., Barb\'a, R., 
         Borges Fernandes, M., \& Brandi, E. 
         2012, MNRAS, 427, L80
\bibitem[Wheatley et al. (2003)]{wheatley+03}
         Wheatley, P. J., Mukai, K., \& de Martino, D. 
         2003, MNRAS, 346, 855
\bibitem[Yaron et al.(2005)]{yaron+05}
         Yaron, O., Prialnik, D., Shara, M. M., \& Kovetz, A. 
         2005, ApJ, 623, 398
%
\end{thebibliography}
\end{document}